\documentclass{article}
\usepackage{iclr2027_conference,times}
\usepackage[T1]{fontenc}
\usepackage{mathptmx}

\usepackage{amsmath,amsfonts,bm}

\def\eqref#1{equation~\ref{#1}}

\def\1{\bm{1}}

\DeclareMathAlphabet{\mathsfit}{\encodingdefault}{\sfdefault}{m}{sl}
\SetMathAlphabet{\mathsfit}{bold}{\encodingdefault}{\sfdefault}{bx}{n}

\usepackage{hyperref}
\usepackage{url}
\definecolor{linkgreen}{RGB}{0,170,0}
\hypersetup{colorlinks=true, breaklinks=true, linkcolor=linkgreen, citecolor=linkgreen, urlcolor=linkgreen}
\usepackage{booktabs}
\usepackage{enumitem}
\usepackage{longtable}
\usepackage{multirow}
\usepackage{float}
\usepackage{graphicx}
\usepackage{wrapfig}
\usepackage{needspace}
\usepackage{afterpage}
\usepackage{xcolor}
\usepackage{colortbl}
\definecolor{tabgreen}{HTML}{87D787}
\usepackage{amsmath,amssymb}

\newcommand{\editm}{\textsuperscript{E}}
\newcommand{\presm}{\textsuperscript{P}}
\newcommand{\bothm}{\textsuperscript{E+P}}

\title{ProCraft: Instruction-Driven Music Production from Free-Form Text}

\author{Yuan-Chiao Cheng\\
Georgia Institute\\
of Technology
\And
Yi-Hsuan Yang\\
AI-CoRE, National\\
Taiwan University
\And
Vincent K.M. Cheung\\
Sony Computer Science\\
Laboratories, Tokyo}

\iclrfinalcopy

\begin{document}

\maketitle
\lhead{Preprint. Under review.}

\begin{abstract}
Music production typically involves multiple editing operations that each require specialized tools and expertise. Existing text-guided music editing systems support few operations or instruments, operate on short clips, or handle only one operation per instruction, unlike real workflows where several changes are requested at once. Here, we present ProCraft, a lightweight 0.5B-parameter autoregressive codec language model that edits 30s music excerpts end-to-end from free-form instructions, including compound requests. ProCraft supports seven operations across eight instruments, from tasks with a unique target (e.g., track removal) to creative ones (e.g., accompaniment generation). We align intermediate representations with pretrained music embeddings and pitch and instrument targets to preserve content outside the edit, and gated cross-attention in every decoder layer to keep generation tied to the instruction. To train on compound requests, we introduce On-the-fly Editing-pair Synthesis, which generates paired instructions and audio combining track-level operations with effect changes. ProCraft performs competitively against task-specific systems on single-task editing, and outperforms four free-text editors on compound instructions for track removal, extraction and denoising on real (MoisesDB) and synthetic (Slakh) recordings. In a user study (16 listeners), ProCraft was rated highest on removal and denoising and on-par with a 16-times larger model on extraction and instrument transfer. Ablations further show that representation alignment and cross-attention reduce leakage from unwanted tracks and raise editing success rates, especially on out-of-domain data.
\end{abstract}

\section{Introduction}
\label{sec:intro}

Editing a music recording often requires changing both what is played and how it sounds. Task-specific models have made individual operations increasingly capable, including source separation \citep{rouard2023htdemucs,liu2023audiosep,bie2024sdcodec}, accompaniment generation \citep{mariani2024msdm,rouard2025musicgenstem}, instrument transfer \citep{tsai2024apadapter,tsai2025musecontrollite}, and audio effects \citep{doh2026llm2fxtools}, but each is designed for a single type of edit, whereas production workflows typically request several changes at once.

Text-guided editing lets users express such requests directly in natural language, yet existing music editors remain limited in scope (Table~\ref{tab:functionality}): most cover only a handful of operations or instruments, do not accept free-form text, operate on clips of a few seconds, or perform one operation per instruction. For example, recent models such as Audio-Omni \citep{tian2026audioomni} edit full mixtures from free-form text but struggle with stem-level operations such as removing a single instrument, whilst LLM2Fx-Tools \citep{doh2026llm2fxtools} handles multi-step instructions but predicts audio effect parameters rather than generating edited audio.

Performing diverse edits from free-form text in a single generative model raises two challenges. First, the model must infer (possibly several) target stems and operations from loosely worded instructions rather than from fixed options. Second, it must preserve everything the instruction does not target. This is especially hard for creative operations such as accompaniment generation, where the model must generate new material while preserve context from the rest of the mix faithfully. One promising direction is unified audio-text codec language models. In speech, SpeechX \citep{wang2023speechx} reads a text query and audio as a single stream to perform tasks such as noise suppression and target-speaker extraction. Music, however, is harder: It contains several concurrent stems rather than one dominant source; its creative edits must respect harmonic and rhythmic content rather than only remove or isolate existing sounds, and its longer excerpts place the instruction and source further from later output frames.

Here, we propose ProCraft, a 0.5B-parameter autoregressive codec language model initialized from the music generation model InspireMusic \citep{zhang2025inspiremusic}. Given a 30\,s mixture and a free-form instruction, ProCraft generates the edited audio in an end-to-end manner. It supports seven edit types across eight instrument families, applied alone or together: stem removal, stem extraction, accompaniment generation, single-instrument cover, instrument transfer, denoising, and audio effects. In particular, we introduce two mechanisms to address the challenges above. A \emph{persistent edit head} summarizes the instruction and source audio and is attended by every decoder layer through gated cross-attention \citep{alayrac2022flamingo} to ensure instruction accessibility over long sequences. \emph{Intermediate-layer music alignment} aligns an intermediate layer with pretrained music embeddings (MERT \citep{li2024mert}, StemFX \citep{cheng2026stemfx}) and with pitch and instrument targets, following representation alignment \citep{yu2024repa}, so that the model's representation of the mixture carries the stem-level structure needed to preserve untargeted stems. Because compound editing triplets do not exist at scale, we also introduce On-the-fly Editing-pair Synthesis, in which a tool-calling language model \citep{yang2025qwen3} writes each instruction and specifies audio effects as executable function calls on multitrack stems.

In summary, our contributions\footnote{We plan to release model weights and code for the model, training, and data generation.} are as follows:
\begin{itemize}[itemsep=2pt,parsep=0pt,topsep=2pt,partopsep=0pt]
\item \textbf{A single model for diverse and compound music editing.} We formulate diverse and compound edits from free-form instructions as one task, evaluated on both the requested change and the preservation of untargeted content, so that neither editing the wrong stem nor returning the input unchanged scores well. ProCraft remains competitive with task-specific systems on single operations and outperforms four free-text editors on compound instructions on MoisesDB and Slakh.
\item \textbf{Two mechanisms for conditioning long codec sequences.} A persistent edit head keeps the instruction accessible to every decoder layer, and intermediate-layer music alignment preserves stem-level structure. Ablations on real recordings show that both reduce leakage from untargeted stems and improve editing success.
\item \textbf{On-the-fly Editing-pair Synthesis.} A tool-calling language model generates source, instruction, and target triplets for training without manual annotation.
\end{itemize}

\providecommand{\fyes}{{\color{green!50!black}$\bullet$}}
\providecommand{\fpart}{{\color{orange!95!black}$\blacktriangle$}}
\providecommand{\fno}{{\color{black!45}$\times$}}
\begin{table*}[!t]
\centering
\scriptsize
\setlength{\tabcolsep}{2pt}
\renewcommand{\arraystretch}{1.08}
\caption{Supported music editing operations per system. \fyes{}\,= supported, \fpart{}\,= partial or indirect, \fno{}\,= not supported; judgments argued in Appendix~\ref{app:functionality}.}
\label{tab:functionality}
\begin{tabular}{@{}l c cccccc c@{}}
\toprule
\textbf{Model} & Free phrasing & Remove & Extract & Accomp. & Cover & Instr. transfer & Denoise & Effects \\
\midrule
\textbf{ProCraft (ours)}        & \fyes  & \fyes  & \fyes  & \fyes  & \fyes  & \fyes  & \fyes  & \fyes  \\
\midrule
Instruct-MusicGen \citep{zhang2024instructmusicgen} & \fno   & \fyes  & \fyes  & \fyes  & \fno   & \fno   & \fno   & \fno   \\
Audio-Omni \citep{tian2026audioomni}       & \fyes  & \fno   & \fno & \fyes  & \fyes  & \fyes  & \fno   & \fno   \\
AudioSep \citep{liu2023audiosep}            & \fpart & \fpart & \fyes  & \fno   & \fno   & \fno   & \fpart & \fno   \\
SD-Codec \citep{bie2024sdcodec}             & \fno   & \fno   & \fpart & \fno   & \fno   & \fno   & \fyes  & \fno   \\
MSDM \citep{mariani2024msdm}                & \fno   & \fno & \fno & \fyes  & \fno   & \fno   & \fno   & \fno   \\
MusicGen-Stem \citep{rouard2025musicgenstem} & \fpart & \fno & \fno & \fyes  & \fno   & \fno   & \fno   & \fno   \\
Pop2Piano \citep{choi2023pop2piano}         & \fno   & \fno   & \fno   & \fno   & \fyes  & \fno   & \fno   & \fno   \\
PiCoGen2 \citep{tan2024picogen2}            & \fno   & \fno   & \fno   & \fno   & \fyes  & \fno   & \fno   & \fno   \\
AP-Adapter \citep{tsai2024apadapter}        & \fyes  & \fno   & \fno   & \fpart & \fpart & \fpart & \fno   & \fno   \\
MuseControlLite \citep{tsai2025musecontrollite} & \fyes  & \fno   & \fno   & \fpart & \fpart & \fyes  & \fno   & \fno   \\
ZETA / DDPM-inv. \citep{manor2024zeta}      & \fyes  & \fpart & \fno   & \fno   & \fpart & \fyes  & \fno   & \fpart \\
SteerMusic. \citep{niu2025steermusic}       & \fyes  & \fpart & \fno   & \fno   & \fpart & \fpart  & \fno   & \fpart \\
LLM2Fx-Tools \citep{doh2026llm2fxtools}     & \fyes  & \fno   & \fno   & \fno   & \fno   & \fno   & \fno   & \fyes  \\
\bottomrule
\end{tabular}
\end{table*}

\section{ProCraft}
\label{sec:method}

Given a source mix $x$ and a free-form instruction $c$, the goal is to use our model $f$ to generate an edited mix $\hat{y} = f(x, c)$ that applies the requested change whilst leaving all other content intact. A codec encoder $E$ maps audio to discrete codes, such that $a^{\mathrm{src}} = E(x)$ and $a^{\mathrm{tgt}} = E(y)$, and $t$ denotes the tokenized instruction. We model the codes of the edited mix autoregressively with $f$ as follows:
\begin{equation}
\label{eq:lm}
p_\theta\big(a^{\mathrm{tgt}} \mid t, a^{\mathrm{src}}\big) = \prod_{i=1}^{T} p_\theta\big(a^{\mathrm{tgt}}_i \mid a^{\mathrm{tgt}}_{<i},\, t,\, a^{\mathrm{src}}\big).
\end{equation}

The model is trained with teacher forcing, and at inference, $a^{\mathrm{tgt}}$ is sampled one code at a time and decoded into $\hat{y}$.
Importantly, this formulation is shared across all editing operations, from those with a unique target such as track removal to creative tasks such as accompaniment generation.

\begin{figure}[!b]
\centering
\includegraphics[width=0.99\linewidth]{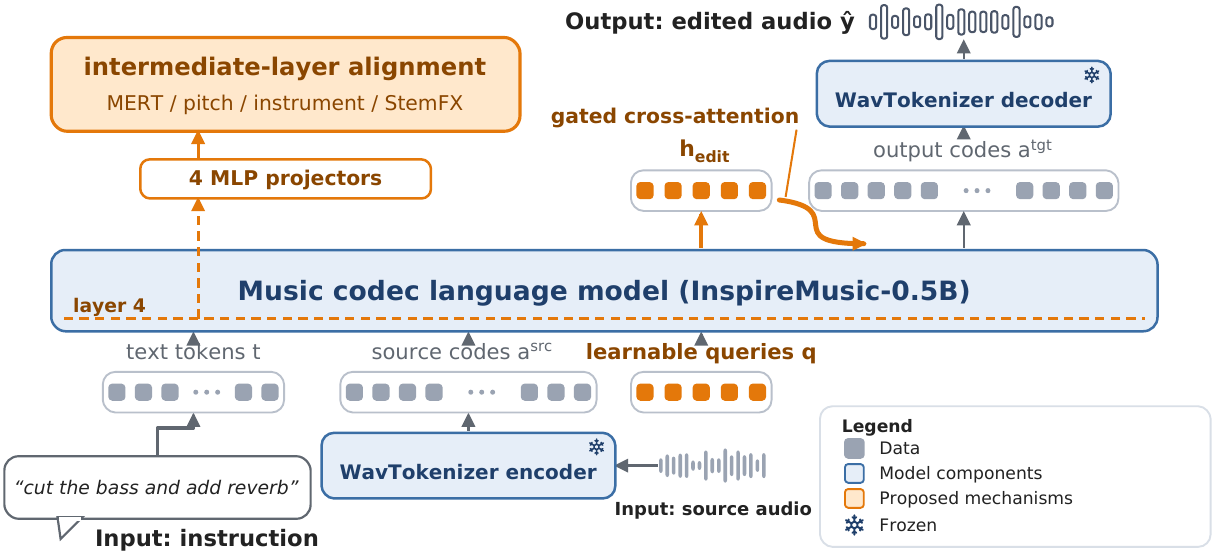}
\caption{Architecture of ProCraft. The instruction, source codes, and $K$ learnable queries precede the target codes in a single causal sequence. The final hidden states of the queries form the edit head $h_{\mathrm{edit}}$, which is injected into every backbone layer through gated cross-attention. During training, four MLP projectors align layer~4 with MERT, pitch, instrument, and StemFX targets.}
\label{fig:arch}
\end{figure}

\subsection{Backbone, tokenization, and sequence layout}
\label{sec:codec}
As illustrated in Figure~\ref{fig:arch}, ProCraft is initialized from the InspireMusic language model \citep{zhang2025inspiremusic}, a 0.5B-parameter codec language model pretrained for text-to-music generation on the Qwen-2 architecture \citep{yang2024qwen2} ($d{=}896$, 24 layers). We chose InspireMusic because it already models text tokens and audio codes as a single sequence, which provides a natural starting point for text-guided editing. All weights are fine-tuned, and we add gated cross-attention layers (Section~\ref{sec:edithead}) and alignment projectors (Section~\ref{sec:align}) to the backbone.

To obtain audio codes, a frozen WavTokenizer codec \citep{ji2024wavtokenizer} encodes a 24\,kHz mono downmix at 75 frames per second into a single stream, such that $a = E(x) \in \{1,\dots,V\}^{T}$, with $\qquad V = 4096$ and $T = 2250$ for a 30\,s clip.%
Instructions are tokenized using the text tokenizer of the backbone, and the output head predicts over the $V$ audio codes plus one end-of-sequence token.

To place the instruction and audio in the same context, the backbone reads the instruction tokens $t$, the source codes $a^{\mathrm{src}}$, and $K{=}32$ learnable query embeddings $q$, followed by the target codes, as one causal sequence:
\begin{equation}
\label{eq:layout}
s = \big[\, t \,;\; a^{\mathrm{src}} \,;\; q \,;\; a^{\mathrm{tgt}} \,\big].
\end{equation}
Under causal attention, this ordering determines what each part of the sequence $f$ can access. In particular, the audio source codes are encoded with the instruction already in context, and the queries attend to both the instruction and the source to summarize the requested edit before any target code is produced. These allow the each target code to attend to the full preceding context. 

At inference, the sampled target codes terminated by the end-of-sequence token are first decoded into waveform using WavTokenizer then upsampled to 48\,kHz mono waveform using the frozen flow-matching super-resolution stage and HifiCodec vocoder of InspireMusic \citep{zhang2025inspiremusic}. Please see Appendix~\ref{app:impl} for sampling and other implementation details.

\subsection{Persistent edit head and gated cross-attention}
\label{sec:edithead}

In the layout of Eq.~\ref{eq:layout}, the instruction appears only at the start, so its share of the causal context shrinks as generation proceeds and its influence on later target codes may weaken.
We therefore summarize the requested edit into a few vectors and pass them directly to every decoder layer with the $K$ learnable queries $q$ in Eq.~\ref{eq:layout}.
The queries are shared across examples, but because they attend to the instruction and source audio, their final-layer hidden states (the \emph{edit head} $h_{\mathrm{edit}} \in \mathbb{R}^{K \times d}$) describe the edit requested for each input.
This follows BLIP-2 \citep{li2023blip2}, with one difference: BLIP-2 queries replace the image tokens, whereas ours are added alongside the source audio, which remains in full in the context.

Each backbone layer gains a cross-attention path from the target positions to $h_{\mathrm{edit}}$.
As in Flamingo \citep{alayrac2022flamingo}, the output of this path is scaled by $\tanh(\alpha)$, with a learnable scalar $\alpha$ initialized at zero, so the model starts from the pretrained backbone and learns during fine-tuning how much to rely on the new path. Because $h_{\mathrm{edit}}$ comes from the final layer but is used in every layer, it cannot be computed in the single teacher-forced pass normally used to train codec language models.
We therefore train with two passes.
The first encodes the prefix $[\,t\,;\,a^{\mathrm{src}}\,;\,q\,]$ and returns $h_{\mathrm{edit}}$ and the prefix key-value cache.
The second processes $a^{\mathrm{tgt}}$ with teacher forcing, attending to the cache through self-attention and to $h_{\mathrm{edit}}$ through cross-attention, so that the generation loss back-propagates into the queries and the prefix.
Since each token is processed only once, the added cost is small, and training matches inference, where the prefix is encoded once and $h_{\mathrm{edit}}$ is held fixed while target codes are sampled.

\subsection{Intermediate-layer music alignment}
\label{sec:align}

The edit head keeps the instruction accessible, but editing some stems while preserving the others also requires the backbone to represent the stem-level content of the mix.
REPA \citep{yu2024repa} showed that aligning early hidden states of image diffusion transformers with pretrained self-supervised representations speeds up training and improves generation quality.
We apply this idea to a codec language model, aligning layer 4 of 24 at the target positions with four targets computed from the target audio:
frozen MERT-330M \mbox{\citep{li2024mert}} embeddings, which capture musical structure in mixtures;
frame-level pitch and instrument activity, which MERT does not encode explicitly;
and StemFX \citep{cheng2026stemfx} embeddings for audio effects.

Each target has its own MLP projector from the layer-4 hidden states, used only during training.
The MERT projector is trained with per-frame cosine distance to the target MERT embeddings ($\mathcal{L}_{\mathrm{mert}}$).
The pitch and instrument projectors predict which of 88 pitched-note and 47 drum-note classes, and which of the eight instrument classes, are active at each frame, trained with binary cross-entropy ($\mathcal{L}_{\mathrm{roll}}$, $\mathcal{L}_{\mathrm{inst}}$) on frames with note labels.
The effect projector mean-pools the hidden states over the target positions and is trained with cosine distance to the target StemFX embedding ($\mathcal{L}_{\mathrm{fx}}$).

\subsection{Training objective}
\label{sec:objective}

The model is trained with the next-code cross-entropy $\mathcal{L}_{\mathrm{CE}}$ of Eq.~\ref{eq:lm} over the target codes and the end-of-sequence token.
To stabilize training, we further add a z-loss \citep{chowdhery2023palm} that keeps the log partition function of the pre-softmax logits $\ell$ close to zero, averaged over target positions:
\begin{equation}
\mathcal{L}_{z} = \Big(\log \textstyle\sum_{v} e^{\ell_v}\Big)^2 .
\end{equation}
The final training objective is the weighted sum of these terms and the alignment losses (Section~\ref{sec:align}):
\begin{equation}
\label{eq:total}
\mathcal{L} = \mathcal{L}_{\mathrm{CE}} + \lambda_z \mathcal{L}_z + \lambda_{\mathrm{mert}} \mathcal{L}_{\mathrm{mert}} + \lambda_{\mathrm{p}} \mathcal{L}_{\mathrm{roll}} + \lambda_{\mathrm{i}} \mathcal{L}_{\mathrm{inst}} + \lambda_{\mathrm{f}} \mathcal{L}_{\mathrm{fx}},
\end{equation}
with $\lambda_z{=}10^{-4}$, $\lambda_{\mathrm{mert}}{=}0.5$, $\lambda_{\mathrm{p}}{=}1.0$, $\lambda_{\mathrm{i}}{=}0.5$, and $\lambda_{\mathrm{f}}{=}0.5$.
Appendix~\ref{app:training} gives the training settings.

\section{On-the-fly editing-pair synthesis}
\label{sec:data}

Training a text-guided editor requires triplets of a source recording, an instruction, and the edited result.
No existing corpus provides these at scale, particularly for compound instructions that combine track-level operations with audio effects.
We therefore synthesize them from multitrack stems (Figure~\ref{fig:pipeline}): instructions are generated once in advance (see Appendix~\ref{app:instruction-examples}), and the matching audio is rendered on-the-fly each time an item is drawn during training.

\begin{figure}[!t]
\centering
\includegraphics[width=0.99\linewidth]{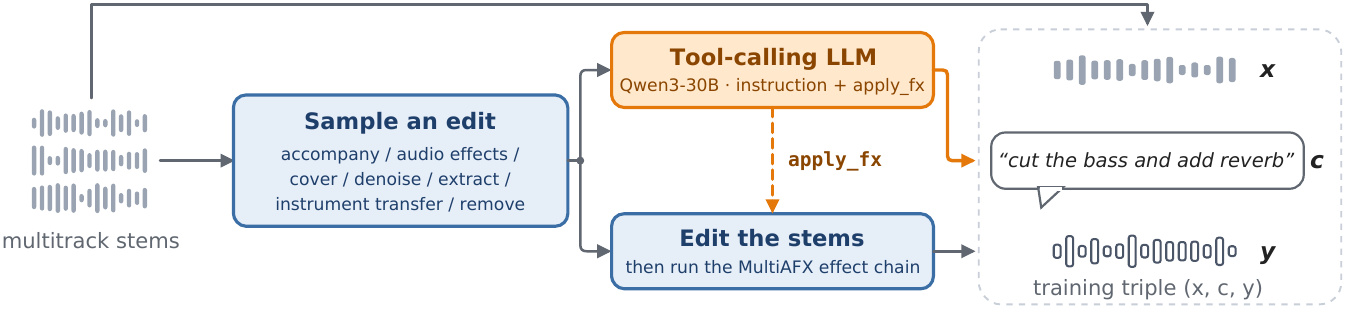}
\caption{On-the-fly editing-pair synthesis. For each training item, a track-level task and/or audio effects are applied to the stems of a multitrack recording. A tool-calling language model writes the instruction and specifies audio effects as executable function calls, which are rendered on the stems to produce the target mix.}
\label{fig:pipeline}
\end{figure}

Each audio sample is a random 30\,s excerpt from a song in Slakh2100 \citep{manilow2019slakh}, whose stems and MIDI allow every task to be rendered with a known target.
We cover eight instrument families: bass, drums, guitar, piano, organ, brass, reed, and string ensemble.
An item combines a track-level task with audio effects (40\%), uses a track-level task alone (40\%), or uses audio effects alone (20\%).
Track-level tasks are drawn uniformly from six operations, which together with audio effects make up the seven operations of ProCraft.

For removal and extraction, the target omits or keeps only the requested stem.
For accompaniment, the requested instrument is withheld from the source, and the target contains the new part alone.
For cover generation, the whole excerpt is re-voiced on a single instrument (guitar, piano, organ, or string ensemble), using a score reduction method for piano covers \citep{nakamura2015reduction} to merge all parts into one playable part.
For instrument transfer, the MIDI of one track is re-rendered with the soundfont of the new instrument, so only the timbre changes.
For denoising, an environmental sound from ESC-50 \citep{piczak2015esc50}, such as rain or an alarm, is repeated throughout the excerpt, and the clean mix is the target.
Appendix~\ref{app:data} gives further details.

Qwen3-30B-A3B-Instruct-2507 \citep{yang2025qwen3} writes each instruction from the sampled task, the instruments present, and one of 30 editing motivations (e.g., ``make it sound vintage''; Appendix~\ref{app:styles}), which diversifies both wording and effect choices beyond fixed templates.
When effects are included, the model also specifies them as tool calls, which we execute on the stems with MultiAFX \citep{cheng2026stemfx}, following LLM2Fx-Tools \citep{doh2026llm2fxtools}.
Because the effects are executed rather than only described, the target matches the instruction by construction.

\section{Evaluation}
\label{sec:eval}

\subsection{Metrics}
\label{sec:metrics}

\paragraph{General metrics.}
\emph{CLAP-T out} and \emph{CLAP-A} are the cosine similarities between the CLAP embeddings \citep{wu2023clap} of the output and, respectively, the instruction and the source audio.
\emph{KL} is the divergence between the AudioSet label posteriors predicted by PaSST \citep{koutini2022passt} for the output and the target.
\emph{FAD} and \emph{KAD} compare outputs with a fixed reference set of 390 real recordings (240 from MoisesDB, 150 from MUSDB18-HQ) \citep{rafii2019musdb18hq,kilgour2019fad,gui2024fadtk,chung2025kad}.
These metrics cannot adequately assess editing or preservation: CLAP-A is maximized by returning the source unchanged, and CLAP-T may reward a removal output that still contains the named instrument.
We therefore add task-specific metrics for Editing and Preservation.

\paragraph{Operations with a deterministic target.}
For removal, extraction, denoising and audio effects, we measure Editing and Preservation with \emph{mel-L1} on log-mel spectrograms and multi-resolution STFT distance (\emph{MRSTFT}, four window lengths), and additionally \emph{SI-SDR} for all but audio effects.
For removal and extraction, \emph{edit rate} and signed \emph{edit strength} \citep{zhang2024instructmusicgen} measure whether, and by how much, the named stem was removed or isolated, using stem energy from HTDemucs-6s \citep{rouard2023htdemucs}; we evaluate only drums, bass, guitar and piano, the instruments our baselines support.
For audio effects, \emph{AFx-Rep cosine} \citep{steinmetz2024stito} compares the effect style of output and target, and \emph{audio-feature distance} \citep{vanka2024diffmst} compares their loudness and spectral descriptors.
We use AFx-Rep rather than StemFX to avoid circular evaluation, as StemFX is an alignment target (see Section~\ref{sec:align} on intermediate layer alignment).

\paragraph{Operations without a unique target.}
Accompaniment and cover generation add new material, and in instrument transfer any realistic timbre of the requested instrument is valid, so comparing with a single reference would penalize valid outputs.
For Editing on all three, \emph{instrument match} is the accuracy of PaSST in recognizing the requested instrument, and \emph{isolation strength} tests whether the requested part is output alone rather than within a full mix.
For Preservation, \emph{APA} \citep{grachten2025apa} on accompaniment and cover checks, by comparing FADs, that outputs fit their own source better than a randomly paired one;
\emph{StructDist} \citep{liu2024medic} (accompaniment) and \emph{melody chroma accuracy} \citep{tan2024picogen} (cover and transfer) test whether the output follows the source's structure and melody;
and for transfer, the \emph{frame-F1} of a MuScriptor transcription \citep{rouard2026muscriptor} against the ground-truth MIDI measures pitch and timing.
Details are in Appendix~\ref{app:metrics}.

\subsection{Test sets and baselines}
\label{sec:testsets}

We evaluate all seven operations on both synthetic and real recordings. Synthetic test items are drawn from the official test split of Slakh2100 \citep{manilow2019slakh}, and all items from MoisesDB \citep{pereira2023moisesdb}. As ProCraft is trained only on Slakh, MoisesDB provides an out-of-domain test on real recordings. Each operation has 150 test items per dataset, each consisting of a 30\,s source excerpt, an instruction, and, where applicable, a target.

We compare ProCraft with task-specific systems (AudioSep \citep{liu2023audiosep}, HTDemucs \citep{rouard2023htdemucs}, InstructMusicGen \citep{zhang2024instructmusicgen}, InstructFX2FX \citep{yu2026instructfx2fx}, MG-Stem \citep{rouard2025musicgenstem}, MuseControlLite \citep{tsai2025musecontrollite}, PiCoGen2 \citep{tan2024picogen2}) on each single operation, and with four free-text editors (ZETA and a DDIM-inversion editor \citep{manor2024zeta}, SteerMusic \citep{niu2025steermusic}, and Audio-Omni \citep{tian2026audioomni}) on compound instructions. All baselines are run in their native input formats (see Appendix~\ref{app:baselines} for checkpoints and settings, and further baselines), and metrics that do not apply to a system are omitted.

\section{Results}
\label{sec:results}

\subsection{Main results}
\label{sec:main}

Figure~\ref{fig:headline} summarizes model performance for each operation (see full results in Tables~\ref{tab:main-slakh} and~\ref{tab:main-real}). Overall, ProCraft was most competitive when the target contains material absent from the source, but trailed dedicated systems when ground-truth target exists. 

\textbf{Creative operations.}
On accompaniment, cover, and instrument transfer, task-specific systems tended to favor either Editing or Preservation. On Slakh instrument transfer, for example, TokenSynth best preserved pitch and timing (frame-F1 of 0.641) but rendered the requested instrument less reliably (instrument match of 0.420), whereas the DDPM-inversion editor shows the opposite (0.326 and 0.653). ProCraft was the only system above 0.55 on instrument match, frame-F1, and melody chroma. Similarly, piano-cover systems follow the source melody at most slightly more closely than ProCraft (Pop2Piano on Slakh, PiCoGen2 on MoisesDB) but fit their source less well (APA of 0.394--0.712 vs 0.775--0.820). The exception is in accompaniment, where MusicGen-Stem led on instrument match and APA, and AP-Adapter on instrument transfer on MoisesDB.

\textbf{Audio effects.}
InstructFX2FX was closest to the processed target in both datasets. Applying DSP rather than regenerating the audio does not fully explain this, however, as ProCraft outperformed the DSP-based FxSearcher and Text2FX on mel-L1, MRSTFT, and AFx-Rep (e.g., AFx-Rep cosine of 0.195 vs 0.157 and 0.092 on MoisesDB). This suggests that inferring the correct effect from the instruction mattered just as much as the signal processing.

\textbf{Removal, extraction, and denoising.}
Here, dedicated source separator models led SI-SDR by a wide margin (e.g., $+8.31$ vs $-2.96$\,dB for HTDemucs and ProCraft on Slakh removal). To explain this gap, we encoded and decoded the ground-truth targets with WavTokenizer, which yielded $+1.05$\,dB, so 7.3\,dB of this 11.3\,dB gap stems from the low-bitrate (0.9\,kbps) single-codebook codec rather than the model. Nevertheless, SI-SDR does not check whether the edit actually took place: on MoisesDB extraction, AudioSep scored higher than ProCraft ($-3.69$ vs $-15.14$\,dB) yet isolated the requested stem in only 7.3\% of items, whereas ProCraft succeeded in 82.0\% of items.

\subsection{Ablation study}
\label{sec:ablation}

\begin{figure}[t]
\centering
\includegraphics[width=\linewidth]{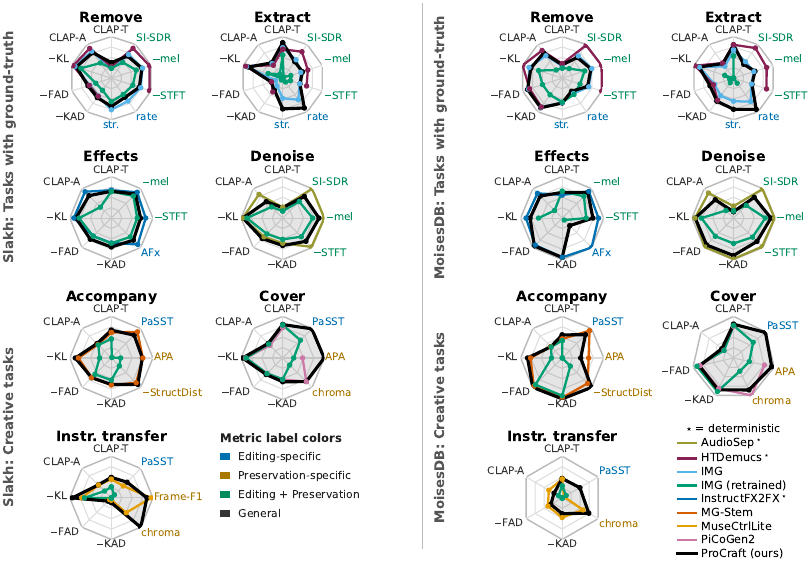}
\caption{Model performance on single-task music editing from free-form text, tested on synthetic recordings in Slakh (left) and real recordings in MoisesDB (right) over diverse metrics and model baselines. We multiply a metric by -1 if the raw metric value is better when lower so larger radii always indicate better results. See full results in Appendix~\ref{app:fulltables}.}
\label{fig:headline}
\end{figure}

\begin{figure}[b]
\centering
\includegraphics[width=0.99\linewidth]{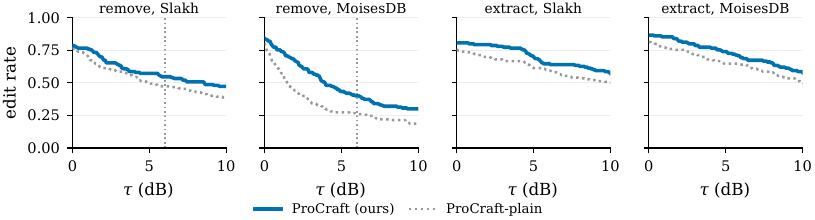}
\caption{Fraction of items whose edit strength exceeds a threshold $\tau$, for ProCraft and ProCraft-plain (150 items per panel). The dotted line marks the 6\,dB threshold used for edit rate.}
\label{fig:ablation-threshold}
\end{figure}

We tested the importance of joint instruction-audio conditioning, as well as our cross-attention and alignment mechanisms by retraining Instruct-MusicGen on our data (IMG retrained), and training ProCraft-plain, which removes the gated cross-attention and alignment losses.

\textbf{Joint conditioning.}
IMG (retrained) trailed ProCraft on every operation (Figure~\ref{fig:headline}). Whereas Instruct-MusicGen conditions on the instruction and source audio through separate paths \citep{zhang2024instructmusicgen}, ProCraft reads both in one sequence, so every output code can attend to them together. This suggests improvements due to joint audio-text conditioning and model backbone.

\textbf{Cross-attention and alignment.}
ProCraft and ProCraft-plain performed similarly on accompaniment, cover, and instrument transfer, but differed on operations that must remove content whilst keeping the rest intact. On MoisesDB removal, ProCraft raised edit rate from 26.0\% to 38.0\% and edit strength from 4.9 to 7.4\,dB, and on extraction it improved SI-SDR in both datasets (e.g., $-15.1$ vs $-17.4$\,dB on MoisesDB).
Figure~\ref{fig:ablation-threshold} further shows that the advantage of ProCraft does not depend on the 6\,dB threshold used to define edit rate, as it led at every threshold in both datasets. These demonstrate the improvement from cross-attention and alignment, particularly on out-of-domain, real datasets.

\subsection{Compound edit-and-effects instructions}
\label{sec:fullcase}

Production requests often combine an edit with a change in sound, such as ``remove the bass and make the piano brighter and more spacious''.
Here, we paired each of the six track-level operations with audio effects and compared ProCraft with the four free-text editors (see Section~\ref{sec:testsets}).
\par
\Needspace{14\baselineskip}
\begin{wraptable}[12]{R}{0.45\linewidth}
\vspace{-6pt}
\centering
\footnotesize
\setlength{\tabcolsep}{3.8pt}
\caption{SI-SDR (dB, higher is better) on compound instructions in Slakh.}
\label{tab:fullcase-fixed-summary}
\begin{tabular}{lrrr}
\toprule
System & Removal & Extraction & Denoising \\
\midrule
\textbf{ProCraft (ours)} & \textbf{$-7.39$} & \textbf{$-15.73$} & \textbf{$-4.74$} \\
ZETA & $-46.18$ & $-38.03$ & $-35.42$ \\
DDIM inv. & $-44.83$ & $-41.26$ & $-32.18$ \\
SteerMusic & $-25.60$ & $-27.20$ & $-29.90$ \\
Audio-Omni & $-24.88$ & $-23.98$ & $-26.02$ \\
\bottomrule
\end{tabular}
\end{wraptable}

\textbf{Removal, extraction, and denoising.}
ProCraft outperformed all four editors on Slakh by at least 8.25\,dB SI-SDR (Table~\ref{tab:fullcase-fixed-summary}); on removal, for example, it reached $-7.39$\,dB against $-24.88$\,dB for the next best system.
ProCraft also led on MoisesDB by at least $6.33$\,dB.

\begin{figure}[!ht]
\centering
\includegraphics[width=0.84\linewidth]{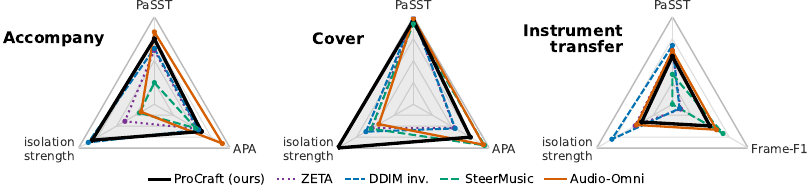}
\caption{Model performance on music editing from compound instructions on accompaniment, cover, and instrument transfer on Slakh. Larger radii indicate better performance; isolation strength shares the same range across panels. See Appendix~\ref{app:fullcase} for full results including MoisesDB.}
\label{fig:fullcase-summary}
\end{figure}

\textbf{Accompaniment, cover, and instrument transfer.}
Each baseline fell short on at least one aspect (Figure~\ref{fig:fullcase-summary}).
Audio-Omni scored higher than ProCraft on APA for accompaniment and cover, but its negative isolation strength shows that it often returned a full mix instead of the requested part.
The DDIM-inversion editor isolated the part more strongly on some operations, but fit the source far less well.
SteerMusic rarely rendered the requested instrument on accompaniment and barely isolated the part on transfer.
ProCraft was the only editor without a large drop on any metric across the three operations, although its isolation on instrument transfer remains imperfect.
Appendix~\ref{app:fullcase} gives full results of this analysis.

\subsection{Listening test}
\label{sec:listening}

We ran a MUSHRA-style listening test \citep{itu2015mushra} to compare how listeners judged ProCraft and the four free-text editors on compound instructions in MoisesDB.
For each of the six operations, we prepared three recordings with an instruction.
Each of the sixteen listeners heard one of these per operation, together with the outputs of all five systems in random order, and rated each output from 0 to 5 on Editing, Preservation, and audio quality.
Appendix~\ref{app:listening} gives the protocol, participant details, and statistical tests (task-wise repeated-measures ANOVA with Holm-corrected paired comparisons).

\begin{figure}[!ht]
\centering
\includegraphics[width=\linewidth]{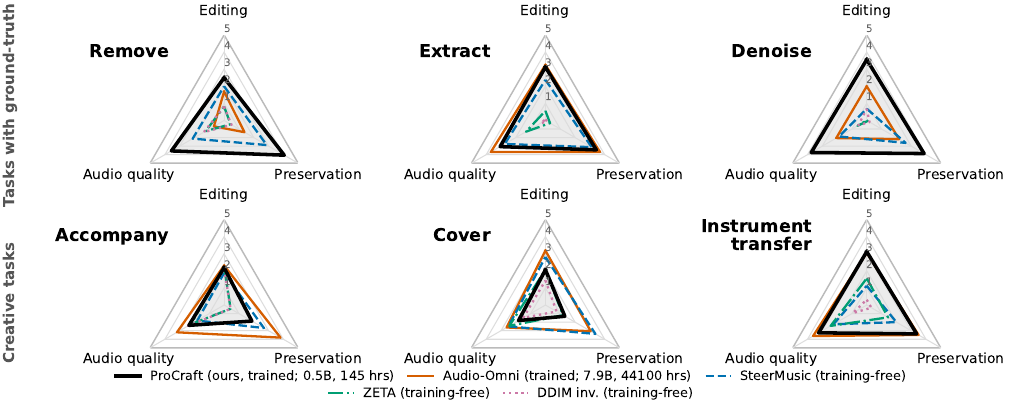}
\caption{MUSHRA-style user study on perceived audio from music editing with compound instructions. Each triangle shows a task's mean Editing, Preservation, and Audio-quality ratings (0 to 5) from sixteen raters; See Table~\ref{tab:listening} and Appendix~\ref{app:listening} for results, statistical inference, and protocol details.}
\label{fig:listening}
\end{figure}

\textbf{Removal and denoising.}
ProCraft received the highest mean ratings on all three aspects (Figure~\ref{fig:listening}).
It was rated significantly higher than all four editors on every aspect for denoising, and on Preservation and audio quality for removal.

\textbf{Extraction and instrument transfer.}
ProCraft and Audio-Omni were rated highest, with no significant difference between them on any aspect.
This is notable given that Audio-Omni has 16 times more parameters (7.9B vs 0.5B) and was trained on about 300 times more audio (44,100 vs 145 hours).

\textbf{Accompaniment and cover.}
Audio-Omni and SteerMusic received higher mean Preservation ratings than ProCraft, significantly so on cover.
However, both systems often returned the original mix rather than only the requested part (negative isolation strength in Table~\ref{tab:fullcase-real}), which would make their outputs sound closer to the source.
Audio-Omni also received higher audio-quality ratings on both operations, and although these differences were not significant, they suggest that audio quality is the main area for ProCraft to improve.

\section{Conclusion}
\label{sec:conclusion}

Music production typically involves several changes to a recording at once, yet existing editors handle only a fixed type of edit or one operation per instruction.
Here, we presented ProCraft, a single 0.5B-parameter codec language model that performs seven editing operations across eight instrument families on 30\,s excerpts from free-form instructions, including compound requests that combine stem-level operations with audio effects.
Trained only on triplets from On-the-fly Editing-pair Synthesis, ProCraft outperformed four free-text editors on compound instructions, and was rated highest by listeners on removal and denoising, and on par with the 16-times larger Audio-Omni on extraction and instrument transfer.
Ablations on real recordings showed that the persistent edit head and intermediate-layer music alignment together reduce leakage from untargeted stems and raise editing success rates. Nevertheless, our generative model still trailed behind source-separator models such as HTDemucs in tasks with ground truth (such as track removal), and listeners generally preferred larger models such as Audio-Omni on creative tasks like accompaniment and cover generation. Producing high-quality audio edits from natural language input whilst preserving the source in a lightweight model thus remains an open challenge.

\subsubsection*{AI use statement}
A large language model is part of the data pipeline, which is a required disclosure.
Qwen3-30B-A3B-Instruct-2507 writes the freely phrased instructions and the typed audio effect chains for every training pair, so the instruction distribution and the audio effect vocabulary both inherit that model's biases.

A large language model also assisted in polishing the text of this paper and in writing code.
Every number reported here is produced by that code from the stored model outputs, and each table and figure was checked against the underlying results files.
It was not used for the research idea, the experimental design, the choice of what to claim, or the analysis of results, all of which are the authors' own work.
The authors take responsibility for all claims in the paper.

\bibliography{references}

\begin{thebibliography}{52}
\providecommand{\natexlab}[1]{#1}
\providecommand{\url}[1]{\texttt{#1}}
\expandafter\ifx\csname urlstyle\endcsname\relax
  \providecommand{\doi}[1]{doi: #1}\else
  \providecommand{\doi}{doi: \begingroup \urlstyle{rm}\Url}\fi

\bibitem[Alayrac et~al.(2022)Alayrac, Donahue, Luc, Miech, Barr, Hasson,
  et~al.]{alayrac2022flamingo}
Jean-Baptiste Alayrac, Jeff Donahue, Pauline Luc, Antoine Miech, Iain Barr,
  Yana Hasson, et~al.
\newblock Flamingo: a visual language model for few-shot learning.
\newblock In \emph{Advances in Neural Information Processing Systems
  (NeurIPS)}, 2022.

\bibitem[Bie et~al.(2025)Bie, Liu, and Richard]{bie2024sdcodec}
Xiaoyu Bie, Xubo Liu, and Ga{\"e}l Richard.
\newblock Learning source disentanglement in neural audio codec.
\newblock In \emph{Proceedings of the IEEE International Conference on
  Acoustics, Speech and Signal Processing (ICASSP)}, 2025.

\bibitem[Cheng et~al.(2026)Cheng, Wu, Chen, Yeh, Chen, and
  Yang]{cheng2026stemfx}
Yuan-Chiao Cheng, Jui-Te Wu, Brian Chen, Yen-Tung Yeh, Yu-Hua Chen, and
  Yi-Hsuan Yang.
\newblock {StemFX}: Learning mixing style representations via autoregressive fx
  chain prediction on source-separated stems.
\newblock In \emph{Proceedings of the 27th International Society for Music
  Information Retrieval Conference (ISMIR)}, 2026.

\bibitem[Choi \& Lee(2023)Choi and Lee]{choi2023pop2piano}
Jongho Choi and Kyogu Lee.
\newblock {Pop2Piano}: Pop audio-based piano cover generation.
\newblock In \emph{Proceedings of the IEEE International Conference on
  Acoustics, Speech and Signal Processing (ICASSP)}, 2023.

\bibitem[Chowdhery et~al.(2023)Chowdhery, Narang, Devlin, Bosma, Mishra,
  Roberts, Barham, Chung, Sutton, Gehrmann, et~al.]{chowdhery2023palm}
Aakanksha Chowdhery, Sharan Narang, Jacob Devlin, Maarten Bosma, Gaurav Mishra,
  Adam Roberts, Paul Barham, Hyung~Won Chung, Charles Sutton, Sebastian
  Gehrmann, et~al.
\newblock {PaLM}: Scaling language modeling with pathways.
\newblock \emph{Journal of Machine Learning Research}, 24\penalty0
  (240):\penalty0 1--113, 2023.

\bibitem[Chu et~al.(2025)Chu, O'Reilly, Barnett, and Pardo]{chu2024text2fx}
Annie Chu, Patrick O'Reilly, Julia Barnett, and Bryan Pardo.
\newblock {Text2FX}: Harnessing {CLAP} embeddings for text-guided audio
  effects.
\newblock In \emph{Proceedings of the IEEE International Conference on
  Acoustics, Speech and Signal Processing (ICASSP)}, 2025.

\bibitem[Chung et~al.(2025)Chung, Eu, Lee, Choi, Nam, and Chon]{chung2025kad}
Yoonjin Chung, Pilsun Eu, Junwon Lee, Keunwoo Choi, Juhan Nam, and Ben~Sangbae
  Chon.
\newblock {KAD}: No more {FAD}! an effective and efficient evaluation metric
  for audio generation.
\newblock In \emph{ICML 2025 Workshop on Machine Learning for Audio}, 2025.

\bibitem[Copet et~al.(2023)Copet, Kreuk, Gat, Remez, Kant, Synnaeve, Adi, and
  D{\'e}fossez]{copet2023musicgen}
Jade Copet, Felix Kreuk, Itai Gat, Tal Remez, David Kant, Gabriel Synnaeve,
  Yossi Adi, and Alexandre D{\'e}fossez.
\newblock Simple and controllable music generation.
\newblock In \emph{Advances in Neural Information Processing Systems
  (NeurIPS)}, 2023.

\bibitem[Doh et~al.(2026)Doh, Koo, Mart{\'i}nez-Ram{\'i}rez, Choi, Liao, Wu,
  Nam, and Mitsufuji]{doh2026llm2fxtools}
Seungheon Doh, Junghyun Koo, Marco~A. Mart{\'i}nez-Ram{\'i}rez, Woosung Choi,
  Wei-Hsiang Liao, Qiyu Wu, Juhan Nam, and Yuki Mitsufuji.
\newblock {LLM2Fx-Tools}: Tool calling for music post-production.
\newblock In \emph{International Conference on Learning Representations
  (ICLR)}, 2026.

\bibitem[Evans et~al.(2025)Evans, Parker, Carr, Zukowski, Taylor, and
  Pons]{evans2024stableaudio}
Zach Evans, Julian~D. Parker, C.~J. Carr, Zack Zukowski, Josiah Taylor, and
  Jordi Pons.
\newblock Stable audio open.
\newblock In \emph{Proceedings of the IEEE International Conference on
  Acoustics, Speech and Signal Processing (ICASSP)}, 2025.

\bibitem[Grachten \& Nistal(2025)Grachten and Nistal]{grachten2025apa}
Maarten Grachten and Javier Nistal.
\newblock Accompaniment prompt adherence: A measure for evaluating music
  accompaniment systems.
\newblock In \emph{IEEE International Conference on Acoustics, Speech and
  Signal Processing (ICASSP)}, 2025.

\bibitem[Gui et~al.(2024)Gui, Gamper, Braun, and Emmanouilidou]{gui2024fadtk}
Azalea Gui, Hannes Gamper, Sebastian Braun, and Dimitra Emmanouilidou.
\newblock Adapting frechet audio distance for generative music evaluation.
\newblock In \emph{IEEE International Conference on Acoustics, Speech and
  Signal Processing (ICASSP)}, 2024.

\bibitem[Hu et~al.(2022)Hu, Shen, Wallis, Allen-Zhu, Li, Wang, Wang, and
  Chen]{hu2022lora}
Edward~J. Hu, Yelong Shen, Phillip Wallis, Zeyuan Allen-Zhu, Yuanzhi Li, Shean
  Wang, Lu~Wang, and Weizhu Chen.
\newblock {LoRA}: Low-rank adaptation of large language models.
\newblock In \emph{International Conference on Learning Representations
  (ICLR)}, 2022.

\bibitem[{ITU-R}(2015)]{itu2015mushra}
{ITU-R}.
\newblock Method for the subjective assessment of intermediate quality level of
  audio systems.
\newblock Technical Report Recommendation BS.1534-3, International
  Telecommunication Union, 2015.

\bibitem[Ji et~al.(2025)Ji, Jiang, Wang, Chen, Fang, Zuo,
  et~al.]{ji2024wavtokenizer}
Shengpeng Ji, Ziyue Jiang, Wen Wang, Yifu Chen, Minghui Fang, Jialong Zuo,
  et~al.
\newblock {WavTokenizer}: An efficient acoustic discrete codec tokenizer for
  audio language modeling.
\newblock In \emph{International Conference on Learning Representations
  (ICLR)}, 2025.

\bibitem[Ki et~al.(2026)Ki, Kim, Kwon, and Kim]{ki2025fxsearcher}
Hojoon Ki, Jongsuk Kim, Minchan Kwon, and Junmo Kim.
\newblock {FxSearcher}: Gradient-free text-driven audio transformation.
\newblock In \emph{IEEE International Conference on Acoustics, Speech and
  Signal Processing (ICASSP)}, 2026.

\bibitem[Kilgour et~al.(2019)Kilgour, Zuluaga, Roblek, and
  Sharifi]{kilgour2019fad}
Kevin Kilgour, Mauricio Zuluaga, Dominik Roblek, and Matthew Sharifi.
\newblock Fr{\'e}chet audio distance: A reference-free metric for evaluating
  music enhancement algorithms.
\newblock In \emph{Interspeech}, pp.\  2350--2354, 2019.

\bibitem[Kim et~al.(2025)Kim, Koo, Lee, Joung, and Lee]{kim2025tokensynth}
Kyungsu Kim, Junghyun Koo, Sungho Lee, Haesun Joung, and Kyogu Lee.
\newblock {TokenSynth}: A token-based neural synthesizer for instrument cloning
  and text-to-instrument.
\newblock In \emph{IEEE International Conference on Acoustics, Speech and
  Signal Processing (ICASSP)}, 2025.

\bibitem[Koutini et~al.(2022)Koutini, Schl{\"u}ter, Eghbal-zadeh, and
  Widmer]{koutini2022passt}
Khaled Koutini, Jan Schl{\"u}ter, Hamid Eghbal-zadeh, and Gerhard Widmer.
\newblock Efficient training of audio transformers with patchout.
\newblock In \emph{Interspeech}, pp.\  2753--2757, 2022.

\bibitem[Li et~al.(2023)Li, Li, Savarese, and Hoi]{li2023blip2}
Junnan Li, Dongxu Li, Silvio Savarese, and Steven Hoi.
\newblock {BLIP-2}: Bootstrapping language-image pre-training with frozen image
  encoders and large language models.
\newblock In \emph{Proceedings of the 40th International Conference on Machine
  Learning (ICML)}, 2023.

\bibitem[Li et~al.(2024)Li, Yuan, Zhang, Ma, Chen, Yin, et~al.]{li2024mert}
Yizhi Li, Ruibin Yuan, Ge~Zhang, Yinghao Ma, Xingran Chen, Hanzhi Yin, et~al.
\newblock {MERT}: Acoustic music understanding model with large-scale
  self-supervised training.
\newblock In \emph{International Conference on Learning Representations
  (ICLR)}, 2024.

\bibitem[Liu et~al.(2024)Liu, Yuan, Liu, Mei, Kong, Tian, Wang, Wang, Wang, and
  Plumbley]{liu2024audioldm2}
Haohe Liu, Yi~Yuan, Xubo Liu, Xinhao Mei, Qiuqiang Kong, Qiao Tian, Yuping
  Wang, Wenwu Wang, Yuxuan Wang, and Mark~D. Plumbley.
\newblock {AudioLDM} 2: Learning holistic audio generation with self-supervised
  pretraining.
\newblock \emph{IEEE/ACM Transactions on Audio, Speech, and Language
  Processing}, 32, 2024.

\bibitem[Liu et~al.(2025{\natexlab{a}})Liu, Wang, Li, Wang, Chen, Huang,
  et~al.]{liu2024medic}
Huadai Liu, Jialei Wang, Xiangtai Li, Wen Wang, Qian Chen, Rongjie Huang,
  et~al.
\newblock {MelodyEdit}: Zero-shot music editing with disentangled inversion
  control.
\newblock In \emph{Proceedings of the 33rd ACM International Conference on
  Multimedia}, 2025{\natexlab{a}}.

\bibitem[Liu et~al.(2025{\natexlab{b}})Liu, Kong, Zhao, Liu, Yuan, Liu,
  et~al.]{liu2023audiosep}
Xubo Liu, Qiuqiang Kong, Yan Zhao, Haohe Liu, Yi~Yuan, Yuzhuo Liu, et~al.
\newblock Separate anything you describe.
\newblock \emph{IEEE Transactions on Audio, Speech and Language Processing},
  33:\penalty0 458--471, 2025{\natexlab{b}}.

\bibitem[Manilow et~al.(2019)Manilow, Wichern, Seetharaman, and
  Le~Roux]{manilow2019slakh}
Ethan Manilow, Gordon Wichern, Prem Seetharaman, and Jonathan Le~Roux.
\newblock Cutting music source separation some {Slakh}: A dataset to study the
  impact of training data quality and quantity.
\newblock In \emph{IEEE Workshop on Applications of Signal Processing to Audio
  and Acoustics (WASPAA)}, 2019.

\bibitem[Manor \& Michaeli(2024)Manor and Michaeli]{manor2024zeta}
Hila Manor and Tomer Michaeli.
\newblock Zero-shot unsupervised and text-based audio editing using {DDPM}
  inversion.
\newblock In \emph{Proceedings of the 41st International Conference on Machine
  Learning (ICML)}, 2024.

\bibitem[Mariani et~al.(2024)Mariani, Tallini, Postolache, Mancusi, Cosmo, and
  Rodol{\`a}]{mariani2024msdm}
Giorgio Mariani, Irene Tallini, Emilian Postolache, Michele Mancusi, Luca
  Cosmo, and Emanuele Rodol{\`a}.
\newblock Multi-source diffusion models for simultaneous music generation and
  separation.
\newblock In \emph{International Conference on Learning Representations
  (ICLR)}, 2024.

\bibitem[Nakamura \& Sagayama(2015)Nakamura and
  Sagayama]{nakamura2015reduction}
Eita Nakamura and Shigeki Sagayama.
\newblock Automatic piano reduction from ensemble scores based on merged-output
  hidden markov model.
\newblock In \emph{Proceedings of the International Computer Music Conference
  (ICMC)}, 2015.

\bibitem[Niu et~al.(2026)Niu, Cheuk, Zhang, Murata, Lai, Mancusi,
  et~al.]{niu2025steermusic}
Xinlei Niu, Kin~Wai Cheuk, Jing Zhang, Naoki Murata, Chieh-Hsin Lai, Michele
  Mancusi, et~al.
\newblock {SteerMusic}: Enhanced musical consistency for zero-shot text-guided
  and personalized music editing.
\newblock In \emph{Proceedings of the AAAI Conference on Artificial
  Intelligence}, 2026.

\bibitem[Pereira et~al.(2023)Pereira, Ara{\'u}jo, Korzeniowski, and
  Vogl]{pereira2023moisesdb}
Igor Pereira, Felipe Ara{\'u}jo, Filip Korzeniowski, and Richard Vogl.
\newblock {MoisesDB}: A dataset for source separation beyond 4-stems.
\newblock In \emph{Proceedings of the 24th International Society for Music
  Information Retrieval Conference (ISMIR)}, 2023.

\bibitem[Piczak(2015)]{piczak2015esc50}
Karol~J. Piczak.
\newblock {ESC}: Dataset for environmental sound classification.
\newblock In \emph{Proceedings of the 23rd ACM International Conference on
  Multimedia}, pp.\  1015--1018, 2015.
\newblock \doi{10.1145/2733373.2806390}.

\bibitem[Raffel et~al.(2014)Raffel, McFee, Humphrey, Salamon, Nieto, Liang, and
  Ellis]{raffel2014mireval}
Colin Raffel, Brian McFee, Eric~J. Humphrey, Justin Salamon, Oriol Nieto, Dawen
  Liang, and Daniel P.~W. Ellis.
\newblock {mir\_eval}: A transparent implementation of common {MIR} metrics.
\newblock In \emph{Proceedings of the 15th International Society for Music
  Information Retrieval Conference (ISMIR)}, pp.\  367--372, 2014.

\bibitem[Rafii et~al.(2019)Rafii, Liutkus, St{\"o}ter, Mimilakis, and
  Bittner]{rafii2019musdb18hq}
Zafar Rafii, Antoine Liutkus, Fabian-Robert St{\"o}ter, Stylianos~Ioannis
  Mimilakis, and Rachel Bittner.
\newblock {MUSDB18-HQ} - an uncompressed version of {MUSDB18}.
\newblock \url{https://doi.org/10.5281/zenodo.3338373}, 2019.

\bibitem[Rouard et~al.(2023)Rouard, Massa, and
  D{\'e}fossez]{rouard2023htdemucs}
Simon Rouard, Francisco Massa, and Alexandre D{\'e}fossez.
\newblock Hybrid transformers for music source separation.
\newblock In \emph{IEEE International Conference on Acoustics, Speech and
  Signal Processing (ICASSP)}, pp.\  1--5, 2023.

\bibitem[Rouard et~al.(2025)Rouard, San~Roman, Adi, and
  Roebel]{rouard2025musicgenstem}
Simon Rouard, Robin San~Roman, Yossi Adi, and Axel Roebel.
\newblock {MusicGen-Stem}: Multi-stem music generation and edition through
  autoregressive modeling.
\newblock In \emph{IEEE International Conference on Acoustics, Speech and
  Signal Processing (ICASSP)}, 2025.

\bibitem[Rouard et~al.(2026)Rouard, Krause, Roebel, Simon-Gabriel, and
  D{\'e}fossez]{rouard2026muscriptor}
Simon Rouard, Michael Krause, Axel Roebel, Carl-Johann Simon-Gabriel, and
  Alexandre D{\'e}fossez.
\newblock {MuScriptor}: An open model for multi-instrument music transcription.
\newblock In \emph{International Society for Music Information Retrieval
  Conference (ISMIR)}, 2026.

\bibitem[Steinmetz et~al.(2024)Steinmetz, Singh, Comunit{\`a}, Ibnyahya, Yuan,
  Benetos, and Reiss]{steinmetz2024stito}
Christian~J. Steinmetz, Shubhr Singh, Marco Comunit{\`a}, Ilias Ibnyahya,
  Shanxin Yuan, Emmanouil Benetos, and Joshua~D. Reiss.
\newblock {ST-ITO}: Controlling audio effects for style transfer with
  inference-time optimization.
\newblock In \emph{Proceedings of the International Society for Music
  Information Retrieval Conference (ISMIR)}, 2024.

\bibitem[Tan et~al.(2024{\natexlab{a}})Tan, Ai, Chang, Guan, and
  Yang]{tan2024picogen2}
Chih-Pin Tan, Hsin Ai, Yi-Hsin Chang, Shuen-Huei Guan, and Yi-Hsuan Yang.
\newblock Picogen2: Piano cover generation with transfer learning approach and
  weakly aligned data.
\newblock In \emph{Proceedings of the 25th International Society for Music
  Information Retrieval Conference (ISMIR)}, 2024{\natexlab{a}}.

\bibitem[Tan et~al.(2024{\natexlab{b}})Tan, Guan, and Yang]{tan2024picogen}
Chih-Pin Tan, Shuen-Huei Guan, and Yi-Hsuan Yang.
\newblock Picogen: Generate piano covers with a two-stage approach.
\newblock In \emph{Proceedings of the International Conference on Multimedia
  Retrieval (ICMR)}, 2024{\natexlab{b}}.

\bibitem[Tian et~al.(2026)]{tian2026audioomni}
Zeyue Tian et~al.
\newblock {Audio-Omni}: Extending multi-modal understanding to versatile audio
  generation and editing.
\newblock \emph{arXiv preprint arXiv:2604.10708}, 2026.

\bibitem[Tsai et~al.(2024)Tsai, Wu, Kim, Chen, Cheng, and
  Yang]{tsai2024apadapter}
Fang-Duo Tsai, Shih-Lun Wu, Haven Kim, Bo-Yu Chen, Hao-Chung Cheng, and
  Yi-Hsuan Yang.
\newblock Audio prompt adapter: Unleashing music editing abilities for
  text-to-music with lightweight finetuning.
\newblock In \emph{Proceedings of the 25th International Society for Music
  Information Retrieval Conference (ISMIR)}, 2024.

\bibitem[Tsai et~al.(2025)Tsai, Wu, Lee, Yang, Chen, Cheng, and
  Yang]{tsai2025musecontrollite}
Fang-Duo Tsai, Shih-Lun Wu, Weijaw Lee, Sheng-Ping Yang, Bo-Rui Chen, Hao-Chung
  Cheng, and Yi-Hsuan Yang.
\newblock {MuseControlLite}: Multifunctional music generation with lightweight
  conditioners.
\newblock In \emph{Proceedings of the 42nd International Conference on Machine
  Learning (ICML)}, 2025.

\bibitem[Vanka et~al.(2024)Vanka, Steinmetz, Rolland, Reiss, and
  Fazekas]{vanka2024diffmst}
Soumya~Sai Vanka, Christian~J. Steinmetz, Jean-Baptiste Rolland, Joshua~D.
  Reiss, and George Fazekas.
\newblock {Diff-MST}: Differentiable mixing style transfer.
\newblock In \emph{Proceedings of the International Society for Music
  Information Retrieval Conference (ISMIR)}, 2024.

\bibitem[Wang et~al.(2024)Wang, Thakker, Chen, Kanda, Eskimez, Chen,
  et~al.]{wang2023speechx}
Xiaofei Wang, Manthan Thakker, Zhuo Chen, Naoyuki Kanda, Sefik~Emre Eskimez,
  Sanyuan Chen, et~al.
\newblock {SpeechX}: Neural codec language model as a versatile speech
  transformer.
\newblock \emph{IEEE/ACM Transactions on Audio, Speech, and Language
  Processing}, 32:\penalty0 3355--3364, 2024.

\bibitem[Watcharasupat \& Lerch(2024)Watcharasupat and
  Lerch]{watcharasupat2024banquet}
Karn~N. Watcharasupat and Alexander Lerch.
\newblock A stem-agnostic single-decoder system for music source separation
  beyond four stems.
\newblock In \emph{Proceedings of the International Society for Music
  Information Retrieval Conference (ISMIR)}, 2024.

\bibitem[Wu et~al.(2023)Wu, Chen, Zhang, Hui, Nezhurina, Berg-Kirkpatrick, and
  Dubnov]{wu2023clap}
Yusong Wu, Ke~Chen, Tianyu Zhang, Yuchen Hui, Marianna Nezhurina, Taylor
  Berg-Kirkpatrick, and Shlomo Dubnov.
\newblock Large-scale contrastive language-audio pretraining with feature
  fusion and keyword-to-caption augmentation.
\newblock In \emph{IEEE International Conference on Acoustics, Speech and
  Signal Processing (ICASSP)}, 2023.

\bibitem[Yang et~al.(2024)Yang, Yang, Hui, Zheng, Yu, Zhou,
  et~al.]{yang2024qwen2}
An~Yang, Baosong Yang, Binyuan Hui, Bo~Zheng, Bowen Yu, Chang Zhou, et~al.
\newblock Qwen2 technical report.
\newblock \emph{arXiv preprint arXiv:2407.10671}, 2024.

\bibitem[Yang et~al.(2025)Yang, Li, Yang, Zhang, Hui, Zheng,
  et~al.]{yang2025qwen3}
An~Yang, Anfeng Li, Baosong Yang, Beichen Zhang, Binyuan Hui, Bo~Zheng, et~al.
\newblock Qwen3 technical report.
\newblock \emph{arXiv preprint arXiv:2505.09388}, 2025.

\bibitem[Yu et~al.(2025)Yu, Kwak, Jang, Jeong, Huang, Shin, and
  Xie]{yu2024repa}
Sihyun Yu, Sangkyung Kwak, Huiwon Jang, Jongheon Jeong, Jonathan Huang, Jinwoo
  Shin, and Saining Xie.
\newblock Representation alignment for generation: Training diffusion
  transformers is easier than you think.
\newblock In \emph{International Conference on Learning Representations
  (ICLR)}, 2025.

\bibitem[Yu et~al.(2026)Yu, Dujardin, Cai, Zhang, Cruz, Wagner, and
  Cella]{yu2026instructfx2fx}
Song-Ze Yu, Milan~Liessens Dujardin, Yuxuan Cai, Wantong Zhang, Brian Cruz,
  Jeremy Wagner, and Carmine-Emanuele Cella.
\newblock {InstructFX2FX}: A multi-turn text-to-effect system for sequential
  audio effect refinement.
\newblock In \emph{Proceedings of the 29th International Conference on Digital
  Audio Effects (DAFx)}, 2026.

\bibitem[Zhang et~al.(2025{\natexlab{a}})Zhang, Ma, Chen, Wang, Zhao, Pan,
  et~al.]{zhang2025inspiremusic}
Chong Zhang, Yukun Ma, Qian Chen, Wen Wang, Shengkui Zhao, Zexu Pan, et~al.
\newblock {InspireMusic}: Integrating super resolution and large language model
  for high-fidelity long-form music generation.
\newblock \emph{arXiv preprint arXiv:2503.00084}, 2025{\natexlab{a}}.

\bibitem[Zhang et~al.(2025{\natexlab{b}})Zhang, Ikemiya, Choi, Murata,
  Mart{\'i}nez-Ram{\'i}rez, Lin, et~al.]{zhang2024instructmusicgen}
Yixiao Zhang, Yukara Ikemiya, Woosung Choi, Naoki Murata, Marco~A.
  Mart{\'i}nez-Ram{\'i}rez, Liwei Lin, et~al.
\newblock Instruct-musicgen: Unlocking text-to-music editing for music language
  models via instruction tuning.
\newblock In \emph{Proceedings of the 26th International Society for Music
  Information Retrieval Conference (ISMIR)}, 2025{\natexlab{b}}.

\end{thebibliography}
\bibliographystyle{iclr2027_conference}

\clearpage
\appendix

\section{Implementation details}
\label{app:impl}

\paragraph{Autoregressive sampling and classifier-free guidance.}
We sample target codes at temperature 1.0 without top-$k$ or nucleus truncation ($k=0$, $p=1.0$). Generation stops at the first audio end-of-sequence token, with a fallback maximum equal to the source-code length: 2{,}250 decoding steps for a 30\,s clip at 75\,Hz. Instructions are tokenized to at most 192 text tokens. The prefix is encoded once per guidance branch, and decoding uses a key--value cache.
We combine the logits as
\[
\ell = \ell_{\varnothing} + 1.5(\ell_c-\ell_{\varnothing}),
\]
where $c$ is the instruction-conditioned branch. In the null-instruction branch, instruction tokens are replaced with padding tokens that remain attended, matching instruction dropout during training. Both branches retain the same source audio codes and learnable query embeddings. Each branch computes its own edit-head states from its prefix, so the null-instruction queries still see the source audio; these two edit heads remain fixed throughout target generation.

\paragraph{Gated cross-attention.}
All 24 decoder layers contain a cross-attention sublayer after the self-attention residual update and before the MLP normalization and MLP. Each sublayer uses 14 heads of width 64, with separate bias-free $896\times896$ query, key, value, and output projections. The target states receive RMS normalization before supplying the attention queries. A shared memory adapter applies RMS normalization and a bias-free $896\times896$ linear projection to the 32 edit-head states. Each decoder layer has its own scalar $\alpha$, initialized to zero, and scales its cross-attention residual by $\tanh(\alpha)$. The cross-attention sublayers and shared adapter add 77{,}895{,}576 parameters; the 32 learnable query embeddings add 28{,}672, for 77{,}924{,}248 parameters in total (77.92M). This count excludes the training-only alignment projectors and frozen audio encoders and decoders; the 0.5B designation refers to the backbone.

\paragraph{Alignment projectors and temporal resolution.}
The projectors read the layer-4 hidden states at target positions. The MERT projector has three linear layers with widths $896\rightarrow2048\rightarrow2048\rightarrow1024$, with SiLU activations after the first two layers. The pitch projector is $896\rightarrow448\rightarrow135$ with a GELU activation between the two linear layers. The instrument projector uses the same hidden width, $896\rightarrow448\rightarrow8$, with GELU between the two linear layers. For StemFX, target-position states are mean-pooled before a $896\rightarrow512\rightarrow512$ projector with GELU.
Both WavTokenizer and MERT take 24\,kHz mono audio and have a nominal frame rate of 75\,Hz. MERT's convolutional strides have a product of 320 samples, matching WavTokenizer's 320-sample hop. Pitch and instrument rolls are constructed from the target MIDI on the 75\,Hz codec grid, with cover labels taken from the reduced score used to render the target.
StemFX instead takes 44.1\,kHz audio grouped into vocals, bass, drums, and other stems. We compute a 512-dimensional embedding for each of three consecutive 10\,s segments, average the segment embeddings, and normalize the result to unit length. This is a clip-level target and does not require frame-rate matching.

\paragraph{Super-resolution.}
The frozen InspireMusic flow model conditions directly on the sampled WavTokenizer codes to generate HifiCodec latents; the HifiCodec generator renders the 48\,kHz mono waveform. The length regulator linearly interpolates the encoded conditioning sequence to twice its original number of frames, followed by its convolutional refinement layers. Flow matching uses 10 Euler integration steps. Thus the 48\,kHz output is generated through the pretrained latent model and vocoder rather than by waveform resampling alone.

\paragraph{Inference cost.}
On one RTX 5090, processing a 30\,s clip takes approximately 38\,s at batch size 1, or 40\,s with super-resolution. At batch size 8, the amortized time is approximately 5\,s per clip, or 7\,s with super-resolution. These times include source encoding, autoregressive generation, and waveform decoding. We use the paper checkpoint in bf16 with CFG scale 1.5 and generate 2{,}250 codes per clip; each timing is measured on a second run after a warm-up pass.

\clearpage

\section{Full results for individual editing tasks}
\label{app:fulltables}

Tables~\ref{tab:main-slakh} and~\ref{tab:main-real} give the complete task-specific results behind Figure~\ref{fig:headline}.
Their row-wise shading shows the relative range of each metric; it does not make values on different metrics directly comparable.
Some metrics are not reported in several cases, where we mark them as --. Instruct-MusicGen returns a full mix on accompaniment, so we separate its added part with HTDemucs-6s before scoring and omit instrument match. Also, as HTDemucs-6s is used to compute edit rate, edit strength, and isolation strength, the model is omitted on those metrics. Piano-cover systems also always output solo piano, so their instrument match is trivially perfect and is thus omitted.

\begin{table*}[!ht]
\centering
\scriptsize
\setlength{\tabcolsep}{1.5pt}
\renewcommand{\arraystretch}{1.0}
\caption{Full metric set, \textbf{Slakh (synthetic)} domain, N=150 per task. E=editing, P=preservation, E+P=both. Our metrics are defined in Appendix~\ref{app:metrics}; a dash marks an undefined metric. \textbf{Bold}\,= best in column, shaded worst to best; circle and star mark paired tests against ProCraft at $p<0.05$ and $p<0.01$. IMG is Instruct-MusicGen.}
\label{tab:main-slakh}
\resizebox{\textwidth}{!}{%
\begin{tabular}{@{}lrrrrrrrrrr@{}}
\toprule
\textbf{Accompany} & CLAP-T $\uparrow$ & CLAP-A $\uparrow$ & KL $\downarrow$ & FAD $\downarrow$ & KAD $\downarrow$ & PaSST\editm{} $\uparrow$ & APA\presm{} $\uparrow$ & StructDist\presm{} $\downarrow$ &  &  \\
\midrule
ProCraft (ours) & \cellcolor{tabgreen!51}0.200 & \cellcolor{tabgreen!57}0.590 & \cellcolor{tabgreen!98}0.406 & \cellcolor{tabgreen!78}0.476 & \cellcolor{tabgreen!61}1.959 & \cellcolor{tabgreen!87}0.827 & \cellcolor{tabgreen!98}0.614 & \cellcolor{tabgreen!92}216.624 &  &  \\
ProCraft-plain & \cellcolor{tabgreen!53}0.203 & \cellcolor{tabgreen!42}0.581 & \cellcolor{tabgreen!100}0.376 & \cellcolor{tabgreen!82}0.472 & \cellcolor{tabgreen!66}1.937 & \cellcolor{tabgreen!91}0.853 & \cellcolor{tabgreen!89}0.573 & \textbf{\cellcolor{tabgreen!100}211.236} &  &  \\
IMG (released) & \textbf{\cellcolor{tabgreen!100}0.264$^{*}$} & \cellcolor{tabgreen!83}0.604 & \textbf{\cellcolor{tabgreen!100}0.374} & \cellcolor{tabgreen!61}0.492 & \cellcolor{tabgreen!52}2.005 & --- & \cellcolor{tabgreen!87}0.565 & \cellcolor{tabgreen!73}230.702$^{*}$ &  &  \\
IMG (retrained) & \cellcolor{tabgreen!0}0.134$^{*}$ & \cellcolor{tabgreen!0}0.558$^{*}$ & \cellcolor{tabgreen!32}1.600$^{*}$ & \cellcolor{tabgreen!0}0.549 & \cellcolor{tabgreen!0}2.271 & \cellcolor{tabgreen!0}0.240$^{*}$ & \cellcolor{tabgreen!0}0.183 & \cellcolor{tabgreen!28}262.266$^{*}$ &  &  \\
MSDM & \cellcolor{tabgreen!7}0.143$^{*}$ & \textbf{\cellcolor{tabgreen!100}0.614$^{\circ}$} & \cellcolor{tabgreen!0}2.179$^{*}$ & \textbf{\cellcolor{tabgreen!100}0.456} & \textbf{\cellcolor{tabgreen!100}1.761} & \cellcolor{tabgreen!38}0.493$^{*}$ & \cellcolor{tabgreen!94}0.595 & \cellcolor{tabgreen!0}282.190$^{*}$ &  &  \\
MusicGen-Stem & \cellcolor{tabgreen!38}0.183$^{*}$ & \cellcolor{tabgreen!18}0.568$^{\circ}$ & \cellcolor{tabgreen!90}0.558$^{\circ}$ & \cellcolor{tabgreen!79}0.475 & \cellcolor{tabgreen!68}1.924 & \textbf{\cellcolor{tabgreen!100}0.913$^{\circ}$} & \textbf{\cellcolor{tabgreen!100}0.623} & \cellcolor{tabgreen!87}220.629 &  &  \\
\midrule
\textbf{Cover} & CLAP-T $\uparrow$ & CLAP-A $\uparrow$ & KL $\downarrow$ & FAD $\downarrow$ & KAD $\downarrow$ & PaSST\editm{} $\uparrow$ & APA\presm{} $\uparrow$ & chroma\presm{} $\uparrow$ &  &  \\
\midrule
ProCraft (ours) & \textbf{\cellcolor{tabgreen!100}0.243} & \cellcolor{tabgreen!90}0.593 & \textbf{\cellcolor{tabgreen!100}0.202} & \cellcolor{tabgreen!90}0.546 & \cellcolor{tabgreen!94}2.152 & \cellcolor{tabgreen!98}0.993 & \cellcolor{tabgreen!100}0.820 & \cellcolor{tabgreen!85}0.595 &  &  \\
ProCraft-plain & \cellcolor{tabgreen!93}0.241 & \textbf{\cellcolor{tabgreen!100}0.598} & \cellcolor{tabgreen!95}0.213 & \textbf{\cellcolor{tabgreen!100}0.543} & \textbf{\cellcolor{tabgreen!100}2.141} & \textbf{\cellcolor{tabgreen!100}1.000} & \textbf{\cellcolor{tabgreen!100}0.822} & \cellcolor{tabgreen!94}0.618 &  &  \\
IMG (retrained) & \cellcolor{tabgreen!84}0.239 & \cellcolor{tabgreen!65}0.581$^{\circ}$ & \cellcolor{tabgreen!23}0.373$^{*}$ & \cellcolor{tabgreen!7}0.565 & \cellcolor{tabgreen!0}2.327 & \cellcolor{tabgreen!0}0.693$^{*}$ & \cellcolor{tabgreen!0}0.218 & \cellcolor{tabgreen!0}0.370$^{*}$ &  &  \\
Pop2Piano & \cellcolor{tabgreen!12}0.224$^{*}$ & \cellcolor{tabgreen!85}0.591 & \cellcolor{tabgreen!0}0.425$^{*}$ & \cellcolor{tabgreen!43}0.556 & \cellcolor{tabgreen!30}2.271 & --- & \cellcolor{tabgreen!43}0.475 & \textbf{\cellcolor{tabgreen!100}0.634$^{\circ}$} &  &  \\
PiCoGen2 & \cellcolor{tabgreen!0}0.221$^{*}$ & \cellcolor{tabgreen!0}0.549$^{*}$ & \cellcolor{tabgreen!33}0.351$^{*}$ & \cellcolor{tabgreen!0}0.566 & \cellcolor{tabgreen!8}2.312 & --- & \cellcolor{tabgreen!29}0.394 & \cellcolor{tabgreen!85}0.594 &  &  \\
\midrule
\textbf{Instrument transfer} & CLAP-T $\uparrow$ & CLAP-A $\uparrow$ & KL $\downarrow$ & FAD $\downarrow$ & KAD $\downarrow$ & PaSST\editm{} $\uparrow$ & Frame-F1\presm{} $\uparrow$ & chroma\presm{} $\uparrow$ &  &  \\
\midrule
ProCraft (ours) & \cellcolor{tabgreen!72}0.145 & \cellcolor{tabgreen!98}0.557 & \textbf{\cellcolor{tabgreen!100}0.256} & \cellcolor{tabgreen!73}0.834 & \cellcolor{tabgreen!55}3.498 & \cellcolor{tabgreen!96}0.640 & \cellcolor{tabgreen!86}0.558 & \textbf{\cellcolor{tabgreen!100}0.673} &  &  \\
ProCraft-plain & \cellcolor{tabgreen!79}0.152 & \textbf{\cellcolor{tabgreen!100}0.560} & \cellcolor{tabgreen!95}0.288 & \cellcolor{tabgreen!49}0.856 & \cellcolor{tabgreen!38}3.603 & \cellcolor{tabgreen!92}0.627 & \cellcolor{tabgreen!83}0.546 & \cellcolor{tabgreen!95}0.654 &  &  \\
IMG (retrained) & \cellcolor{tabgreen!0}0.073$^{*}$ & \cellcolor{tabgreen!0}0.415$^{*}$ & \cellcolor{tabgreen!0}0.852$^{*}$ & \cellcolor{tabgreen!0}0.901 & \cellcolor{tabgreen!0}3.841 & \cellcolor{tabgreen!0}0.307$^{*}$ & \cellcolor{tabgreen!0}0.071 & \cellcolor{tabgreen!1}0.278$^{*}$ &  &  \\
AP-Adapter & \textbf{\cellcolor{tabgreen!100}0.173$^{*}$} & \cellcolor{tabgreen!57}0.498$^{*}$ & \cellcolor{tabgreen!6}0.815$^{*}$ & \cellcolor{tabgreen!49}0.856 & \cellcolor{tabgreen!61}3.461 & \cellcolor{tabgreen!98}0.647 & \cellcolor{tabgreen!42}0.311 & \cellcolor{tabgreen!59}0.512$^{*}$ &  &  \\
MuseCtrlLite & \cellcolor{tabgreen!58}0.131 & \cellcolor{tabgreen!95}0.553 & \cellcolor{tabgreen!22}0.720$^{*}$ & \cellcolor{tabgreen!35}0.869 & \cellcolor{tabgreen!26}3.676 & \cellcolor{tabgreen!67}0.540$^{*}$ & \cellcolor{tabgreen!95}0.614 & \cellcolor{tabgreen!50}0.476$^{*}$ &  &  \\
DDPM-inv & \cellcolor{tabgreen!73}0.146 & \cellcolor{tabgreen!94}0.551 & \cellcolor{tabgreen!45}0.582$^{*}$ & \textbf{\cellcolor{tabgreen!100}0.810} & \textbf{\cellcolor{tabgreen!100}3.213} & \textbf{\cellcolor{tabgreen!100}0.653} & \cellcolor{tabgreen!45}0.326 & \cellcolor{tabgreen!83}0.606$^{*}$ &  &  \\
TokenSynth & \cellcolor{tabgreen!70}0.143 & \cellcolor{tabgreen!75}0.523$^{*}$ & \cellcolor{tabgreen!70}0.434$^{*}$ & \cellcolor{tabgreen!12}0.890 & \cellcolor{tabgreen!54}3.504 & \cellcolor{tabgreen!33}0.420$^{*}$ & \textbf{\cellcolor{tabgreen!100}0.641} & \cellcolor{tabgreen!0}0.276$^{*}$ &  &  \\
\midrule
\textbf{Effects} & CLAP-T $\uparrow$ & CLAP-A $\uparrow$ & KL $\downarrow$ & FAD $\downarrow$ & KAD $\downarrow$ & mel-L1\bothm{} $\downarrow$ & MRSTFT\bothm{} $\downarrow$ & AFx\editm{} $\uparrow$ &  &  \\
\midrule
ProCraft (ours) & \cellcolor{tabgreen!1}0.193 & \cellcolor{tabgreen!75}0.776 & \cellcolor{tabgreen!67}0.437 & \textbf{\cellcolor{tabgreen!100}0.445} & \cellcolor{tabgreen!99}1.792 & \cellcolor{tabgreen!82}0.871 & \cellcolor{tabgreen!71}1.889 & \cellcolor{tabgreen!57}0.393 &  &  \\
ProCraft-plain & \cellcolor{tabgreen!11}0.203$^{*}$ & \cellcolor{tabgreen!74}0.774 & \cellcolor{tabgreen!75}0.411 & \cellcolor{tabgreen!100}0.446 & \textbf{\cellcolor{tabgreen!100}1.787} & \cellcolor{tabgreen!84}0.857 & \cellcolor{tabgreen!73}1.865 & \cellcolor{tabgreen!64}0.402$^{\circ}$ &  &  \\
IMG (retrained) & \cellcolor{tabgreen!0}0.191 & \cellcolor{tabgreen!2}0.511$^{*}$ & \cellcolor{tabgreen!20}0.591$^{*}$ & \cellcolor{tabgreen!67}0.510 & \cellcolor{tabgreen!69}2.061 & \cellcolor{tabgreen!61}1.024$^{*}$ & \cellcolor{tabgreen!52}2.178$^{*}$ & \cellcolor{tabgreen!26}0.350$^{*}$ &  &  \\
InstructFX2FX & \cellcolor{tabgreen!9}0.201 & \textbf{\cellcolor{tabgreen!100}0.867$^{*}$} & \textbf{\cellcolor{tabgreen!100}0.331$^{*}$} & \cellcolor{tabgreen!66}0.512 & \cellcolor{tabgreen!67}2.075 & \textbf{\cellcolor{tabgreen!100}0.743$^{*}$} & \textbf{\cellcolor{tabgreen!100}1.458$^{*}$} & \textbf{\cellcolor{tabgreen!100}0.452$^{*}$} &  &  \\
FxSearcher & \textbf{\cellcolor{tabgreen!100}0.304$^{*}$} & \cellcolor{tabgreen!0}0.505$^{*}$ & \cellcolor{tabgreen!0}0.657$^{*}$ & \cellcolor{tabgreen!45}0.553 & \cellcolor{tabgreen!81}1.953 & \cellcolor{tabgreen!0}1.467$^{*}$ & \cellcolor{tabgreen!0}2.943$^{*}$ & \cellcolor{tabgreen!42}0.373 &  &  \\
Text2FX & \cellcolor{tabgreen!68}0.269$^{*}$ & \cellcolor{tabgreen!9}0.537$^{*}$ & \cellcolor{tabgreen!42}0.519$^{*}$ & \cellcolor{tabgreen!0}0.641 & \cellcolor{tabgreen!0}2.672 & \cellcolor{tabgreen!34}1.220$^{*}$ & \cellcolor{tabgreen!41}2.327$^{*}$ & \cellcolor{tabgreen!0}0.314$^{*}$ &  &  \\
\midrule
\textbf{Remove} & CLAP-T $\uparrow$ & CLAP-A $\uparrow$ & KL $\downarrow$ & FAD $\downarrow$ & KAD $\downarrow$ & SI-SDR\bothm{} $\uparrow$ & mel-L1\bothm{} $\downarrow$ & MRSTFT\bothm{} $\downarrow$ & edit rate\editm{} $\uparrow$ & edit strength\editm{} $\uparrow$ \\
\midrule
ProCraft (ours) & \cellcolor{tabgreen!92}0.108 & \cellcolor{tabgreen!56}0.774 & \cellcolor{tabgreen!58}0.367 & \cellcolor{tabgreen!95}0.527 & \cellcolor{tabgreen!97}2.143 & \cellcolor{tabgreen!46}-2.961 & \cellcolor{tabgreen!52}0.975 & \cellcolor{tabgreen!36}1.874 & \cellcolor{tabgreen!42}0.527 & \cellcolor{tabgreen!59}10.153 \\
ProCraft-plain & \textbf{\cellcolor{tabgreen!100}0.112} & \cellcolor{tabgreen!56}0.773 & \cellcolor{tabgreen!49}0.403 & \textbf{\cellcolor{tabgreen!100}0.520} & \textbf{\cellcolor{tabgreen!100}2.123} & \cellcolor{tabgreen!44}-3.370 & \cellcolor{tabgreen!52}0.981 & \cellcolor{tabgreen!36}1.881 & \cellcolor{tabgreen!8}0.473 & \cellcolor{tabgreen!26}7.983$^{\circ}$ \\
IMG (released) & \cellcolor{tabgreen!11}0.068$^{*}$ & \cellcolor{tabgreen!64}0.803$^{*}$ & \cellcolor{tabgreen!75}0.298 & \cellcolor{tabgreen!74}0.554 & \cellcolor{tabgreen!77}2.272 & \cellcolor{tabgreen!56}-0.872$^{*}$ & \cellcolor{tabgreen!55}0.944 & \cellcolor{tabgreen!44}1.751$^{\circ}$ & \textbf{\cellcolor{tabgreen!100}0.620} & \textbf{\cellcolor{tabgreen!100}12.890$^{\circ}$} \\
IMG (retrained) & \cellcolor{tabgreen!7}0.066$^{*}$ & \cellcolor{tabgreen!0}0.564$^{*}$ & \cellcolor{tabgreen!0}0.599$^{*}$ & \cellcolor{tabgreen!0}0.654 & \cellcolor{tabgreen!0}2.779 & \cellcolor{tabgreen!0}-12.431$^{*}$ & \cellcolor{tabgreen!21}1.294$^{*}$ & \cellcolor{tabgreen!0}2.452$^{*}$ & \cellcolor{tabgreen!0}0.460 & \cellcolor{tabgreen!0}6.267$^{*}$ \\
AudioSep & \cellcolor{tabgreen!0}0.062$^{*}$ & \cellcolor{tabgreen!23}0.650$^{*}$ & \cellcolor{tabgreen!66}0.332 & \cellcolor{tabgreen!35}0.607 & \cellcolor{tabgreen!57}2.405 & \cellcolor{tabgreen!29}-6.317$^{*}$ & \cellcolor{tabgreen!0}1.500$^{*}$ & \cellcolor{tabgreen!9}2.302$^{*}$ & \cellcolor{tabgreen!17}0.487 & \cellcolor{tabgreen!45}9.271 \\
HTDemucs-6s & \cellcolor{tabgreen!33}0.078$^{*}$ & \cellcolor{tabgreen!79}0.859$^{*}$ & \textbf{\cellcolor{tabgreen!100}0.198$^{*}$} & \cellcolor{tabgreen!80}0.547 & \cellcolor{tabgreen!80}2.255 & \textbf{\cellcolor{tabgreen!100}8.312$^{*}$} & \cellcolor{tabgreen!95}0.544$^{*}$ & \cellcolor{tabgreen!92}0.973$^{*}$ & --- & --- \\
Banquet & \cellcolor{tabgreen!81}0.103 & \textbf{\cellcolor{tabgreen!100}0.937$^{*}$} & \cellcolor{tabgreen!98}0.205$^{*}$ & \cellcolor{tabgreen!55}0.580 & \cellcolor{tabgreen!51}2.443 & \cellcolor{tabgreen!96}7.565$^{*}$ & \textbf{\cellcolor{tabgreen!100}0.494$^{*}$} & \textbf{\cellcolor{tabgreen!100}0.852$^{*}$} & \cellcolor{tabgreen!12}0.480 & \cellcolor{tabgreen!73}11.109 \\
\midrule
\textbf{Extract} & CLAP-T $\uparrow$ & CLAP-A $\uparrow$ & KL $\downarrow$ & FAD $\downarrow$ & KAD $\downarrow$ & SI-SDR\bothm{} $\uparrow$ & mel-L1\bothm{} $\downarrow$ & MRSTFT\bothm{} $\downarrow$ & edit rate\editm{} $\uparrow$ & edit strength\editm{} $\uparrow$ \\
\midrule
ProCraft (ours) & \cellcolor{tabgreen!96}0.258 & \cellcolor{tabgreen!42}0.568 & \cellcolor{tabgreen!88}0.402 & \cellcolor{tabgreen!81}0.719 & \cellcolor{tabgreen!82}2.827 & \cellcolor{tabgreen!61}-11.346 & \cellcolor{tabgreen!69}1.577 & \cellcolor{tabgreen!59}2.827 & \textbf{\cellcolor{tabgreen!100}0.787} & \textbf{\cellcolor{tabgreen!100}12.176} \\
ProCraft-plain & \textbf{\cellcolor{tabgreen!100}0.262} & \cellcolor{tabgreen!43}0.573 & \cellcolor{tabgreen!85}0.435 & \cellcolor{tabgreen!76}0.729 & \cellcolor{tabgreen!77}2.885 & \cellcolor{tabgreen!57}-12.743$^{*}$ & \cellcolor{tabgreen!66}1.607 & \cellcolor{tabgreen!60}2.815 & \cellcolor{tabgreen!89}0.707$^{\circ}$ & \cellcolor{tabgreen!93}10.463 \\
IMG (released) & \cellcolor{tabgreen!56}0.219$^{*}$ & \cellcolor{tabgreen!39}0.553 & \cellcolor{tabgreen!96}0.310$^{\circ}$ & \cellcolor{tabgreen!47}0.791 & \cellcolor{tabgreen!50}3.163 & \cellcolor{tabgreen!68}-8.488$^{*}$ & \cellcolor{tabgreen!59}1.672$^{\circ}$ & \cellcolor{tabgreen!58}2.848 & \cellcolor{tabgreen!72}0.587$^{*}$ & \cellcolor{tabgreen!65}3.480$^{*}$ \\
IMG (retrained) & \cellcolor{tabgreen!0}0.165$^{*}$ & \cellcolor{tabgreen!0}0.337$^{*}$ & \cellcolor{tabgreen!0}1.423$^{*}$ & \cellcolor{tabgreen!0}0.890 & \cellcolor{tabgreen!0}3.684 & \cellcolor{tabgreen!0}-33.954$^{*}$ & \cellcolor{tabgreen!0}2.233$^{*}$ & \cellcolor{tabgreen!0}3.677$^{*}$ & \cellcolor{tabgreen!13}0.160$^{*}$ & \cellcolor{tabgreen!0}-12.369$^{*}$ \\
AudioSep & \cellcolor{tabgreen!27}0.192$^{*}$ & \textbf{\cellcolor{tabgreen!100}0.889$^{*}$} & \cellcolor{tabgreen!63}0.692$^{*}$ & \textbf{\cellcolor{tabgreen!100}0.679} & \cellcolor{tabgreen!69}2.967 & \cellcolor{tabgreen!82}-3.525$^{*}$ & \cellcolor{tabgreen!59}1.675 & \cellcolor{tabgreen!55}2.891 & \cellcolor{tabgreen!0}0.067$^{*}$ & \cellcolor{tabgreen!27}-5.802$^{*}$ \\
HTDemucs-6s & \cellcolor{tabgreen!44}0.208$^{*}$ & \cellcolor{tabgreen!33}0.519$^{*}$ & \textbf{\cellcolor{tabgreen!100}0.259$^{*}$} & \cellcolor{tabgreen!78}0.725 & \textbf{\cellcolor{tabgreen!100}2.642} & \textbf{\cellcolor{tabgreen!100}3.356$^{*}$} & \textbf{\cellcolor{tabgreen!100}1.280$^{*}$} & \textbf{\cellcolor{tabgreen!100}2.239$^{*}$} & --- & --- \\
Banquet & \cellcolor{tabgreen!46}0.210$^{*}$ & \cellcolor{tabgreen!25}0.474$^{*}$ & \cellcolor{tabgreen!87}0.410 & \cellcolor{tabgreen!40}0.806 & \cellcolor{tabgreen!61}3.044 & \cellcolor{tabgreen!90}-0.421$^{*}$ & \cellcolor{tabgreen!59}1.671 & \cellcolor{tabgreen!75}2.599$^{*}$ & \cellcolor{tabgreen!77}0.620$^{*}$ & \cellcolor{tabgreen!86}8.658$^{\circ}$ \\
\midrule
\textbf{Denoise} & CLAP-T $\uparrow$ & CLAP-A $\uparrow$ & KL $\downarrow$ & FAD $\downarrow$ & KAD $\downarrow$ & SI-SDR\bothm{} $\uparrow$ & mel-L1\bothm{} $\downarrow$ & MRSTFT\bothm{} $\downarrow$ &  &  \\
\midrule
ProCraft (ours) & \cellcolor{tabgreen!95}0.069 & \cellcolor{tabgreen!42}0.637 & \cellcolor{tabgreen!94}0.223 & \cellcolor{tabgreen!93}0.459 & \cellcolor{tabgreen!88}1.893 & \cellcolor{tabgreen!40}0.804 & \cellcolor{tabgreen!54}0.753 & \cellcolor{tabgreen!45}1.584 &  &  \\
ProCraft-plain & \cellcolor{tabgreen!95}0.069 & \cellcolor{tabgreen!41}0.633 & \cellcolor{tabgreen!90}0.230 & \cellcolor{tabgreen!95}0.457 & \cellcolor{tabgreen!91}1.877 & \cellcolor{tabgreen!40}0.782 & \cellcolor{tabgreen!50}0.777$^{\circ}$ & \cellcolor{tabgreen!43}1.614$^{\circ}$ &  &  \\
IMG (retrained) & \cellcolor{tabgreen!31}0.043$^{*}$ & \cellcolor{tabgreen!0}0.516$^{*}$ & \cellcolor{tabgreen!0}0.415$^{*}$ & \cellcolor{tabgreen!0}0.561 & \cellcolor{tabgreen!0}2.332 & \cellcolor{tabgreen!0}-7.825$^{*}$ & \cellcolor{tabgreen!0}1.089$^{*}$ & \cellcolor{tabgreen!0}2.137$^{*}$ &  &  \\
AudioSep & \textbf{\cellcolor{tabgreen!100}0.071} & \textbf{\cellcolor{tabgreen!100}0.805$^{*}$} & \textbf{\cellcolor{tabgreen!100}0.210} & \cellcolor{tabgreen!58}0.498 & \cellcolor{tabgreen!55}2.058 & \textbf{\cellcolor{tabgreen!100}13.903$^{*}$} & \textbf{\cellcolor{tabgreen!100}0.471$^{*}$} & \textbf{\cellcolor{tabgreen!100}0.917$^{*}$} &  &  \\
SD-Codec & \cellcolor{tabgreen!0}0.030$^{*}$ & \cellcolor{tabgreen!47}0.651 & \cellcolor{tabgreen!83}0.244 & \textbf{\cellcolor{tabgreen!100}0.451} & \textbf{\cellcolor{tabgreen!100}1.832} & \cellcolor{tabgreen!76}8.627$^{*}$ & \cellcolor{tabgreen!35}0.871$^{*}$ & \cellcolor{tabgreen!14}1.971$^{*}$ &  &  \\
\bottomrule
\end{tabular}}
\end{table*}

\begin{table*}[!ht]
\centering
\scriptsize
\setlength{\tabcolsep}{1.5pt}
\renewcommand{\arraystretch}{1.0}
\caption{Full metric set, \textbf{real audio} domain, N=150 per task. E=editing, P=preservation, E+P=both. Our metrics are defined in Appendix~\ref{app:metrics}; a dash marks an undefined metric. \textbf{Bold}\,= best in column, shaded worst to best; circle and star mark paired tests against ProCraft at $p<0.05$ and $p<0.01$. IMG is Instruct-MusicGen.}
\label{tab:main-real}
\resizebox{\textwidth}{!}{%
\begin{tabular}{@{}lrrrrrrrrrr@{}}
\toprule
\textbf{Accompany} & CLAP-T $\uparrow$ & CLAP-A $\uparrow$ & KL $\downarrow$ & FAD $\downarrow$ & KAD $\downarrow$ & PaSST\editm{} $\uparrow$ & APA\presm{} $\uparrow$ & StructDist\presm{} $\downarrow$ &  &  \\
\midrule
ProCraft (ours) & \cellcolor{tabgreen!33}0.165 & \cellcolor{tabgreen!4}0.518 & \cellcolor{tabgreen!98}0.472 & \textbf{\cellcolor{tabgreen!100}0.264} & \cellcolor{tabgreen!97}0.939 & \cellcolor{tabgreen!85}0.840 & \cellcolor{tabgreen!70}0.366 & \textbf{\cellcolor{tabgreen!100}209.256} &  &  \\
ProCraft-plain & \cellcolor{tabgreen!34}0.166 & \cellcolor{tabgreen!11}0.526 & \cellcolor{tabgreen!98}0.469 & \cellcolor{tabgreen!85}0.272 & \cellcolor{tabgreen!73}0.995 & \cellcolor{tabgreen!85}0.840 & \cellcolor{tabgreen!71}0.369 & \cellcolor{tabgreen!90}216.648$^{\circ}$ &  &  \\
IMG (released) & \textbf{\cellcolor{tabgreen!100}0.257$^{*}$} & \cellcolor{tabgreen!19}0.535$^{\circ}$ & \textbf{\cellcolor{tabgreen!100}0.442} & \cellcolor{tabgreen!100}0.264 & \textbf{\cellcolor{tabgreen!100}0.933} & --- & \cellcolor{tabgreen!78}0.409 & \cellcolor{tabgreen!89}217.349$^{\circ}$ &  &  \\
IMG (retrained) & \cellcolor{tabgreen!1}0.121$^{*}$ & \cellcolor{tabgreen!0}0.513 & \cellcolor{tabgreen!0}1.654$^{*}$ & \cellcolor{tabgreen!0}0.323 & \cellcolor{tabgreen!0}1.160 & \cellcolor{tabgreen!0}0.227$^{*}$ & \cellcolor{tabgreen!0}0.000 & \cellcolor{tabgreen!25}262.031$^{*}$ &  &  \\
MSDM & \cellcolor{tabgreen!0}0.119$^{*}$ & \textbf{\cellcolor{tabgreen!100}0.629$^{*}$} & \cellcolor{tabgreen!6}1.584$^{*}$ & \cellcolor{tabgreen!14}0.315 & \cellcolor{tabgreen!17}1.121 & \cellcolor{tabgreen!46}0.560$^{*}$ & \cellcolor{tabgreen!61}0.321 & \cellcolor{tabgreen!0}279.847$^{*}$ &  &  \\
MusicGen-Stem & \cellcolor{tabgreen!11}0.134$^{*}$ & \cellcolor{tabgreen!6}0.520 & \cellcolor{tabgreen!85}0.622$^{\circ}$ & \cellcolor{tabgreen!62}0.286 & \cellcolor{tabgreen!61}1.022 & \textbf{\cellcolor{tabgreen!100}0.947$^{*}$} & \textbf{\cellcolor{tabgreen!100}0.524} & \cellcolor{tabgreen!86}219.187$^{*}$ &  &  \\
\midrule
\textbf{Cover} & CLAP-T $\uparrow$ & CLAP-A $\uparrow$ & FAD $\downarrow$ & KAD $\downarrow$ & PaSST\editm{} $\uparrow$ & APA\presm{} $\uparrow$ & chroma\presm{} $\uparrow$ &  &  &  \\
\midrule
ProCraft (ours) & \cellcolor{tabgreen!90}0.246 & \cellcolor{tabgreen!40}0.501 & \cellcolor{tabgreen!21}0.364 & \cellcolor{tabgreen!43}1.331 & \cellcolor{tabgreen!96}0.987 & \textbf{\cellcolor{tabgreen!100}0.775} & \cellcolor{tabgreen!78}0.614 &  &  &  \\
ProCraft-plain & \textbf{\cellcolor{tabgreen!100}0.250} & \cellcolor{tabgreen!43}0.504 & \cellcolor{tabgreen!24}0.363 & \cellcolor{tabgreen!49}1.318 & \textbf{\cellcolor{tabgreen!100}1.000} & \cellcolor{tabgreen!95}0.751 & \cellcolor{tabgreen!75}0.608 &  &  &  \\
IMG (retrained) & \cellcolor{tabgreen!65}0.236 & \cellcolor{tabgreen!0}0.455$^{*}$ & \textbf{\cellcolor{tabgreen!100}0.326} & \textbf{\cellcolor{tabgreen!100}1.202} & \cellcolor{tabgreen!0}0.693$^{*}$ & \cellcolor{tabgreen!0}0.321 & \cellcolor{tabgreen!0}0.415$^{*}$ &  &  &  \\
Pop2Piano & \cellcolor{tabgreen!0}0.210$^{*}$ & \textbf{\cellcolor{tabgreen!100}0.569$^{*}$} & \cellcolor{tabgreen!52}0.350 & \cellcolor{tabgreen!40}1.338 & --- & \cellcolor{tabgreen!86}0.712 & \cellcolor{tabgreen!6}0.429$^{*}$ &  &  &  \\
PiCoGen2 & \cellcolor{tabgreen!52}0.231$^{*}$ & \cellcolor{tabgreen!4}0.459$^{*}$ & \cellcolor{tabgreen!0}0.374 & \cellcolor{tabgreen!0}1.428 & --- & \cellcolor{tabgreen!68}0.628 & \textbf{\cellcolor{tabgreen!100}0.672$^{*}$} &  &  &  \\
\midrule
\textbf{Instrument transfer} & CLAP-T $\uparrow$ & CLAP-A $\uparrow$ & FAD $\downarrow$ & KAD $\downarrow$ & PaSST\editm{} $\uparrow$ & chroma\presm{} $\uparrow$ &  &  &  &  \\
\midrule
ProCraft (ours) & \cellcolor{tabgreen!51}0.136 & \cellcolor{tabgreen!46}0.411 & \cellcolor{tabgreen!80}0.704 & \cellcolor{tabgreen!80}2.721 & \cellcolor{tabgreen!88}0.607 & \cellcolor{tabgreen!82}0.569 &  &  &  &  \\
ProCraft-plain & \cellcolor{tabgreen!75}0.163$^{*}$ & \cellcolor{tabgreen!72}0.457$^{*}$ & \cellcolor{tabgreen!75}0.716 & \cellcolor{tabgreen!74}2.797 & \cellcolor{tabgreen!98}0.647 & \cellcolor{tabgreen!83}0.570 &  &  &  &  \\
IMG (retrained) & \cellcolor{tabgreen!0}0.081$^{*}$ & \cellcolor{tabgreen!0}0.327$^{*}$ & \cellcolor{tabgreen!0}0.906 & \cellcolor{tabgreen!0}3.805 & \cellcolor{tabgreen!12}0.313$^{*}$ & \cellcolor{tabgreen!0}0.275$^{*}$ &  &  &  &  \\
AP-Adapter & \textbf{\cellcolor{tabgreen!100}0.190$^{*}$} & \cellcolor{tabgreen!76}0.465$^{*}$ & \cellcolor{tabgreen!62}0.748 & \cellcolor{tabgreen!56}3.040 & \textbf{\cellcolor{tabgreen!100}0.653} & \textbf{\cellcolor{tabgreen!100}0.632$^{*}$} &  &  &  &  \\
MuseCtrlLite & \cellcolor{tabgreen!47}0.132 & \textbf{\cellcolor{tabgreen!100}0.509$^{*}$} & \textbf{\cellcolor{tabgreen!100}0.652} & \textbf{\cellcolor{tabgreen!100}2.442} & \cellcolor{tabgreen!0}0.267$^{*}$ & \cellcolor{tabgreen!64}0.505$^{*}$ &  &  &  &  \\
DDPM-inv & \cellcolor{tabgreen!89}0.178$^{*}$ & \cellcolor{tabgreen!84}0.480$^{*}$ & \cellcolor{tabgreen!92}0.671 & \cellcolor{tabgreen!91}2.560 & \cellcolor{tabgreen!53}0.473$^{\circ}$ & \cellcolor{tabgreen!77}0.549 &  &  &  &  \\
\midrule
\textbf{Effects} & CLAP-T $\uparrow$ & CLAP-A $\uparrow$ & KL $\downarrow$ & FAD $\downarrow$ & KAD $\downarrow$ & mel-L1\bothm{} $\downarrow$ & MRSTFT\bothm{} $\downarrow$ & AFx\editm{} $\uparrow$ &  &  \\
\midrule
ProCraft (ours) & \cellcolor{tabgreen!21}0.185 & \cellcolor{tabgreen!84}0.766 & \cellcolor{tabgreen!64}0.598 & \textbf{\cellcolor{tabgreen!100}0.298} & \textbf{\cellcolor{tabgreen!100}1.049} & \cellcolor{tabgreen!88}0.756 & \cellcolor{tabgreen!66}1.761 & \cellcolor{tabgreen!26}0.195 &  &  \\
ProCraft-plain & \cellcolor{tabgreen!16}0.180 & \cellcolor{tabgreen!92}0.797$^{*}$ & \cellcolor{tabgreen!69}0.567 & \cellcolor{tabgreen!94}0.315 & \cellcolor{tabgreen!95}1.112 & \cellcolor{tabgreen!89}0.747 & \cellcolor{tabgreen!67}1.751 & \cellcolor{tabgreen!27}0.202 &  &  \\
IMG (retrained) & \cellcolor{tabgreen!26}0.189 & \cellcolor{tabgreen!0}0.424$^{*}$ & \cellcolor{tabgreen!0}0.951$^{*}$ & \cellcolor{tabgreen!27}0.503 & \cellcolor{tabgreen!26}1.964 & \cellcolor{tabgreen!40}1.049$^{*}$ & \cellcolor{tabgreen!26}2.314$^{*}$ & \cellcolor{tabgreen!6}0.118$^{*}$ &  &  \\
InstructFX2FX & \cellcolor{tabgreen!0}0.165 & \textbf{\cellcolor{tabgreen!100}0.829$^{*}$} & \textbf{\cellcolor{tabgreen!100}0.398$^{*}$} & \cellcolor{tabgreen!94}0.316 & \cellcolor{tabgreen!99}1.056 & \textbf{\cellcolor{tabgreen!100}0.680$^{\circ}$} & \textbf{\cellcolor{tabgreen!100}1.292$^{*}$} & \textbf{\cellcolor{tabgreen!100}0.493$^{*}$} &  &  \\
FxSearcher & \textbf{\cellcolor{tabgreen!100}0.257$^{*}$} & \cellcolor{tabgreen!14}0.481$^{*}$ & \cellcolor{tabgreen!25}0.811$^{*}$ & \cellcolor{tabgreen!57}0.420 & \cellcolor{tabgreen!74}1.375 & \cellcolor{tabgreen!0}1.297$^{*}$ & \cellcolor{tabgreen!0}2.679$^{*}$ & \cellcolor{tabgreen!16}0.157$^{*}$ &  &  \\
Text2FX & \cellcolor{tabgreen!76}0.235$^{*}$ & \cellcolor{tabgreen!9}0.459$^{*}$ & \cellcolor{tabgreen!39}0.733 & \cellcolor{tabgreen!0}0.578 & \cellcolor{tabgreen!0}2.288 & \cellcolor{tabgreen!14}1.208$^{*}$ & \cellcolor{tabgreen!27}2.303$^{*}$ & \cellcolor{tabgreen!0}0.092$^{*}$ &  &  \\
\midrule
\textbf{Remove} & CLAP-T $\uparrow$ & CLAP-A $\uparrow$ & KL $\downarrow$ & FAD $\downarrow$ & KAD $\downarrow$ & SI-SDR\bothm{} $\uparrow$ & mel-L1\bothm{} $\downarrow$ & MRSTFT\bothm{} $\downarrow$ & edit rate\editm{} $\uparrow$ & edit strength\editm{} $\uparrow$ \\
\midrule
ProCraft (ours) & \cellcolor{tabgreen!56}0.091 & \cellcolor{tabgreen!75}0.770 & \cellcolor{tabgreen!48}0.559 & \cellcolor{tabgreen!95}0.344 & \cellcolor{tabgreen!93}1.251 & \cellcolor{tabgreen!48}-5.114 & \cellcolor{tabgreen!51}1.077 & \cellcolor{tabgreen!40}2.100 & \cellcolor{tabgreen!26}0.380 & \cellcolor{tabgreen!19}7.373 \\
ProCraft-plain & \cellcolor{tabgreen!64}0.095 & \cellcolor{tabgreen!82}0.804$^{*}$ & \cellcolor{tabgreen!35}0.665$^{*}$ & \cellcolor{tabgreen!93}0.348 & \cellcolor{tabgreen!93}1.252 & \cellcolor{tabgreen!49}-4.803 & \cellcolor{tabgreen!45}1.144$^{\circ}$ & \cellcolor{tabgreen!35}2.200$^{\circ}$ & \cellcolor{tabgreen!0}0.260$^{*}$ & \cellcolor{tabgreen!0}4.936$^{*}$ \\
IMG (released) & \cellcolor{tabgreen!40}0.084 & \cellcolor{tabgreen!62}0.711$^{*}$ & \cellcolor{tabgreen!60}0.461$^{\circ}$ & \cellcolor{tabgreen!88}0.361 & \cellcolor{tabgreen!89}1.301 & \cellcolor{tabgreen!54}-2.914$^{*}$ & \cellcolor{tabgreen!55}1.028 & \cellcolor{tabgreen!45}2.000 & \cellcolor{tabgreen!40}0.440 & \cellcolor{tabgreen!16}7.017 \\
IMG (retrained) & \cellcolor{tabgreen!0}0.066$^{*}$ & \cellcolor{tabgreen!0}0.422$^{*}$ & \cellcolor{tabgreen!0}0.959$^{*}$ & \cellcolor{tabgreen!0}0.586 & \cellcolor{tabgreen!0}2.382 & \cellcolor{tabgreen!0}-21.513$^{*}$ & \cellcolor{tabgreen!22}1.430$^{*}$ & \cellcolor{tabgreen!7}2.778$^{*}$ & \cellcolor{tabgreen!44}0.460 & \cellcolor{tabgreen!13}6.637 \\
AudioSep & \cellcolor{tabgreen!13}0.072$^{*}$ & \cellcolor{tabgreen!13}0.481$^{*}$ & \cellcolor{tabgreen!46}0.575 & \cellcolor{tabgreen!13}0.552 & \cellcolor{tabgreen!28}2.046 & \cellcolor{tabgreen!30}-11.242$^{*}$ & \cellcolor{tabgreen!0}1.699$^{*}$ & \cellcolor{tabgreen!0}2.926$^{*}$ & \cellcolor{tabgreen!78}0.613$^{*}$ & \cellcolor{tabgreen!57}12.173$^{*}$ \\
HTDemucs-6s & \cellcolor{tabgreen!75}0.100 & \cellcolor{tabgreen!85}0.818$^{*}$ & \textbf{\cellcolor{tabgreen!100}0.130$^{*}$} & \cellcolor{tabgreen!82}0.377 & \cellcolor{tabgreen!87}1.324 & \textbf{\cellcolor{tabgreen!100}12.854$^{*}$} & \textbf{\cellcolor{tabgreen!100}0.478$^{*}$} & \textbf{\cellcolor{tabgreen!100}0.844$^{*}$} & --- & --- \\
Banquet & \textbf{\cellcolor{tabgreen!100}0.111$^{*}$} & \textbf{\cellcolor{tabgreen!100}0.886$^{*}$} & \cellcolor{tabgreen!95}0.170$^{*}$ & \textbf{\cellcolor{tabgreen!100}0.330} & \textbf{\cellcolor{tabgreen!100}1.165} & \cellcolor{tabgreen!97}11.850$^{*}$ & \cellcolor{tabgreen!98}0.503$^{*}$ & \cellcolor{tabgreen!98}0.887$^{*}$ & \textbf{\cellcolor{tabgreen!100}0.713$^{*}$} & \textbf{\cellcolor{tabgreen!100}17.550$^{*}$} \\
\midrule
\textbf{Extract} & CLAP-T $\uparrow$ & CLAP-A $\uparrow$ & KL $\downarrow$ & FAD $\downarrow$ & KAD $\downarrow$ & SI-SDR\bothm{} $\uparrow$ & mel-L1\bothm{} $\downarrow$ & MRSTFT\bothm{} $\downarrow$ & edit rate\editm{} $\uparrow$ & edit strength\editm{} $\uparrow$ \\
\midrule
ProCraft (ours) & \cellcolor{tabgreen!97}0.242 & \cellcolor{tabgreen!44}0.553 & \cellcolor{tabgreen!86}0.395 & \cellcolor{tabgreen!64}0.524 & \cellcolor{tabgreen!74}1.831 & \cellcolor{tabgreen!46}-15.137 & \cellcolor{tabgreen!65}1.486 & \cellcolor{tabgreen!56}2.722 & \cellcolor{tabgreen!93}0.820 & \cellcolor{tabgreen!75}12.508 \\
ProCraft-plain & \cellcolor{tabgreen!84}0.230 & \cellcolor{tabgreen!45}0.558 & \cellcolor{tabgreen!83}0.439 & \cellcolor{tabgreen!64}0.520 & \cellcolor{tabgreen!73}1.861 & \cellcolor{tabgreen!41}-17.360$^{*}$ & \cellcolor{tabgreen!55}1.660$^{*}$ & \cellcolor{tabgreen!49}2.937$^{*}$ & \cellcolor{tabgreen!82}0.733$^{\circ}$ & \cellcolor{tabgreen!71}11.168 \\
IMG (released) & \cellcolor{tabgreen!54}0.205$^{*}$ & \cellcolor{tabgreen!36}0.497$^{*}$ & \cellcolor{tabgreen!85}0.413 & \cellcolor{tabgreen!49}0.599 & \cellcolor{tabgreen!53}2.295 & \cellcolor{tabgreen!55}-11.465$^{*}$ & \cellcolor{tabgreen!61}1.549 & \cellcolor{tabgreen!55}2.763 & \cellcolor{tabgreen!69}0.633$^{*}$ & \cellcolor{tabgreen!55}5.882$^{*}$ \\
IMG (retrained) & \cellcolor{tabgreen!0}0.158$^{*}$ & \cellcolor{tabgreen!0}0.257$^{*}$ & \cellcolor{tabgreen!0}1.720$^{*}$ & \cellcolor{tabgreen!0}0.846 & \cellcolor{tabgreen!0}3.437 & \cellcolor{tabgreen!0}-35.738$^{*}$ & \cellcolor{tabgreen!0}2.689$^{*}$ & \cellcolor{tabgreen!0}4.364$^{*}$ & \cellcolor{tabgreen!14}0.187$^{*}$ & \cellcolor{tabgreen!0}-12.951$^{*}$ \\
AudioSep & \cellcolor{tabgreen!31}0.185$^{*}$ & \textbf{\cellcolor{tabgreen!100}0.928$^{*}$} & \cellcolor{tabgreen!70}0.645$^{*}$ & \textbf{\cellcolor{tabgreen!100}0.340} & \textbf{\cellcolor{tabgreen!100}1.281} & \cellcolor{tabgreen!72}-3.686$^{*}$ & \cellcolor{tabgreen!64}1.506 & \cellcolor{tabgreen!61}2.578 & \cellcolor{tabgreen!0}0.073$^{*}$ & \cellcolor{tabgreen!26}-3.944$^{*}$ \\
HTDemucs-6s & \cellcolor{tabgreen!98}0.243 & \cellcolor{tabgreen!46}0.564 & \textbf{\cellcolor{tabgreen!100}0.185$^{*}$} & \cellcolor{tabgreen!53}0.578 & \cellcolor{tabgreen!71}1.916 & \textbf{\cellcolor{tabgreen!100}8.760$^{*}$} & \textbf{\cellcolor{tabgreen!100}0.832$^{*}$} & \textbf{\cellcolor{tabgreen!100}1.427$^{*}$} & --- & --- \\
Banquet & \textbf{\cellcolor{tabgreen!100}0.245} & \cellcolor{tabgreen!30}0.461$^{*}$ & \cellcolor{tabgreen!95}0.268$^{*}$ & \cellcolor{tabgreen!33}0.678 & \cellcolor{tabgreen!51}2.347 & \cellcolor{tabgreen!96}6.859$^{*}$ & \cellcolor{tabgreen!77}1.268$^{*}$ & \cellcolor{tabgreen!83}1.914$^{*}$ & \textbf{\cellcolor{tabgreen!100}0.880} & \textbf{\cellcolor{tabgreen!100}21.081$^{*}$} \\
\midrule
\textbf{Denoise} & CLAP-T $\uparrow$ & CLAP-A $\uparrow$ & KL $\downarrow$ & FAD $\downarrow$ & KAD $\downarrow$ & SI-SDR\bothm{} $\uparrow$ & mel-L1\bothm{} $\downarrow$ & MRSTFT\bothm{} $\downarrow$ &  &  \\
\midrule
ProCraft (ours) & \cellcolor{tabgreen!54}0.039 & \cellcolor{tabgreen!62}0.682 & \cellcolor{tabgreen!96}0.292 & \cellcolor{tabgreen!78}0.319 & \cellcolor{tabgreen!75}1.185 & \cellcolor{tabgreen!49}-0.765 & \cellcolor{tabgreen!61}0.693 & \cellcolor{tabgreen!46}1.563 &  &  \\
ProCraft-plain & \cellcolor{tabgreen!59}0.043 & \cellcolor{tabgreen!59}0.669 & \cellcolor{tabgreen!86}0.326 & \cellcolor{tabgreen!82}0.307 & \cellcolor{tabgreen!79}1.131 & \cellcolor{tabgreen!46}-1.679$^{\circ}$ & \cellcolor{tabgreen!47}0.768$^{\circ}$ & \cellcolor{tabgreen!36}1.681$^{*}$ &  &  \\
IMG (retrained) & \cellcolor{tabgreen!60}0.044 & \cellcolor{tabgreen!0}0.433$^{*}$ & \cellcolor{tabgreen!0}0.596$^{*}$ & \cellcolor{tabgreen!0}0.509 & \cellcolor{tabgreen!0}2.042 & \cellcolor{tabgreen!0}-14.367$^{*}$ & \cellcolor{tabgreen!0}0.999$^{*}$ & \cellcolor{tabgreen!0}2.086$^{*}$ &  &  \\
AudioSep & \textbf{\cellcolor{tabgreen!100}0.080$^{*}$} & \textbf{\cellcolor{tabgreen!100}0.836$^{*}$} & \textbf{\cellcolor{tabgreen!100}0.280} & \cellcolor{tabgreen!95}0.277 & \cellcolor{tabgreen!95}0.947 & \textbf{\cellcolor{tabgreen!100}13.186$^{*}$} & \textbf{\cellcolor{tabgreen!100}0.502$^{*}$} & \textbf{\cellcolor{tabgreen!100}0.945$^{*}$} &  &  \\
SD-Codec & \cellcolor{tabgreen!0}-0.010$^{*}$ & \cellcolor{tabgreen!65}0.694 & \cellcolor{tabgreen!68}0.381$^{*}$ & \textbf{\cellcolor{tabgreen!100}0.264} & \textbf{\cellcolor{tabgreen!100}0.892} & \cellcolor{tabgreen!80}7.566$^{*}$ & \cellcolor{tabgreen!38}0.812$^{*}$ & \cellcolor{tabgreen!9}1.983$^{*}$ &  &  \\
\bottomrule
\end{tabular}}
\end{table*}

\clearpage
\section{Full results for combined editing and audio effects}
\label{app:fullcase}
Tables~\ref{tab:fullcase-slakh} and~\ref{tab:fullcase-real} give the complete results for the six editing tasks paired with audio effects, one table each for Slakh and MoisesDB.
They provide the underlying values and additional metrics for the summaries in Section~\ref{sec:fullcase}.

\begin{table*}[!ht]
\centering
\scriptsize
\setlength{\tabcolsep}{1.5pt}
\renewcommand{\arraystretch}{1.0}
\caption{Combined edit and audio effects benchmark, \textbf{Slakh (synthetic)} domain, 150 samples per task; the target is the edited and processed audio. E=editing, P=preservation, E+P=both. \textbf{Bold}\,= best in column, shaded worst to best.}
\label{tab:fullcase-slakh}
\resizebox{\textwidth}{!}{%
\begin{tabular}{@{}lrrrrrrrrrrr@{}}
\toprule
\textbf{Accompany} & CLAP-T $\uparrow$ & CLAP-A in-context $\uparrow$ & KL $\downarrow$ & FAD $\downarrow$ & KAD $\downarrow$ & PaSST\editm{} $\uparrow$ & APA\presm{} $\uparrow$ & StructDist\presm{} $\downarrow$ & Audio Feat.\editm{} $\downarrow$ & isolation\editm{} $\uparrow$ &  \\
\midrule
ProCraft (ours) & \cellcolor{tabgreen!22}0.170 & \cellcolor{tabgreen!15}0.565 & \textbf{\cellcolor{tabgreen!100}0.681} & \cellcolor{tabgreen!63}0.446 & \cellcolor{tabgreen!61}1.782 & \cellcolor{tabgreen!87}0.747 & \cellcolor{tabgreen!62}0.621 & \cellcolor{tabgreen!47}208.784 & \cellcolor{tabgreen!40}0.852 & \cellcolor{tabgreen!93}7.471 &  \\
ProCraft-plain & \textbf{\cellcolor{tabgreen!100}0.323$^{*}$} & \cellcolor{tabgreen!3}0.553 & \cellcolor{tabgreen!96}0.723 & \cellcolor{tabgreen!74}0.422 & \cellcolor{tabgreen!71}1.677 & \cellcolor{tabgreen!87}0.747 & \cellcolor{tabgreen!59}0.598 & \cellcolor{tabgreen!46}211.537 & \cellcolor{tabgreen!58}0.809 & \textbf{\cellcolor{tabgreen!100}8.880} &  \\
IMG (retrained) & \cellcolor{tabgreen!7}0.139$^{*}$ & \cellcolor{tabgreen!0}0.550 & \cellcolor{tabgreen!0}1.677$^{*}$ & \cellcolor{tabgreen!0}0.579 & \cellcolor{tabgreen!2}2.389 & \cellcolor{tabgreen!0}0.193$^{*}$ & \cellcolor{tabgreen!0}0.170 & \textbf{\cellcolor{tabgreen!100}106.944$^{*}$} & \cellcolor{tabgreen!0}0.945$^{\circ}$ & \cellcolor{tabgreen!0}-11.929$^{*}$ &  \\
ZETA & \cellcolor{tabgreen!77}0.277$^{*}$ & \cellcolor{tabgreen!52}0.605$^{*}$ & \cellcolor{tabgreen!43}1.247$^{*}$ & \cellcolor{tabgreen!9}0.561 & \cellcolor{tabgreen!7}2.334 & \cellcolor{tabgreen!67}0.617$^{\circ}$ & \cellcolor{tabgreen!54}0.559 & \cellcolor{tabgreen!0}299.294$^{*}$ & \cellcolor{tabgreen!35}0.862 & \cellcolor{tabgreen!46}-2.330$^{*}$ &  \\
DDIM inv. & \cellcolor{tabgreen!99}0.322$^{*}$ & \cellcolor{tabgreen!20}0.571 & \cellcolor{tabgreen!49}1.185$^{*}$ & \cellcolor{tabgreen!0}0.579 & \cellcolor{tabgreen!0}2.409 & \cellcolor{tabgreen!70}0.638$^{\circ}$ & \cellcolor{tabgreen!57}0.583 & \cellcolor{tabgreen!13}275.083$^{*}$ & \cellcolor{tabgreen!41}0.850 & \cellcolor{tabgreen!99}8.607 &  \\
SteerMusic & \cellcolor{tabgreen!90}0.304$^{*}$ & \textbf{\cellcolor{tabgreen!100}0.655$^{*}$} & \cellcolor{tabgreen!14}1.536$^{*}$ & \cellcolor{tabgreen!21}0.534 & \cellcolor{tabgreen!25}2.153 & \cellcolor{tabgreen!8}0.247$^{*}$ & \cellcolor{tabgreen!52}0.546 & \cellcolor{tabgreen!95}117.297$^{*}$ & \textbf{\cellcolor{tabgreen!100}0.712$^{*}$} & \cellcolor{tabgreen!25}-6.677$^{*}$ &  \\
Audio-Omni & \cellcolor{tabgreen!0}0.126$^{*}$ & \cellcolor{tabgreen!16}0.566 & \cellcolor{tabgreen!75}0.931$^{*}$ & \textbf{\cellcolor{tabgreen!100}0.368} & \textbf{\cellcolor{tabgreen!100}1.377} & \textbf{\cellcolor{tabgreen!100}0.827} & \textbf{\cellcolor{tabgreen!100}0.897} & \cellcolor{tabgreen!76}153.319$^{*}$ & \cellcolor{tabgreen!51}0.826 & \cellcolor{tabgreen!23}-7.204$^{*}$ &  \\
\midrule
\textbf{Cover} & CLAP-T $\uparrow$ & CLAP-A in-context $\uparrow$ & KL $\downarrow$ & FAD $\downarrow$ & KAD $\downarrow$ & PaSST\editm{} $\uparrow$ & APA\presm{} $\uparrow$ & chroma\presm{} $\uparrow$ & Audio Feat.\editm{} $\downarrow$ & isolation\editm{} $\uparrow$ &  \\
\midrule
ProCraft (ours) & \cellcolor{tabgreen!14}0.243 & \cellcolor{tabgreen!22}0.573 & \textbf{\cellcolor{tabgreen!100}0.435} & \cellcolor{tabgreen!100}0.493 & \textbf{\cellcolor{tabgreen!100}1.894} & \cellcolor{tabgreen!95}0.960 & \cellcolor{tabgreen!69}0.752 & \cellcolor{tabgreen!87}0.574 & \cellcolor{tabgreen!53}0.681 & \cellcolor{tabgreen!89}11.495 &  \\
ProCraft-plain & \cellcolor{tabgreen!47}0.305$^{*}$ & \cellcolor{tabgreen!0}0.548$^{*}$ & \cellcolor{tabgreen!98}0.445 & \cellcolor{tabgreen!93}0.497 & \cellcolor{tabgreen!99}1.897 & \cellcolor{tabgreen!95}0.960 & \cellcolor{tabgreen!65}0.723 & \cellcolor{tabgreen!91}0.592 & \cellcolor{tabgreen!61}0.663 & \textbf{\cellcolor{tabgreen!100}13.896} &  \\
IMG (retrained) & \cellcolor{tabgreen!0}0.217$^{\circ}$ & \cellcolor{tabgreen!24}0.575 & \cellcolor{tabgreen!0}1.021$^{*}$ & \cellcolor{tabgreen!51}0.526 & \cellcolor{tabgreen!37}2.137 & \cellcolor{tabgreen!0}0.227$^{*}$ & \cellcolor{tabgreen!0}0.329 & \cellcolor{tabgreen!0}0.191$^{*}$ & \cellcolor{tabgreen!0}0.803$^{*}$ & \cellcolor{tabgreen!0}-7.261$^{*}$ &  \\
ZETA & \cellcolor{tabgreen!97}0.400$^{*}$ & \cellcolor{tabgreen!12}0.562 & \cellcolor{tabgreen!22}0.892$^{*}$ & \cellcolor{tabgreen!13}0.551 & \cellcolor{tabgreen!8}2.249 & \textbf{\cellcolor{tabgreen!100}1.000$^{\circ}$} & \cellcolor{tabgreen!35}0.542 & \cellcolor{tabgreen!63}0.467$^{*}$ & \cellcolor{tabgreen!39}0.712 & \cellcolor{tabgreen!40}1.222$^{*}$ &  \\
DDIM inv. & \textbf{\cellcolor{tabgreen!100}0.405$^{*}$} & \cellcolor{tabgreen!9}0.559 & \cellcolor{tabgreen!25}0.874$^{*}$ & \cellcolor{tabgreen!0}0.560 & \cellcolor{tabgreen!0}2.278 & \cellcolor{tabgreen!100}1.000$^{\circ}$ & \cellcolor{tabgreen!37}0.552 & \cellcolor{tabgreen!50}0.412$^{*}$ & \cellcolor{tabgreen!44}0.702 & \cellcolor{tabgreen!49}3.039$^{*}$ &  \\
SteerMusic & \cellcolor{tabgreen!70}0.349$^{*}$ & \cellcolor{tabgreen!73}0.632$^{*}$ & \cellcolor{tabgreen!67}0.626$^{*}$ & \textbf{\cellcolor{tabgreen!100}0.493} & \cellcolor{tabgreen!78}1.980 & \cellcolor{tabgreen!90}0.920 & \textbf{\cellcolor{tabgreen!100}0.938} & \cellcolor{tabgreen!97}0.616$^{\circ}$ & \textbf{\cellcolor{tabgreen!100}0.571$^{*}$} & \cellcolor{tabgreen!41}1.494$^{*}$ &  \\
Audio-Omni & \cellcolor{tabgreen!9}0.234 & \textbf{\cellcolor{tabgreen!100}0.664$^{*}$} & \cellcolor{tabgreen!70}0.610$^{*}$ & \cellcolor{tabgreen!26}0.542 & \cellcolor{tabgreen!13}2.228 & \cellcolor{tabgreen!97}0.980 & \cellcolor{tabgreen!98}0.923 & \textbf{\cellcolor{tabgreen!100}0.631$^{*}$} & \cellcolor{tabgreen!34}0.725 & \cellcolor{tabgreen!30}-0.852$^{*}$ &  \\
\midrule
\textbf{Instrument transfer} & CLAP-T $\uparrow$ & CLAP-A in-context $\uparrow$ & KL $\downarrow$ & FAD $\downarrow$ & KAD $\downarrow$ & PaSST\editm{} $\uparrow$ & Frame-F1\presm{} $\uparrow$ & chroma\presm{} $\uparrow$ & Audio Feat.\editm{} $\downarrow$ & isolation\editm{} $\uparrow$ &  \\
\midrule
ProCraft (ours) & \cellcolor{tabgreen!35}0.244 & \cellcolor{tabgreen!74}0.521 & \textbf{\cellcolor{tabgreen!100}0.504} & \cellcolor{tabgreen!96}0.744 & \cellcolor{tabgreen!95}3.037 & \cellcolor{tabgreen!74}0.547 & \cellcolor{tabgreen!70}0.495 & \cellcolor{tabgreen!64}0.581 & \cellcolor{tabgreen!86}0.773 & \cellcolor{tabgreen!53}-1.960 &  \\
ProCraft-plain & \cellcolor{tabgreen!22}0.225$^{*}$ & \cellcolor{tabgreen!70}0.505 & \cellcolor{tabgreen!95}0.546 & \textbf{\cellcolor{tabgreen!100}0.726} & \textbf{\cellcolor{tabgreen!100}2.946} & \cellcolor{tabgreen!86}0.607 & \cellcolor{tabgreen!70}0.497 & \cellcolor{tabgreen!67}0.603 & \cellcolor{tabgreen!81}0.791 & \cellcolor{tabgreen!59}-0.831 &  \\
IMG (retrained) & \cellcolor{tabgreen!0}0.192$^{*}$ & \cellcolor{tabgreen!32}0.346$^{*}$ & \cellcolor{tabgreen!31}1.070$^{*}$ & \cellcolor{tabgreen!67}0.872 & \cellcolor{tabgreen!58}3.651 & \cellcolor{tabgreen!0}0.187$^{*}$ & \cellcolor{tabgreen!17}0.191 & \cellcolor{tabgreen!0}0.131$^{*}$ & \cellcolor{tabgreen!22}1.020$^{*}$ & \cellcolor{tabgreen!0}-11.866$^{*}$ &  \\
ZETA & \cellcolor{tabgreen!37}0.247 & \cellcolor{tabgreen!2}0.218$^{*}$ & \cellcolor{tabgreen!0}1.319$^{*}$ & \cellcolor{tabgreen!0}1.164 & \cellcolor{tabgreen!0}4.630 & \cellcolor{tabgreen!88}0.613 & \cellcolor{tabgreen!3}0.109 & \cellcolor{tabgreen!96}0.802$^{*}$ & \cellcolor{tabgreen!21}1.024$^{*}$ & \cellcolor{tabgreen!63}-0.014 &  \\
DDIM inv. & \cellcolor{tabgreen!49}0.264 & \cellcolor{tabgreen!0}0.209$^{*}$ & \cellcolor{tabgreen!10}1.241$^{*}$ & \cellcolor{tabgreen!11}1.116 & \cellcolor{tabgreen!15}4.377 & \textbf{\cellcolor{tabgreen!100}0.673$^{*}$} & \cellcolor{tabgreen!0}0.095 & \textbf{\cellcolor{tabgreen!100}0.830$^{*}$} & \cellcolor{tabgreen!0}1.103$^{*}$ & \textbf{\cellcolor{tabgreen!100}6.943$^{*}$} &  \\
SteerMusic & \textbf{\cellcolor{tabgreen!100}0.341$^{*}$} & \textbf{\cellcolor{tabgreen!100}0.632$^{*}$} & \cellcolor{tabgreen!57}0.852$^{*}$ & \cellcolor{tabgreen!47}0.958 & \cellcolor{tabgreen!48}3.829 & \cellcolor{tabgreen!32}0.342$^{*}$ & \textbf{\cellcolor{tabgreen!100}0.668} & \cellcolor{tabgreen!37}0.391$^{*}$ & \textbf{\cellcolor{tabgreen!100}0.720} & \cellcolor{tabgreen!4}-11.137$^{*}$ &  \\
Audio-Omni & \cellcolor{tabgreen!50}0.266$^{\circ}$ & \cellcolor{tabgreen!88}0.580$^{*}$ & \cellcolor{tabgreen!52}0.896$^{*}$ & \cellcolor{tabgreen!51}0.940 & \cellcolor{tabgreen!52}3.756 & \cellcolor{tabgreen!88}0.613 & \cellcolor{tabgreen!85}0.582 & \cellcolor{tabgreen!84}0.722$^{*}$ & \cellcolor{tabgreen!85}0.779 & \cellcolor{tabgreen!61}-0.457 &  \\
\midrule
\textbf{Remove} & CLAP-T $\uparrow$ & CLAP-A in-context $\uparrow$ & KL $\downarrow$ & FAD $\downarrow$ & KAD $\downarrow$ & SI-SDR\bothm{} $\uparrow$ & mel-L1\bothm{} $\downarrow$ & MRSTFT\bothm{} $\downarrow$ & edit rate\editm{} $\uparrow$ & edit strength\editm{} $\uparrow$ & Audio Feat.\editm{} $\downarrow$ \\
\midrule
ProCraft (ours) & \cellcolor{tabgreen!73}0.264 & \cellcolor{tabgreen!99}0.714 & \textbf{\cellcolor{tabgreen!100}0.546} & \cellcolor{tabgreen!95}0.466 & \cellcolor{tabgreen!93}1.817 & \textbf{\cellcolor{tabgreen!100}-7.386} & \cellcolor{tabgreen!100}1.154 & \cellcolor{tabgreen!96}2.493 & \cellcolor{tabgreen!93}0.660 & \cellcolor{tabgreen!76}12.911 & \cellcolor{tabgreen!82}0.568 \\
ProCraft-plain & \cellcolor{tabgreen!70}0.257 & \textbf{\cellcolor{tabgreen!100}0.718} & \cellcolor{tabgreen!100}0.550 & \textbf{\cellcolor{tabgreen!100}0.440} & \textbf{\cellcolor{tabgreen!100}1.725} & \cellcolor{tabgreen!99}-7.786 & \textbf{\cellcolor{tabgreen!100}1.144} & \textbf{\cellcolor{tabgreen!100}2.288} & \cellcolor{tabgreen!93}0.660 & \cellcolor{tabgreen!70}11.956 & \cellcolor{tabgreen!87}0.552 \\
IMG (retrained) & \cellcolor{tabgreen!54}0.224$^{*}$ & \cellcolor{tabgreen!83}0.613$^{*}$ & \cellcolor{tabgreen!95}0.669$^{*}$ & \cellcolor{tabgreen!57}0.644 & \cellcolor{tabgreen!17}2.737 & \cellcolor{tabgreen!87}-12.504$^{*}$ & \cellcolor{tabgreen!95}1.273$^{*}$ & \cellcolor{tabgreen!96}2.487 & \cellcolor{tabgreen!67}0.553$^{\circ}$ & \cellcolor{tabgreen!40}7.443$^{*}$ & \cellcolor{tabgreen!86}0.554 \\
ZETA & \cellcolor{tabgreen!29}0.172$^{*}$ & \cellcolor{tabgreen!9}0.145$^{*}$ & \cellcolor{tabgreen!0}2.964$^{*}$ & \cellcolor{tabgreen!0}0.915 & \cellcolor{tabgreen!17}2.744 & \cellcolor{tabgreen!0}-46.183$^{*}$ & \cellcolor{tabgreen!0}3.559$^{*}$ & \cellcolor{tabgreen!0}7.650$^{*}$ & \cellcolor{tabgreen!97}0.673 & \cellcolor{tabgreen!98}16.348$^{\circ}$ & \cellcolor{tabgreen!11}0.791$^{*}$ \\
DDIM inv. & \cellcolor{tabgreen!0}0.110$^{*}$ & \cellcolor{tabgreen!0}0.087$^{*}$ & \cellcolor{tabgreen!27}2.302$^{*}$ & \cellcolor{tabgreen!0}0.916 & \cellcolor{tabgreen!0}2.951 & \cellcolor{tabgreen!3}-44.831$^{*}$ & \cellcolor{tabgreen!13}3.256$^{*}$ & \cellcolor{tabgreen!12}7.011$^{*}$ & \textbf{\cellcolor{tabgreen!100}0.687} & \textbf{\cellcolor{tabgreen!100}16.599$^{\circ}$} & \cellcolor{tabgreen!0}0.827$^{*}$ \\
SteerMusic & \textbf{\cellcolor{tabgreen!100}0.322$^{*}$} & \cellcolor{tabgreen!86}0.628$^{*}$ & \cellcolor{tabgreen!88}0.832$^{*}$ & \cellcolor{tabgreen!67}0.597 & \cellcolor{tabgreen!46}2.386 & \cellcolor{tabgreen!53}-25.596$^{*}$ & \cellcolor{tabgreen!80}1.619$^{*}$ & \cellcolor{tabgreen!43}5.336$^{*}$ & \cellcolor{tabgreen!0}0.287$^{*}$ & \cellcolor{tabgreen!6}2.213$^{*}$ & \textbf{\cellcolor{tabgreen!100}0.511} \\
Audio-Omni & \cellcolor{tabgreen!64}0.244$^{\circ}$ & \cellcolor{tabgreen!62}0.478$^{*}$ & \cellcolor{tabgreen!95}0.665$^{*}$ & \cellcolor{tabgreen!72}0.573 & \cellcolor{tabgreen!53}2.305 & \cellcolor{tabgreen!55}-24.875$^{*}$ & \cellcolor{tabgreen!87}1.449$^{*}$ & \cellcolor{tabgreen!73}3.730$^{*}$ & \cellcolor{tabgreen!3}0.300$^{*}$ & \cellcolor{tabgreen!0}1.281$^{*}$ & \cellcolor{tabgreen!89}0.546 \\
\midrule
\textbf{Extract} & CLAP-T $\uparrow$ & CLAP-A in-context $\uparrow$ & KL $\downarrow$ & FAD $\downarrow$ & KAD $\downarrow$ & SI-SDR\bothm{} $\uparrow$ & mel-L1\bothm{} $\downarrow$ & MRSTFT\bothm{} $\downarrow$ & edit rate\editm{} $\uparrow$ & edit strength\editm{} $\uparrow$ & Audio Feat.\editm{} $\downarrow$ \\
\midrule
ProCraft (ours) & \cellcolor{tabgreen!77}0.280 & \cellcolor{tabgreen!54}0.440 & \textbf{\cellcolor{tabgreen!100}0.528} & \cellcolor{tabgreen!79}0.671 & \cellcolor{tabgreen!92}2.534 & \textbf{\cellcolor{tabgreen!100}-15.725} & \textbf{\cellcolor{tabgreen!100}1.851} & \cellcolor{tabgreen!98}3.654 & \cellcolor{tabgreen!97}0.680 & \cellcolor{tabgreen!97}8.526 & \cellcolor{tabgreen!37}0.799 \\
ProCraft-plain & \cellcolor{tabgreen!72}0.273 & \cellcolor{tabgreen!60}0.468$^{\circ}$ & \cellcolor{tabgreen!95}0.595$^{\circ}$ & \cellcolor{tabgreen!81}0.667 & \cellcolor{tabgreen!87}2.577 & \cellcolor{tabgreen!93}-17.546$^{*}$ & \cellcolor{tabgreen!93}1.935 & \textbf{\cellcolor{tabgreen!100}3.598} & \textbf{\cellcolor{tabgreen!100}0.700} & \textbf{\cellcolor{tabgreen!100}9.210} & \cellcolor{tabgreen!29}0.816 \\
IMG (retrained) & \cellcolor{tabgreen!0}0.163$^{*}$ & \cellcolor{tabgreen!36}0.352$^{*}$ & \cellcolor{tabgreen!31}1.455$^{*}$ & \cellcolor{tabgreen!0}0.835 & \cellcolor{tabgreen!0}3.424 & \cellcolor{tabgreen!51}-28.327$^{*}$ & \cellcolor{tabgreen!55}2.410$^{*}$ & \cellcolor{tabgreen!80}4.225$^{*}$ & \cellcolor{tabgreen!0}0.067$^{*}$ & \cellcolor{tabgreen!0}-14.241$^{*}$ & \cellcolor{tabgreen!0}0.878$^{\circ}$ \\
ZETA & \cellcolor{tabgreen!89}0.298 & \cellcolor{tabgreen!16}0.255$^{*}$ & \cellcolor{tabgreen!4}1.828$^{*}$ & \cellcolor{tabgreen!9}0.815 & \cellcolor{tabgreen!83}2.623 & \cellcolor{tabgreen!13}-38.028$^{*}$ & \cellcolor{tabgreen!7}2.990$^{*}$ & \cellcolor{tabgreen!0}6.732$^{*}$ & \cellcolor{tabgreen!60}0.447$^{*}$ & \cellcolor{tabgreen!60}-0.120$^{*}$ & \cellcolor{tabgreen!25}0.824 \\
DDIM inv. & \cellcolor{tabgreen!43}0.228$^{*}$ & \cellcolor{tabgreen!0}0.177$^{*}$ & \cellcolor{tabgreen!0}1.878$^{*}$ & \cellcolor{tabgreen!13}0.807 & \cellcolor{tabgreen!87}2.577 & \cellcolor{tabgreen!0}-41.259$^{*}$ & \cellcolor{tabgreen!0}3.082$^{*}$ & \cellcolor{tabgreen!7}6.515$^{*}$ & \cellcolor{tabgreen!49}0.380$^{*}$ & \cellcolor{tabgreen!40}-4.820$^{*}$ & \cellcolor{tabgreen!2}0.873 \\
SteerMusic & \textbf{\cellcolor{tabgreen!100}0.314$^{*}$} & \cellcolor{tabgreen!73}0.532$^{*}$ & \cellcolor{tabgreen!72}0.900$^{*}$ & \cellcolor{tabgreen!92}0.644 & \textbf{\cellcolor{tabgreen!100}2.456} & \cellcolor{tabgreen!55}-27.196$^{*}$ & \cellcolor{tabgreen!88}2.004$^{\circ}$ & \cellcolor{tabgreen!40}5.480$^{*}$ & \cellcolor{tabgreen!73}0.527$^{*}$ & \cellcolor{tabgreen!73}2.907$^{*}$ & \textbf{\cellcolor{tabgreen!100}0.665$^{*}$} \\
Audio-Omni & \cellcolor{tabgreen!54}0.244$^{*}$ & \textbf{\cellcolor{tabgreen!100}0.663$^{*}$} & \cellcolor{tabgreen!92}0.632$^{\circ}$ & \textbf{\cellcolor{tabgreen!100}0.628} & \cellcolor{tabgreen!100}2.458 & \cellcolor{tabgreen!68}-23.980$^{*}$ & \cellcolor{tabgreen!98}1.876 & \cellcolor{tabgreen!95}3.769 & \cellcolor{tabgreen!75}0.540$^{\circ}$ & \cellcolor{tabgreen!79}4.287$^{*}$ & \cellcolor{tabgreen!58}0.755 \\
\midrule
\textbf{Denoise} & CLAP-T $\uparrow$ & CLAP-A in-context $\uparrow$ & KL $\downarrow$ & FAD $\downarrow$ & KAD $\downarrow$ & SI-SDR\bothm{} $\uparrow$ & mel-L1\bothm{} $\downarrow$ & MRSTFT\bothm{} $\downarrow$ & Audio Feat.\editm{} $\downarrow$ &  &  \\
\midrule
ProCraft (ours) & \cellcolor{tabgreen!3}0.092 & \textbf{\cellcolor{tabgreen!100}0.509} & \textbf{\cellcolor{tabgreen!100}0.661} & \textbf{\cellcolor{tabgreen!100}0.503} & \textbf{\cellcolor{tabgreen!100}1.948} & \textbf{\cellcolor{tabgreen!100}-4.735} & \textbf{\cellcolor{tabgreen!100}1.133} & \textbf{\cellcolor{tabgreen!100}2.387} & \cellcolor{tabgreen!0}0.660 &  &  \\
ProCraft-plain & \cellcolor{tabgreen!4}0.093 & \cellcolor{tabgreen!98}0.502 & \cellcolor{tabgreen!99}0.669 & \cellcolor{tabgreen!99}0.507 & \cellcolor{tabgreen!99}1.975 & \cellcolor{tabgreen!99}-4.972 & \cellcolor{tabgreen!97}1.179 & \cellcolor{tabgreen!100}2.394 & \cellcolor{tabgreen!9}0.648 &  &  \\
IMG (retrained) & \cellcolor{tabgreen!0}0.088 & \cellcolor{tabgreen!98}0.503 & \cellcolor{tabgreen!84}0.784$^{*}$ & \cellcolor{tabgreen!88}0.594 & \cellcolor{tabgreen!86}2.488 & \cellcolor{tabgreen!88}-8.515$^{*}$ & \cellcolor{tabgreen!90}1.269$^{*}$ & \cellcolor{tabgreen!97}2.528 & \cellcolor{tabgreen!48}0.593 &  &  \\
ZETA & \cellcolor{tabgreen!65}0.180$^{*}$ & \cellcolor{tabgreen!0}0.108$^{*}$ & \cellcolor{tabgreen!40}1.118$^{*}$ & \cellcolor{tabgreen!0}1.290 & \cellcolor{tabgreen!0}5.676 & \cellcolor{tabgreen!0}-35.418$^{*}$ & \cellcolor{tabgreen!6}2.477$^{*}$ & \cellcolor{tabgreen!4}6.306$^{*}$ & \textbf{\cellcolor{tabgreen!100}0.520$^{*}$} &  &  \\
DDIM inv. & \cellcolor{tabgreen!65}0.180$^{*}$ & \cellcolor{tabgreen!15}0.166$^{*}$ & \cellcolor{tabgreen!0}1.421$^{*}$ & \cellcolor{tabgreen!10}1.211 & \cellcolor{tabgreen!7}5.407 & \cellcolor{tabgreen!11}-32.180$^{*}$ & \cellcolor{tabgreen!0}2.562$^{*}$ & \cellcolor{tabgreen!0}6.488$^{*}$ & \cellcolor{tabgreen!16}0.638 &  &  \\
SteerMusic & \textbf{\cellcolor{tabgreen!100}0.230$^{*}$} & \cellcolor{tabgreen!77}0.418$^{*}$ & \cellcolor{tabgreen!57}0.986$^{*}$ & \cellcolor{tabgreen!83}0.635 & \cellcolor{tabgreen!91}2.275 & \cellcolor{tabgreen!18}-29.896$^{*}$ & \cellcolor{tabgreen!62}1.671$^{*}$ & \cellcolor{tabgreen!29}5.293$^{*}$ & \cellcolor{tabgreen!6}0.652 &  &  \\
Audio-Omni & \cellcolor{tabgreen!48}0.156$^{*}$ & \cellcolor{tabgreen!82}0.438$^{*}$ & \cellcolor{tabgreen!59}0.971$^{*}$ & \cellcolor{tabgreen!84}0.625 & \cellcolor{tabgreen!86}2.482 & \cellcolor{tabgreen!31}-26.023$^{*}$ & \cellcolor{tabgreen!68}1.594$^{*}$ & \cellcolor{tabgreen!68}3.718$^{*}$ & \cellcolor{tabgreen!0}0.660 &  &  \\
\bottomrule
\end{tabular}}
\end{table*}

\begin{table*}[!ht]
\centering
\scriptsize
\setlength{\tabcolsep}{1.5pt}
\renewcommand{\arraystretch}{1.0}
\caption{Combined edit and audio effects benchmark, \textbf{real audio} domain, 150 samples per task; the target is the edited and processed audio. E=editing, P=preservation, E+P=both. \textbf{Bold}\,= best in column, shaded worst to best.}
\label{tab:fullcase-real}
\resizebox{\textwidth}{!}{%
\begin{tabular}{@{}lrrrrrrrrrrr@{}}
\toprule
\textbf{Accompany} & CLAP-T $\uparrow$ & CLAP-A in-context $\uparrow$ & KL $\downarrow$ & FAD $\downarrow$ & KAD $\downarrow$ & PaSST\editm{} $\uparrow$ & APA\presm{} $\uparrow$ & StructDist\presm{} $\downarrow$ & Audio Feat.\editm{} $\downarrow$ & isolation\editm{} $\uparrow$ &  \\
\midrule
ProCraft (ours) & \cellcolor{tabgreen!95}0.303 & \cellcolor{tabgreen!24}0.483 & \cellcolor{tabgreen!96}0.959 & \cellcolor{tabgreen!99}0.256 & \cellcolor{tabgreen!98}0.852 & \cellcolor{tabgreen!87}0.693 & \cellcolor{tabgreen!46}0.366 & \cellcolor{tabgreen!36}123.701 & \cellcolor{tabgreen!75}0.846 & \cellcolor{tabgreen!81}3.873 &  \\
ProCraft-plain & \textbf{\cellcolor{tabgreen!100}0.316} & \cellcolor{tabgreen!33}0.500$^{\circ}$ & \textbf{\cellcolor{tabgreen!100}0.841} & \cellcolor{tabgreen!97}0.260 & \cellcolor{tabgreen!93}0.884 & \textbf{\cellcolor{tabgreen!100}0.773} & \cellcolor{tabgreen!58}0.461 & \cellcolor{tabgreen!37}123.593 & \cellcolor{tabgreen!77}0.839 & \textbf{\cellcolor{tabgreen!100}8.704$^{*}$} &  \\
IMG (retrained) & \cellcolor{tabgreen!19}0.106$^{*}$ & \cellcolor{tabgreen!29}0.491 & \cellcolor{tabgreen!69}1.879$^{*}$ & \cellcolor{tabgreen!28}0.404 & \cellcolor{tabgreen!18}1.450 & \cellcolor{tabgreen!4}0.180$^{*}$ & \cellcolor{tabgreen!0}0.000 & \cellcolor{tabgreen!53}113.529$^{*}$ & \cellcolor{tabgreen!30}0.991$^{*}$ & \cellcolor{tabgreen!22}-11.355$^{*}$ &  \\
ZETA & \cellcolor{tabgreen!22}0.113$^{*}$ & \cellcolor{tabgreen!82}0.594$^{*}$ & \cellcolor{tabgreen!0}4.138$^{*}$ & \cellcolor{tabgreen!36}0.387 & \cellcolor{tabgreen!64}1.104 & \cellcolor{tabgreen!22}0.287$^{*}$ & \cellcolor{tabgreen!52}0.417 & \cellcolor{tabgreen!79}96.880$^{*}$ & \cellcolor{tabgreen!0}1.089$^{*}$ & \cellcolor{tabgreen!14}-13.347$^{*}$ &  \\
DDIM inv. & \cellcolor{tabgreen!0}0.055$^{*}$ & \cellcolor{tabgreen!0}0.435$^{*}$ & \cellcolor{tabgreen!22}3.426$^{*}$ & \cellcolor{tabgreen!0}0.464 & \cellcolor{tabgreen!0}1.590 & \cellcolor{tabgreen!8}0.200$^{*}$ & \cellcolor{tabgreen!30}0.237 & \cellcolor{tabgreen!0}146.692$^{*}$ & \cellcolor{tabgreen!27}1.002$^{*}$ & \cellcolor{tabgreen!0}-17.029$^{*}$ &  \\
SteerMusic & \cellcolor{tabgreen!69}0.234$^{*}$ & \textbf{\cellcolor{tabgreen!100}0.630$^{*}$} & \cellcolor{tabgreen!71}1.781$^{*}$ & \cellcolor{tabgreen!55}0.348 & \cellcolor{tabgreen!57}1.159 & \cellcolor{tabgreen!0}0.153$^{*}$ & \cellcolor{tabgreen!34}0.270 & \textbf{\cellcolor{tabgreen!100}83.552$^{*}$} & \textbf{\cellcolor{tabgreen!100}0.764} & \cellcolor{tabgreen!14}-13.333$^{*}$ &  \\
Audio-Omni & \cellcolor{tabgreen!29}0.131$^{*}$ & \cellcolor{tabgreen!42}0.517$^{*}$ & \cellcolor{tabgreen!92}1.101$^{\circ}$ & \textbf{\cellcolor{tabgreen!100}0.253} & \textbf{\cellcolor{tabgreen!100}0.834} & \cellcolor{tabgreen!74}0.613 & \textbf{\cellcolor{tabgreen!100}0.797} & \cellcolor{tabgreen!37}123.593 & \cellcolor{tabgreen!66}0.876 & \cellcolor{tabgreen!45}-5.432$^{*}$ &  \\
\midrule
\textbf{Cover} & CLAP-T $\uparrow$ & CLAP-A in-context $\uparrow$ & FAD $\downarrow$ & KAD $\downarrow$ & PaSST\editm{} $\uparrow$ & APA\presm{} $\uparrow$ & chroma\presm{} $\uparrow$ & isolation\editm{} $\uparrow$ &  &  &  \\
\midrule
ProCraft (ours) & \textbf{\cellcolor{tabgreen!100}0.308} & \cellcolor{tabgreen!21}0.492 & \cellcolor{tabgreen!75}0.344 & \cellcolor{tabgreen!62}1.212 & \cellcolor{tabgreen!96}0.953 & \cellcolor{tabgreen!67}0.700 & \cellcolor{tabgreen!78}0.574 & \cellcolor{tabgreen!93}13.185 &  &  &  \\
ProCraft-plain & \cellcolor{tabgreen!96}0.295$^{\circ}$ & \cellcolor{tabgreen!18}0.488 & \cellcolor{tabgreen!82}0.333 & \cellcolor{tabgreen!69}1.162 & \textbf{\cellcolor{tabgreen!100}0.980} & \cellcolor{tabgreen!62}0.672 & \cellcolor{tabgreen!75}0.560 & \textbf{\cellcolor{tabgreen!100}14.948} &  &  &  \\
IMG (retrained) & \cellcolor{tabgreen!64}0.196$^{*}$ & \cellcolor{tabgreen!18}0.488 & \cellcolor{tabgreen!86}0.326 & \cellcolor{tabgreen!68}1.168 & \cellcolor{tabgreen!0}0.267$^{*}$ & \cellcolor{tabgreen!0}0.301 & \cellcolor{tabgreen!0}0.184$^{*}$ & \cellcolor{tabgreen!0}-10.217$^{*}$ &  &  &  \\
ZETA & \cellcolor{tabgreen!44}0.133$^{*}$ & \textbf{\cellcolor{tabgreen!100}0.613$^{*}$} & \cellcolor{tabgreen!90}0.319 & \textbf{\cellcolor{tabgreen!100}0.949} & \cellcolor{tabgreen!38}0.540$^{*}$ & \cellcolor{tabgreen!89}0.835 & \cellcolor{tabgreen!58}0.473$^{*}$ & \cellcolor{tabgreen!24}-4.222$^{*}$ &  &  &  \\
DDIM inv. & \cellcolor{tabgreen!0}-0.004$^{*}$ & \cellcolor{tabgreen!0}0.461$^{\circ}$ & \cellcolor{tabgreen!0}0.472 & \cellcolor{tabgreen!0}1.638 & \cellcolor{tabgreen!27}0.460$^{*}$ & \cellcolor{tabgreen!69}0.712 & \cellcolor{tabgreen!29}0.328$^{*}$ & \cellcolor{tabgreen!4}-9.282$^{*}$ &  &  &  \\
SteerMusic & \cellcolor{tabgreen!90}0.278$^{\circ}$ & \cellcolor{tabgreen!78}0.579$^{*}$ & \textbf{\cellcolor{tabgreen!100}0.302} & \cellcolor{tabgreen!98}0.964 & \cellcolor{tabgreen!59}0.687$^{*}$ & \cellcolor{tabgreen!85}0.806 & \cellcolor{tabgreen!67}0.517$^{*}$ & \cellcolor{tabgreen!18}-5.800$^{*}$ &  &  &  \\
Audio-Omni & \cellcolor{tabgreen!87}0.268$^{*}$ & \cellcolor{tabgreen!82}0.585$^{*}$ & \cellcolor{tabgreen!76}0.344 & \cellcolor{tabgreen!57}1.245 & \cellcolor{tabgreen!95}0.947 & \textbf{\cellcolor{tabgreen!100}0.899} & \textbf{\cellcolor{tabgreen!100}0.682$^{*}$} & \cellcolor{tabgreen!22}-4.701$^{*}$ &  &  &  \\
\midrule
\textbf{Instrument transfer} & CLAP-T $\uparrow$ & CLAP-A in-context $\uparrow$ & FAD $\downarrow$ & KAD $\downarrow$ & PaSST\editm{} $\uparrow$ & chroma\presm{} $\uparrow$ & isolation\editm{} $\uparrow$ &  &  &  &  \\
\midrule
ProCraft (ours) & \cellcolor{tabgreen!46}0.216 & \cellcolor{tabgreen!40}0.368 & \cellcolor{tabgreen!89}0.641 & \cellcolor{tabgreen!82}2.466 & \cellcolor{tabgreen!88}0.547 & \cellcolor{tabgreen!62}0.451 & \cellcolor{tabgreen!88}-0.651 &  &  &  &  \\
ProCraft-plain & \cellcolor{tabgreen!47}0.218 & \cellcolor{tabgreen!44}0.387 & \textbf{\cellcolor{tabgreen!100}0.602} & \textbf{\cellcolor{tabgreen!100}2.259} & \cellcolor{tabgreen!95}0.587 & \cellcolor{tabgreen!77}0.527$^{*}$ & \textbf{\cellcolor{tabgreen!100}2.435} &  &  &  &  \\
IMG (retrained) & \cellcolor{tabgreen!14}0.166$^{*}$ & \cellcolor{tabgreen!43}0.381 & \cellcolor{tabgreen!37}0.825 & \cellcolor{tabgreen!0}3.409 & \cellcolor{tabgreen!18}0.147$^{*}$ & \cellcolor{tabgreen!0}0.134$^{*}$ & \cellcolor{tabgreen!43}-12.336$^{*}$ &  &  &  &  \\
ZETA & \cellcolor{tabgreen!28}0.188 & \cellcolor{tabgreen!5}0.177$^{*}$ & \cellcolor{tabgreen!4}0.938 & \cellcolor{tabgreen!44}2.903 & \cellcolor{tabgreen!56}0.367$^{*}$ & \cellcolor{tabgreen!73}0.508 & \cellcolor{tabgreen!61}-7.554$^{*}$ &  &  &  &  \\
DDIM inv. & \cellcolor{tabgreen!0}0.145$^{*}$ & \cellcolor{tabgreen!0}0.149$^{*}$ & \cellcolor{tabgreen!0}0.953 & \cellcolor{tabgreen!16}3.223 & \cellcolor{tabgreen!46}0.307$^{*}$ & \cellcolor{tabgreen!48}0.382 & \cellcolor{tabgreen!25}-16.860$^{*}$ &  &  &  &  \\
SteerMusic & \textbf{\cellcolor{tabgreen!100}0.300$^{*}$} & \textbf{\cellcolor{tabgreen!100}0.693$^{*}$} & \cellcolor{tabgreen!77}0.683 & \cellcolor{tabgreen!95}2.316 & \cellcolor{tabgreen!0}0.047$^{*}$ & \cellcolor{tabgreen!57}0.426 & \cellcolor{tabgreen!0}-23.328$^{*}$ &  &  &  &  \\
Audio-Omni & \cellcolor{tabgreen!77}0.264$^{*}$ & \cellcolor{tabgreen!79}0.580$^{*}$ & \cellcolor{tabgreen!83}0.661 & \cellcolor{tabgreen!73}2.565 & \textbf{\cellcolor{tabgreen!100}0.613} & \textbf{\cellcolor{tabgreen!100}0.646$^{*}$} & \cellcolor{tabgreen!64}-6.753$^{*}$ &  &  &  &  \\
\midrule
\textbf{Remove} & CLAP-T $\uparrow$ & CLAP-A in-context $\uparrow$ & KL $\downarrow$ & FAD $\downarrow$ & KAD $\downarrow$ & SI-SDR\bothm{} $\uparrow$ & mel-L1\bothm{} $\downarrow$ & MRSTFT\bothm{} $\downarrow$ & edit rate\editm{} $\uparrow$ & edit strength\editm{} $\uparrow$ & Audio Feat.\editm{} $\downarrow$ \\
\midrule
ProCraft (ours) & \cellcolor{tabgreen!69}0.228 & \cellcolor{tabgreen!96}0.685 & \cellcolor{tabgreen!99}0.815 & \cellcolor{tabgreen!97}0.351 & \cellcolor{tabgreen!97}1.176 & \cellcolor{tabgreen!98}-12.112 & \cellcolor{tabgreen!97}1.183 & \cellcolor{tabgreen!99}2.418 & \cellcolor{tabgreen!53}0.507 & \cellcolor{tabgreen!35}8.793 & \cellcolor{tabgreen!58}0.616 \\
ProCraft-plain & \cellcolor{tabgreen!66}0.222 & \textbf{\cellcolor{tabgreen!100}0.713$^{\circ}$} & \textbf{\cellcolor{tabgreen!100}0.795} & \textbf{\cellcolor{tabgreen!100}0.334} & \textbf{\cellcolor{tabgreen!100}1.125} & \textbf{\cellcolor{tabgreen!100}-11.491} & \textbf{\cellcolor{tabgreen!100}1.136} & \textbf{\cellcolor{tabgreen!100}2.361} & \cellcolor{tabgreen!36}0.407$^{\circ}$ & \cellcolor{tabgreen!26}7.221 & \cellcolor{tabgreen!58}0.616 \\
IMG (retrained) & \cellcolor{tabgreen!57}0.208$^{\circ}$ & \cellcolor{tabgreen!59}0.455$^{*}$ & \cellcolor{tabgreen!88}1.110$^{*}$ & \cellcolor{tabgreen!59}0.586 & \cellcolor{tabgreen!32}2.397 & \cellcolor{tabgreen!76}-20.292$^{*}$ & \cellcolor{tabgreen!88}1.359$^{*}$ & \cellcolor{tabgreen!93}2.686$^{*}$ & \cellcolor{tabgreen!43}0.447 & \cellcolor{tabgreen!27}7.415 & \cellcolor{tabgreen!81}0.557$^{\circ}$ \\
ZETA & \cellcolor{tabgreen!15}0.141$^{*}$ & \cellcolor{tabgreen!6}0.117$^{*}$ & \cellcolor{tabgreen!0}3.463$^{*}$ & \cellcolor{tabgreen!0}0.956 & \cellcolor{tabgreen!2}2.962 & \cellcolor{tabgreen!0}-47.959$^{*}$ & \cellcolor{tabgreen!0}2.981$^{*}$ & \cellcolor{tabgreen!0}6.939$^{*}$ & \textbf{\cellcolor{tabgreen!100}0.787$^{*}$} & \textbf{\cellcolor{tabgreen!100}21.095$^{*}$} & \cellcolor{tabgreen!1}0.766$^{*}$ \\
DDIM inv. & \cellcolor{tabgreen!0}0.117$^{*}$ & \cellcolor{tabgreen!0}0.081$^{*}$ & \cellcolor{tabgreen!19}2.960$^{*}$ & \cellcolor{tabgreen!6}0.918 & \cellcolor{tabgreen!0}2.997 & \cellcolor{tabgreen!6}-45.645$^{*}$ & \cellcolor{tabgreen!7}2.853$^{*}$ & \cellcolor{tabgreen!6}6.675$^{*}$ & \cellcolor{tabgreen!100}0.787$^{*}$ & \cellcolor{tabgreen!95}20.177$^{*}$ & \cellcolor{tabgreen!0}0.768$^{*}$ \\
SteerMusic & \textbf{\cellcolor{tabgreen!100}0.277$^{*}$} & \cellcolor{tabgreen!82}0.596$^{*}$ & \cellcolor{tabgreen!93}0.985$^{*}$ & \cellcolor{tabgreen!85}0.425 & \cellcolor{tabgreen!85}1.397 & \cellcolor{tabgreen!57}-27.321$^{*}$ & \cellcolor{tabgreen!70}1.695$^{*}$ & \cellcolor{tabgreen!34}5.398$^{*}$ & \cellcolor{tabgreen!0}0.193$^{*}$ & \cellcolor{tabgreen!0}2.311$^{*}$ & \textbf{\cellcolor{tabgreen!100}0.508$^{*}$} \\
Audio-Omni & \cellcolor{tabgreen!78}0.242 & \cellcolor{tabgreen!61}0.467$^{*}$ & \cellcolor{tabgreen!92}1.005$^{*}$ & \cellcolor{tabgreen!73}0.500 & \cellcolor{tabgreen!54}1.980 & \cellcolor{tabgreen!59}-26.489$^{*}$ & \cellcolor{tabgreen!73}1.627$^{*}$ & \cellcolor{tabgreen!85}3.046$^{*}$ & \cellcolor{tabgreen!30}0.369$^{\circ}$ & \cellcolor{tabgreen!26}7.119 & \cellcolor{tabgreen!61}0.610 \\
\midrule
\textbf{Extract} & CLAP-T $\uparrow$ & CLAP-A in-context $\uparrow$ & KL $\downarrow$ & FAD $\downarrow$ & KAD $\downarrow$ & SI-SDR\bothm{} $\uparrow$ & mel-L1\bothm{} $\downarrow$ & MRSTFT\bothm{} $\downarrow$ & edit rate\editm{} $\uparrow$ & edit strength\editm{} $\uparrow$ & Audio Feat.\editm{} $\downarrow$ \\
\midrule
ProCraft (ours) & \cellcolor{tabgreen!86}0.273 & \cellcolor{tabgreen!71}0.454 & \cellcolor{tabgreen!98}0.742 & \cellcolor{tabgreen!73}0.551 & \cellcolor{tabgreen!75}1.869 & \textbf{\cellcolor{tabgreen!100}-19.038} & \cellcolor{tabgreen!94}1.764 & \cellcolor{tabgreen!97}3.328 & \cellcolor{tabgreen!97}0.767 & \cellcolor{tabgreen!90}8.590 & \cellcolor{tabgreen!38}0.810 \\
ProCraft-plain & \textbf{\cellcolor{tabgreen!100}0.291$^{*}$} & \cellcolor{tabgreen!70}0.452 & \textbf{\cellcolor{tabgreen!100}0.702} & \cellcolor{tabgreen!76}0.539 & \cellcolor{tabgreen!75}1.880 & \cellcolor{tabgreen!89}-21.592$^{*}$ & \cellcolor{tabgreen!92}1.783 & \cellcolor{tabgreen!98}3.305 & \textbf{\cellcolor{tabgreen!100}0.787} & \textbf{\cellcolor{tabgreen!100}10.884$^{\circ}$} & \cellcolor{tabgreen!48}0.776 \\
IMG (retrained) & \cellcolor{tabgreen!0}0.157$^{*}$ & \cellcolor{tabgreen!39}0.303$^{*}$ & \cellcolor{tabgreen!43}1.699$^{*}$ & \cellcolor{tabgreen!15}0.779 & \cellcolor{tabgreen!0}3.167 & \cellcolor{tabgreen!42}-32.572$^{*}$ & \cellcolor{tabgreen!37}2.507$^{*}$ & \cellcolor{tabgreen!66}4.379$^{*}$ & \cellcolor{tabgreen!0}0.040$^{*}$ & \cellcolor{tabgreen!0}-12.499$^{*}$ & \cellcolor{tabgreen!0}0.949$^{*}$ \\
ZETA & \cellcolor{tabgreen!64}0.243 & \cellcolor{tabgreen!9}0.162$^{*}$ & \cellcolor{tabgreen!0}2.445$^{*}$ & \cellcolor{tabgreen!6}0.815 & \cellcolor{tabgreen!45}2.392 & \cellcolor{tabgreen!13}-39.465$^{*}$ & \cellcolor{tabgreen!17}2.779$^{*}$ & \cellcolor{tabgreen!0}6.568$^{*}$ & \cellcolor{tabgreen!55}0.453$^{*}$ & \cellcolor{tabgreen!54}0.130$^{*}$ & \cellcolor{tabgreen!45}0.785 \\
DDIM inv. & \cellcolor{tabgreen!30}0.196$^{*}$ & \cellcolor{tabgreen!0}0.117$^{*}$ & \cellcolor{tabgreen!3}2.386$^{*}$ & \cellcolor{tabgreen!0}0.837 & \cellcolor{tabgreen!36}2.538 & \cellcolor{tabgreen!0}-42.502$^{*}$ & \cellcolor{tabgreen!0}2.996$^{*}$ & \cellcolor{tabgreen!0}6.556$^{*}$ & \cellcolor{tabgreen!58}0.473$^{*}$ & \cellcolor{tabgreen!52}-0.411$^{*}$ & \cellcolor{tabgreen!48}0.774 \\
SteerMusic & \cellcolor{tabgreen!90}0.278 & \cellcolor{tabgreen!80}0.501$^{*}$ & \cellcolor{tabgreen!78}1.083$^{*}$ & \textbf{\cellcolor{tabgreen!100}0.447} & \textbf{\cellcolor{tabgreen!100}1.440} & \cellcolor{tabgreen!54}-29.762$^{*}$ & \cellcolor{tabgreen!86}1.864 & \cellcolor{tabgreen!35}5.413$^{*}$ & \cellcolor{tabgreen!65}0.527$^{*}$ & \cellcolor{tabgreen!63}2.282$^{*}$ & \textbf{\cellcolor{tabgreen!100}0.587$^{*}$} \\
Audio-Omni & \cellcolor{tabgreen!66}0.245$^{*}$ & \textbf{\cellcolor{tabgreen!100}0.595$^{*}$} & \cellcolor{tabgreen!99}0.718 & \cellcolor{tabgreen!81}0.522 & \cellcolor{tabgreen!81}1.765 & \cellcolor{tabgreen!73}-25.364$^{*}$ & \textbf{\cellcolor{tabgreen!100}1.681} & \textbf{\cellcolor{tabgreen!100}3.227} & \cellcolor{tabgreen!93}0.733 & \cellcolor{tabgreen!81}6.426 & \cellcolor{tabgreen!56}0.747 \\
\midrule
\textbf{Denoise} & CLAP-T $\uparrow$ & CLAP-A in-context $\uparrow$ & KL $\downarrow$ & FAD $\downarrow$ & KAD $\downarrow$ & SI-SDR\bothm{} $\uparrow$ & mel-L1\bothm{} $\downarrow$ & MRSTFT\bothm{} $\downarrow$ & Audio Feat.\editm{} $\downarrow$ &  &  \\
\midrule
ProCraft (ours) & \cellcolor{tabgreen!15}0.081 & \cellcolor{tabgreen!96}0.502 & \textbf{\cellcolor{tabgreen!100}0.815} & \cellcolor{tabgreen!94}0.435 & \cellcolor{tabgreen!95}1.458 & \textbf{\cellcolor{tabgreen!100}-10.141} & \cellcolor{tabgreen!98}1.255 & \cellcolor{tabgreen!99}2.587 & \cellcolor{tabgreen!77}0.793 &  &  \\
ProCraft-plain & \cellcolor{tabgreen!15}0.081 & \textbf{\cellcolor{tabgreen!100}0.521} & \cellcolor{tabgreen!99}0.854 & \textbf{\cellcolor{tabgreen!100}0.403} & \textbf{\cellcolor{tabgreen!100}1.352} & \cellcolor{tabgreen!99}-10.370 & \textbf{\cellcolor{tabgreen!100}1.199} & \textbf{\cellcolor{tabgreen!100}2.524} & \cellcolor{tabgreen!80}0.778 &  &  \\
IMG (retrained) & \cellcolor{tabgreen!0}0.056$^{\circ}$ & \cellcolor{tabgreen!79}0.409$^{*}$ & \cellcolor{tabgreen!88}1.130$^{*}$ & \cellcolor{tabgreen!73}0.557 & \cellcolor{tabgreen!55}2.243 & \cellcolor{tabgreen!85}-16.107$^{*}$ & \cellcolor{tabgreen!91}1.462$^{*}$ & \cellcolor{tabgreen!93}2.939$^{*}$ & \cellcolor{tabgreen!94}0.695$^{\circ}$ &  &  \\
ZETA & \cellcolor{tabgreen!72}0.175$^{*}$ & \cellcolor{tabgreen!9}0.047$^{*}$ & \cellcolor{tabgreen!0}3.430$^{*}$ & \cellcolor{tabgreen!0}0.968 & \cellcolor{tabgreen!18}2.995 & \cellcolor{tabgreen!0}-50.297$^{*}$ & \cellcolor{tabgreen!0}4.169$^{*}$ & \cellcolor{tabgreen!0}8.783$^{*}$ & \cellcolor{tabgreen!0}1.236$^{*}$ &  &  \\
DDIM inv. & \cellcolor{tabgreen!55}0.147$^{*}$ & \cellcolor{tabgreen!0}-0.001$^{*}$ & \cellcolor{tabgreen!30}2.640$^{*}$ & \cellcolor{tabgreen!1}0.963 & \cellcolor{tabgreen!0}3.345 & \cellcolor{tabgreen!1}-49.838$^{*}$ & \cellcolor{tabgreen!30}3.291$^{*}$ & \cellcolor{tabgreen!31}6.866$^{*}$ & \cellcolor{tabgreen!60}0.893 &  &  \\
SteerMusic & \textbf{\cellcolor{tabgreen!100}0.222$^{*}$} & \cellcolor{tabgreen!79}0.409$^{*}$ & \cellcolor{tabgreen!84}1.241$^{*}$ & \cellcolor{tabgreen!79}0.520 & \cellcolor{tabgreen!83}1.698 & \cellcolor{tabgreen!43}-32.832$^{*}$ & \cellcolor{tabgreen!79}1.817$^{*}$ & \cellcolor{tabgreen!52}5.531$^{*}$ & \textbf{\cellcolor{tabgreen!100}0.663$^{*}$} &  &  \\
Audio-Omni & \cellcolor{tabgreen!63}0.160$^{*}$ & \cellcolor{tabgreen!78}0.408$^{*}$ & \cellcolor{tabgreen!86}1.188$^{*}$ & \cellcolor{tabgreen!56}0.651 & \cellcolor{tabgreen!38}2.596 & \cellcolor{tabgreen!53}-28.934$^{*}$ & \cellcolor{tabgreen!76}1.913$^{*}$ & \cellcolor{tabgreen!84}3.520$^{*}$ & \cellcolor{tabgreen!73}0.819 &  &  \\
\bottomrule
\end{tabular}}
\end{table*}

\clearpage

\section{Training and inference settings}
\label{app:training}

We train for 20{,}000 steps with a 1{,}000-step linear warmup into cosine decay.
The optimizer is AdamW with betas 0.9 and 0.95, weight decay 0.01, and gradient clipping at 1.0.
Peak learning rates are $5{\times}10^{-5}$ for the backbone and $3{\times}10^{-4}$ for projectors and adapters, with an effective batch of 16 sequences in bf16 and uniform instruction dropout of 0.15 as the classifier-free-guidance null signal.
At inference the model encodes the instruction, source codes, and query embeddings once, then samples the target at temperature 1.0 with CFG scale = 1.5.

\section{Full metric definitions}
\label{app:metrics}

We provide the exact definitions and implementation details for all metrics reported in the main tables. The notation follows Sections~\ref{sec:method} and~\ref{sec:metrics}: $x$ denotes the source clip, $c$ the instruction, $\hat{y}$ the system output, and $y$ the target when one is available.

\textbf{AFx-Rep cosine} (Editing, audio effects).
We compute Audio Effects Representation (AFx-Rep) cosine similarity using the representation released with ST-ITO \citep{steinmetz2024stito}. Let $f_{\mathrm{AFx}}(\cdot)$ denote its encoder. We compute
\[ \mathrm{AFxRep}(\hat{y},y)=\frac{f_{\mathrm{AFx}}(\hat{y})^\top f_{\mathrm{AFx}}(y)}{\|f_{\mathrm{AFx}}(\hat{y})\|_2\|f_{\mathrm{AFx}}(y)\|_2}, \]
where $y$ is the ground-truth signal processed with the requested effect chain. Because AFx-Rep is trained to characterize the applied effect chain rather than musical content, the cosine measures whether the requested treatment is present in the output.

\textbf{Audio-feature distance} (Editing).
We compare seven audio features of the output and target, adapted from the audio-feature loss of Diff-MST \citep{vanka2024diffmst}.
That loss compares RMS level, crest factor, stereo width, stereo imbalance, and a Bark spectrum.
Because our audio is mono, we keep RMS level and crest factor and drop the two stereo features.
We replace the Bark spectrum with five scalar spectral features: centroid, bandwidth, 85\% rolloff, flatness, and contrast.
We compute all seven features on the mono signal from a 2048-point STFT with a hop of 512 samples. RMS level, crest factor, and spectral contrast are measured in dB. Centroid, bandwidth, and rolloff are measured in octaves above 20\,Hz. Let $f_k(z)$ denote feature $k$ of signal $z$ and $a_k$ its scaling constant. The distance is
\[ D_{\mathrm{feat}}(\hat{y},y)=\frac{1}{7}\sum_{k=1}^{7}\frac{|f_k(\hat{y})-f_k(y)|}{a_k}. \]
The equal weights and the scaling constants differ from the loss weights of Diff-MST. We set $a_k=6$\,dB for the three features in dB and $a_k=1$ octave for centroid, bandwidth, and rolloff. For spectral flatness, $a_k=0.15$. These values give the seven terms comparable magnitude.
Both signals are RMS-normalized before feature extraction, so the RMS-level term is close to zero. The metric measures whether the requested audio effects characteristics were reached, providing better observation than waveform-level reference metrics, which may be overly sensitive to content differences.

\textbf{CLAP-A vs.\ source.}
We compute Contrastive Language--Audio Pretraining audio--audio similarity (CLAP-A) between the output and source:
\[ \mathrm{CLAP\text{-}A}(\hat{y},x)=\frac{f_{\mathrm{audio}}(\hat{y})^\top f_{\mathrm{audio}}(x)}{\|f_{\mathrm{audio}}(\hat{y})\|_2\|f_{\mathrm{audio}}(x)\|_2}. \]

\textbf{CLAP-T out (pred vs.\ instr).}
We use Contrastive Language--Audio Pretraining text--audio similarity (CLAP-T) \citep{wu2023clap} to measure output-to-instruction adherence. Specifically,
\[ \mathrm{CLAP\text{-}T}(\hat{y},c)=\frac{f_{\mathrm{audio}}(\hat{y})^\top f_{\mathrm{text}}(c)}{\|f_{\mathrm{audio}}(\hat{y})\|_2\|f_{\mathrm{text}}(c)\|_2}. \]

\textbf{Edit rate and edit strength} (Editing; remove and extract).
We first separate each signal into instrument stems using HTDemucs-6s \citep{rouard2023htdemucs}. Let
\[ \mathcal{E}_t(z)=\sqrt{\frac{1}{T}\sum_{n=1}^{T}s_t(z)[n]^2} \]
denote the RMS energy of target-instrument stem $t$ in signal $z$, where $s_t(z)$ is the corresponding HTDemucs-6s stem. Remove and extract strength are respectively
\[ S_{\mathrm{remove}}=10\log_{10}\frac{\mathcal{E}_t(x)}{\mathcal{E}_t(\hat{y})}, \qquad S_{\mathrm{extract}}=10\log_{10}\frac{\mathcal{E}_t(\hat{y})}{\sum_{s\neq t}\mathcal{E}_s(\hat{y})}, \]
with both clipped to $[-30,30]$\,dB. Edit rate thresholds the corresponding quantities at the example level:
\[ \mathrm{ER}_{\mathrm{remove}}=\mathbb{1}\!\left[\mathcal{E}_t(\hat{y})\leq\tfrac{1}{2}\mathcal{E}_t(x)\right], \qquad \mathrm{ER}_{\mathrm{extract}}=\mathbb{1}\!\left[\mathcal{E}_t(\hat{y})>\sum_{s\neq t}\mathcal{E}_s(\hat{y})\right]. \]
The reported edit rate is the mean of these binary indicators over examples. These metrics ask whether the requested stem actually changed rather than how closely the output matches a particular reference. Because HTDemucs-6s supplies the stems used by the metric, it acts as the judge and is not scored on these rows.

\textbf{FAD/KAD} (output naturalness).
We compute Fr\'echet Audio Distance (FAD) \citep{kilgour2019fad} and Kernel Audio Distance (KAD) \citep{chung2025kad} on CLAP-LAION-music embeddings against the same fixed real-music reference pool for every system, following \citet{gui2024fadtk}. Let $(\mu_P,\Sigma_P)$ and $(\mu_Q,\Sigma_Q)$ denote the Gaussian statistics of the output and reference embeddings, respectively. FAD is
\[ \mathrm{FAD}=\|\mu_P-\mu_Q\|_2^2+\operatorname{Tr}\!\left(\Sigma_P+\Sigma_Q-2(\Sigma_P\Sigma_Q)^{1/2}\right). \]
KAD uses the kernel-based maximum mean discrepancy estimator of \citet{chung2025kad}. Lower FAD and KAD indicate distributions closer to the real-music reference pool. We use the same fixed reference pool for every system.

\textbf{Frame-F1} (Preservation, instrument transfer).
We compute frame-level pitch F1 (Frame-F1) between the output's MuScriptor transcription and the source-track Musical Instrument Digital Interface (MIDI) ground truth at 75 frames per second. We accumulate true positives (TP), false positives (FP), and false negatives (FN) over frames and compute
\[ P=\frac{\mathrm{TP}}{\mathrm{TP}+\mathrm{FP}}, \qquad R=\frac{\mathrm{TP}}{\mathrm{TP}+\mathrm{FN}}, \qquad \mathrm{Frame\text{-}F1}=\frac{2PR}{P+R}. \]
We use Frame-F1 only for instrument transfer, where content is preserved by construction, and only on Slakh, since the real-domain evaluation set has no MIDI ground-truth.

\textbf{Isolation strength} (Editing; combined benchmark, accompany, cover and instrument transfer).
Using the RMS stem energy $\mathcal{E}_t(\cdot)$ defined above, isolation strength applies the extract form of edit strength to the output:
\[ S_{\mathrm{iso}}=10\log_{10}\frac{\mathcal{E}_t(\hat{y})}{\sum_{s\neq t}\mathcal{E}_s(\hat{y})}. \]
Higher values indicate that the requested instrument accounts for more of the separated output energy relative to other instruments. The metric is computed from separated stems using HTDemucs-6s as the separator; consequently, HTDemucs-6s itself is not scored on this metric.

\textbf{KL} (output naturalness).
We compute the Kullback--Leibler (KL) divergence between the PaSST \citep{koutini2022passt} audio-tagging posterior distributions of the output and target. If $p_y$ and $p_{\hat{y}}$ denote their posterior vectors, respectively,
\[ D_{\mathrm{KL}}(p_y\|p_{\hat{y}})=\sum_k p_y(k)\log\frac{p_y(k)}{p_{\hat{y}}(k)}. \]
Lower KL indicates that the output induces an audio-tag distribution closer to that of the target.

\textbf{Melody chroma accuracy} (Preservation; cover and instrument transfer).
Following the cover-generation evaluation of \citet{tan2024picogen}, we use melody chroma accuracy defined in \texttt{mir\_eval} \citep{raffel2014mireval} to compare the estimated melody of the output with that of the source frame by frame. For frames in which both estimates are voiced, a prediction is counted as correct when the estimated pitches agree modulo octave within the chroma tolerance used by \texttt{mir\_eval}. Melody chroma accuracy is the fraction of reference voiced frames satisfying this criterion. It therefore measures whether the re-voiced output follows the source melody irrespective of octave. An unchanged copy of the input scores highly by construction, so we interpret this metric together with the Editing metrics rather than in isolation.

\textbf{PaSST instrument match} (Editing).
PaSST \citep{koutini2022passt} predicts AudioSet logits over instrument-related classes. We map the highest-scoring class to our instrument families and report the classification accuracy against the instrument family requested by the instruction. We skip systems for which the judge's own separation determines the classification result, as well as systems that always emit a single fixed instrument family.

\textbf{SI-SDR, mel-L1 and MRSTFT vs.\ ground-truth} (Editing and Preservation).
We compute all three metrics against deterministic ground-truth targets. Scale-Invariant Signal-to-Distortion Ratio (SI-SDR) is used for remove, extract, and denoise, while mel-spectrogram L1 distance (mel-L1) and Multi-Resolution Short-Time Fourier Transform loss (MRSTFT) additionally evaluate audio effects against the processed target. For zero-mean reference $y$ and estimate $\hat{y}$, SI-SDR is
\[ \mathrm{SI\text{-}SDR}(y,\hat{y})=10\log_{10}\frac{\|\alpha y\|_2^2}{\|\hat{y}-\alpha y\|_2^2}, \qquad \alpha=\frac{\langle\hat{y},y\rangle}{\|y\|_2^2}. \]
Let $\mathbf{M}(z)$ denote the log-mel spectrogram of signal $z$. The mel-L1 distance is
\[ \mathcal{L}_{\mathrm{mel}}=\frac{1}{N}\left\|\mathbf{M}(\hat{y})-\mathbf{M}(y)\right\|_1, \]
where $N$ is the number of time--frequency bins. For each Short-Time Fourier Transform (STFT) resolution $r$, with magnitude spectrogram $\mathbf{S}_r(z)$, we compute spectral convergence and log-magnitude error as
\[ \mathcal{L}_{\mathrm{SC}}^{(r)}=\frac{\|\mathbf{S}_r(y)-\mathbf{S}_r(\hat{y})\|_F}{\|\mathbf{S}_r(y)\|_F}, \qquad \mathcal{L}_{\log}^{(r)}=\frac{1}{N_r}\left\|\log(\mathbf{S}_r(y)+\epsilon)-\log(\mathbf{S}_r(\hat{y})+\epsilon)\right\|_1. \]
MRSTFT is
\[ \mathcal{L}_{\mathrm{MRSTFT}}=\sum_{r\in\{2048,1024,512,256\}}\left(\mathcal{L}_{\mathrm{SC}}^{(r)}+\mathcal{L}_{\log}^{(r)}\right). \]
All three metrics are computed on mono audio at 48\,kHz after RMS level matching the system output to the target.

\textbf{StructDist and APA} (Preservation; accompany and cover).
Structural distance (StructDist) compares the temporal self-similarity structure of the output and source. Let $\mathbf{H}(z)\in\mathbb{R}^{T\times D}$ denote frame-level MERT representations of signal $z$, with self-similarity matrix
\[ \mathbf{A}(z)_{ij}=\cos\!\left(\mathbf{H}(z)_i,\mathbf{H}(z)_j\right). \]
StructDist is the RMS distance between the two self-similarity matrices:
\[ \mathrm{StructDist}(\hat{y},x)=\sqrt{\frac{1}{T^2}\left\|\mathbf{A}(\hat{y})-\mathbf{A}(x)\right\|_F^2}. \]
It is computed directly on the generated output to measure the structure distance between the output and the source.

Accompaniment Prompt Adherence (APA) \citep{grachten2025apa} measures how well generated accompaniments fit their source-audio prompts at the distribution level. Let $C$ be the candidate set of context--stem pairs, $R$ a reference set of matched context--stem pairs, and $R'$ a reference set with randomly mismatched stems. APA is defined from their Fr\'echet Audio Distances (FADs) as
\[ \mathrm{APA}=\frac{1}{2}+\frac{\mathrm{FAD}_{C,R'}-\mathrm{FAD}_{C,R}}{2\,\mathrm{FAD}_{R,R'}}, \]
and is clipped to $[0,1]$. A score of $0.5$ indicates that the candidate distribution is equally distant from the matched and mismatched references, while higher values indicate stronger prompt adherence. We apply APA to accompany and cover.

\section{Data pipeline details}
\label{app:data}

\textbf{Corpus and instrument families.}
The training corpus is Slakh2100 multitrack stems at the standard splits, with stems summed into General MIDI families.
Training uses eight families.
Four are the rhythm section and keys: bass and drums, guitar and piano, and the other four are sustained voices: organ and brass, reed and ensemble.
Every edit targets one family.
Input and target windows are 30 seconds at 48\,kHz on disk, downmixed to 24\,kHz mono for the codec.
A window is randomly selected from each audio sample and will be resampled if any target stem is more than 70\% near-silent or peaks below 0.10. We skip an audio sample if there's no valid window in 10 attempts.

\textbf{Effect authoring and validation.}
We build two things through one LLM call: prose a producer might say, and the audio effects settings that realize it.
The prompt offers 3 to 5 audio effect families sampled from the library, and one of the target mixing styles to aim for. Each of them is drawn from a predefined pool.
Sampling the style per item keeps the corpus from collapsing onto a few favored effects, and rotating the examples keeps the model from copying them back.
Chains then execute on the stems, with stereo and spatial effects banned because the evaluation is mono, and each chain's output is RMS-matched to its input so the volume doesn't change while effects are applied.

\textbf{Cover targets.}
The cover reduction collects pitched notes from all non-drum stems, applies salience-based deletion with octave reduction, and renders onto one wide-range target instrument, including piano, guitar, organ, and ensemble.

\textbf{Instrument transfer and event tasks.}
Instrument-transfer items render one pitched track's MIDI from the source and a target instrument sound font, so the pitch is unchanged, and only instrument identity differs.
Denoise items mix one environmental sound into the music, drawn from FSD50K in training and from ESC-50 in the test sets. An ambient sound such as rain or wind is looped to fill the clip, at 9 to 18 dB below the music. A transient sound such as a dog bark or car horn is repeated 3 to 6 times at random positions across the clip, at 3 to 12 dB below the music.

\section{Mixing motivations and example instructions}
\label{app:styles}

\subsection{Mixing motivations}
Each item that carries audio effects samples one of the 30 mixing motivations below uniformly at random. The language model receives it as the aesthetic that the audio effects and the abstract description should aim for; the arrangement part of the instruction stays literal. Items without audio effects do not use a motivation.

\begin{table}[h]
\centering
\small
\begin{tabular}{@{}lll@{}}
\toprule
90s grunge & folk back-porch & reggae dub \\
ambient and dreamy & gospel warmth & shoegaze wall-of-sound \\
arena rock & gritty electronic & smoky jazz club \\
boom-bap hip-hop & industrial cold & spaghetti-western twang \\
bossa nova café & lo-fi bedroom & surf rock \\
chillwave sunset & modern radio pop & synthwave neon \\
cinematic and widescreen & motown groove & trap night-drive \\
crisp and hi-fi & orchestral film score & vintage soul \\
drum-and-bass rollers & psychedelic haze & warm analog tape \\
dubby and cavernous & raw garage punk & witch-house murk \\
\bottomrule
\end{tabular}
\caption{The 30 mixing motivations, in alphabetical order.}
\label{tab:styles}
\end{table}

\subsection{Example instructions}
\label{app:instruction-examples}

Table~\ref{tab:instruction-examples} shows instructions from our demo, covering all seven operations and requests that combine an operation with audio effects. Source audio, model outputs, and baseline comparisons are available on the public demo page at \url{https://barry-mir.github.io/procraft-demo/}.

\begin{table}[ht]
\centering
\small
\setlength{\tabcolsep}{4pt}
\renewcommand{\arraystretch}{1.15}
\begin{tabular}{@{}p{0.20\linewidth}p{0.76\linewidth}@{}}
\toprule
\textbf{Task} & \textbf{Instruction} \\
\midrule
\multicolumn{2}{@{}l}{\textit{Single operations}} \\
Remove & Cut the bass from the mix \\
Extract & Extract only the guitar track. \\
Accompany & Generate a piano part that complements the track \\
Cover & Arrange entire track as solo piano performance \\
Instrument transfer & Replace guitar with bass using same notes. \\
Denoise & Eliminate all non-music sounds. \\
Audio effects & Add rich tube saturation and hall reverb to the piano. \\
\midrule
\multicolumn{2}{@{}l}{\textit{Combined editing and audio effects}} \\
Remove & Remove the bass. Add pumping compression to the piano, airy high-pass filter to the drums, and warm saturation to the guitar. \\
Extract & Extract the bass. Add punchy compression with a subtle saturation to the bass. \\
Accompany & Create a standalone piano part that complements the track. Add gritty bitcrush with analog noise. \\
Cover & Reconfigure full track for solo piano performance. Put on tamed low-end, boosted mid-range, tight delay. \\
Instrument transfer & Swap bass for piano, keep notes identical. Add gritty lo-fi bit crush with noise floor. \\
Denoise & Eliminate the non-musical rain audio. Apply subtle bitcrush haze, soft saturation, gentle noise floor. \\
\bottomrule
\end{tabular}
\caption{Example free-form instructions from the demo page.}
\label{tab:instruction-examples}
\end{table}

\section{Functionality coverage judgments}
\label{app:functionality}

Table~\ref{tab:functionality} distinguishes native support from looser uses of each system; the following judgments explain its cells.
A model is able (\fyes{}) on a functionality when that functionality is a native and reliable mode.
It is partial (\fpart{}) when the capability is indirect, out of distribution, quality-limited, or a loose application.
It is unable (\fno{}) when the model has no such mode.

\textbf{Separators.} AudioSep isolates a described source, so extract is native and remove is partial, needing a query-then-subtract step that empirically guts dense music mixes.
Denoise is partial because AudioSep targets environmental events rather than the complete editing task, and free phrasing is partial because it accepts target names rather than edit instructions.
SD-Codec routes by domain rather than instrument: it can denoise by decoding the music stream, but extracting that whole stream is only a partial match to instrument extraction, and it cannot remove a named instrument.

\textbf{Stem imputers.} MSDM and MusicGen-Stem generate a stem consistent with the others, so accompany is native.
MSDM has no text conditioning, so its free phrasing is unable.
MusicGen-Stem is a text-to-music model whose edits are driven by stem conditioning, so its free phrasing is partial.

\textbf{Cover and instrument transfer systems.} Pop2Piano and PiCoGen2 generate solo-piano covers, so Table~\ref{tab:functionality} marks cover as native and other operations as unsupported; neither takes a free-text instruction.
AP-Adapter supports instrument transfer, cover, and accompaniment, but its timbre edit changes the whole clip rather than a named part, so the corresponding stem-level tasks are partial matches even though it accepts free-form target text.
MuseControlLite conditions on melody and reference audio to transfer a whole clip's instrument; we therefore mark instrument transfer as supported and accompany and cover as partial.

\textbf{Instruction and inversion editors.} Instruct-MusicGen covers remove, extract, and accompany from a fixed Title-case template, so those three are able. It's not trained on cover, instrument transfer, or denoise.
Its templated prompt is not freely phrased text and it has no audio effect vocabulary, so those cells are unable.
Audio-Omni \citep{tian2026audioomni} is a freely phrased instruction editor; in our listening test it is competitive on accompany, cover, and instrument transfer. It falls far behind on remove and denoise, and audio effects are unsupported.
ZETA edits from a target caption, so free phrasing is able.
Instrument transfer is able through the target caption, cover is partial, remove is a partial coarse nudge, and audio effects are a partial non-parametric text request.
Surgical extract, accompany, and denoise are unable.

LLM2Fx-Tools maps language to executable effect chains but does not perform any other edits.
ProCraft covers every column because it reads free-form instructions and performs each evaluated edit including audio effects.

\section{Edit-head probing protocol}
\label{app:probe}

\textbf{Features.} For each item, we run ProCraft once on the input sequence, made of the instruction tokens, the WavTokenizer codes of the 30,s source clip, and the 32 learnable query embeddings. We take the edit head $h_{\mathrm{edit}}$, the last hidden state at the 32 query positions, and average it over those positions into one 896-dimensional vector.

\textbf{Classifier.} We use an MLP with one hidden layer of 128 units, trained on standardized inputs for at most 2000 iterations. Accuracy is computed over the pooled out-of-fold predictions of 5-fold stratified cross-validation, with shuffled folds and a fixed seed. Chance is the frequency of the most common class. Classes with fewer than three items would be dropped, but none of the four probes below has one.

\textbf{Operation and target instrument.} We build 350 edits on the Slakh test split with the pipeline of Section~\ref{sec:data}, using all eight instrument families. The set has 50 edits for each of the seven tasks: remove, extract, accompany, cover, instrument transfer, denoise, and audio effects. The operation probe predicts the task of all 350 edits (7 classes, chance 0.143).
The target-instrument probe uses the 299 edits whose instruction names an instrument, which excludes the 50 denoise edits and one audio effect edit.
Its label is the family that is removed, extracted or added, the cover instrument, the destination of an instrument transfer, or the processed track of an audio effect edit. The eight classes hold between 18 and 49 items, and chance is 0.164.

\textbf{Instrument of a single stem.} We cut 400 clips from isolated Slakh stems, 50 per family (chance 0.125). Each clip is the loudest 30,s window of its stem, searched with a 5,s hop and required to be active in at least 80\% of its seconds. We take at most two stems per track and family, drawing 352 clips from the test split and 48 from the validation split where a family ran short.

\textbf{Pitch of a single note.} We render 343 clips with FluidSynth and the FluidR3 General MIDI soundfont: one for each of the seven pitched families and each MIDI pitch from 36 to 84 (C2 to C6). This gives 49 classes with 7 clips each (chance 0.020). Each clip is 30,s long and repeats one note at velocity 96, sounding for 1.5,s every 2,s, on the family's General MIDI program.

\textbf{Baselines.} The baselines train the same classifier on the mean of the checkpoint's input embeddings for the same items. The two instruction properties use the embeddings of the non-padding instruction tokens, and the two audio properties use the embeddings of the source codec tokens.

\section{Listening test protocol}
\label{app:listening}

The test follows the MUSHRA layout \citep{itu2015mushra} without anchors, since no gold output exists for the generative tasks.
We selected three source-recording-and-instruction pairs per task from the real combined benchmark.
For those eighteen pairs, we collected the outputs of ProCraft, ZETA, DDIM inversion, SteerMusic and Audio-Omni.
We trim every clip to 30\,s and loudness-normalise it to $-23$\,LUFS with a $-1$\,dBTP limit, then fade it over 10\,ms and encode it at 192\,kbps.
Each rater receives one randomly chosen recording-and-instruction pair per task and completes the six trials in random order.
In each trial the rater sees the five stimuli labelled A to E in a random permutation and rates them on three attributes in turn.
Editing asks whether the change the instruction names was made, arrangement and audio effects; Preservation asks whether everything the instruction did not name is intact; audio quality asks for freedom from artifacts, independent of correctness.
Task-specific guidance under each question states what Editing and Preservation mean for that task.
The reference is the source recording with the instruction, and raters must start every clip before advancing; the player switches between stimuli at the same playhead position.
Raters answered a questionnaire on musical training and production experience; the sixteen raters reported between no and more than ten years of training, fourteen of them three years or more, and seven produce music as a hobby.
Table~\ref{tab:listening} lists every cell mean behind Figure~\ref{fig:listening}.
\begin{table}[H]
\centering
\footnotesize
\setlength{\tabcolsep}{4pt}
\caption{Listening test, mean rating (0 to 5) per task, attribute and system over 16 raters. \textbf{Bold} = highest sample mean; \underline{underline} = second-highest sample mean; significance is reported in Table~\ref{tab:listening-pairwise}.}
\label{tab:listening}
\begin{tabular}{@{}llrrrrr@{}}
\toprule
Task & Attribute & ProCraft (ours) & Audio-Omni & SteerMusic & ZETA & DDIM inv. \\
\midrule
Remove & Editing & \textbf{2.50} & 1.69 & \underline{2.00} & 0.81 & 0.81 \\
 & Preservation & \textbf{4.06} & 1.38 & \underline{2.88} & 0.44 & 0.50 \\
 & Audio quality & \textbf{3.56} & 0.69 & \underline{2.12} & 1.25 & 1.38 \\
\midrule
Extract & Editing & \underline{3.12} & \textbf{3.25} & 2.38 & 0.56 & 0.06 \\
 & Preservation & \underline{3.44} & \textbf{3.69} & 3.12 & 0.31 & 0.12 \\
 & Audio quality & \underline{3.06} & \textbf{3.69} & 2.75 & 1.31 & 0.31 \\
\midrule
Accompany & Editing & 2.19 & \textbf{2.31} & 1.88 & \underline{2.25} & 2.12 \\
 & Preservation & 1.88 & \textbf{3.81} & \underline{2.69} & 0.44 & 0.44 \\
 & Audio quality & 2.38 & \textbf{3.19} & 1.81 & 2.25 & \underline{2.44} \\
\midrule
Cover & Editing & 2.06 & \textbf{3.19} & \underline{2.81} & 2.06 & 1.44 \\
 & Preservation & 1.31 & \underline{3.06} & \textbf{3.38} & 1.19 & 0.75 \\
 & Audio quality & 1.81 & \textbf{2.62} & \underline{2.50} & 2.44 & 1.50 \\
\midrule
Instrument transfer & Editing & \textbf{3.12} & \underline{3.06} & 1.12 & 1.56 & 0.31 \\
 & Preservation & \underline{3.38} & \textbf{3.50} & 2.00 & 1.44 & 0.31 \\
 & Audio quality & \underline{3.25} & \textbf{3.62} & 2.25 & 2.44 & 0.94 \\
\midrule
Denoise & Editing & \textbf{3.56} & \underline{2.00} & 0.69 & 0.00 & 0.69 \\
 & Preservation & \textbf{3.88} & 2.19 & \underline{2.62} & 0.06 & 0.19 \\
 & Audio quality & \textbf{3.75} & \underline{2.06} & 1.88 & 0.62 & 0.56 \\
\bottomrule
\end{tabular}
\end{table}

\textbf{Statistical tests.}
Because each rater evaluated all five systems on all six tasks and three attributes, we fit a repeated-measures ANOVA separately for each task, with system and attribute as within-rater factors (Table~\ref{tab:listening-anova}).
We report partial eta squared for each effect.
Every task has a significant system-by-attribute interaction, so we follow each ANOVA with paired $t$-tests comparing ProCraft with each of the four alternatives on each attribute.
Holm's step-down procedure corrects the twelve comparisons within each task; Table~\ref{tab:listening-pairwise} lists all comparisons, including those that are not significant, and the analysis output retains their confidence intervals.
\begin{table}[H]
\centering
\small
\setlength{\tabcolsep}{4pt}
\caption{Repeated-measures ANOVA on the listening ratings, one per task, with system (5 levels) and attribute (3 levels) as within-rater factors over 16 raters. $\eta_p^2$ = partial eta squared.}
\label{tab:listening-anova}
\begin{tabular}{@{}llrrr@{}}
\toprule
Task & Effect & $F$ (df) & $p$ & $\eta_p^2$ \\
\midrule
Remove & system & 19.31 (4, 60) & $<$0.001 & 0.56 \\
 & attribute & 4.39 (2, 30) & 0.021 & 0.23 \\
 & system $\times$ attribute & 6.25 (8, 120) & $<$0.001 & 0.29 \\
\midrule
Extract & system & 62.83 (4, 60) & $<$0.001 & 0.81 \\
 & attribute & 3.06 (2, 30) & 0.062 & 0.17 \\
 & system $\times$ attribute & 2.66 (8, 120) & 0.010 & 0.15 \\
\midrule
Accompany & system & 4.92 (4, 60) & 0.002 & 0.25 \\
 & attribute & 2.46 (2, 30) & 0.103 & 0.14 \\
 & system $\times$ attribute & 11.34 (8, 120) & $<$0.001 & 0.43 \\
\midrule
Cover & system & 9.71 (4, 60) & $<$0.001 & 0.39 \\
 & attribute & 1.95 (2, 30) & 0.160 & 0.11 \\
 & system $\times$ attribute & 4.15 (8, 120) & $<$0.001 & 0.22 \\
\midrule
Instrument transfer & system & 29.75 (4, 60) & $<$0.001 & 0.66 \\
 & attribute & 10.59 (2, 30) & $<$0.001 & 0.41 \\
 & system $\times$ attribute & 2.17 (8, 120) & 0.034 & 0.13 \\
\midrule
Denoise & system & 52.35 (4, 60) & $<$0.001 & 0.78 \\
 & attribute & 5.48 (2, 30) & 0.009 & 0.27 \\
 & system $\times$ attribute & 3.98 (8, 120) & $<$0.001 & 0.21 \\
\bottomrule
\end{tabular}
\end{table}

\begingroup
\small
\setlength{\tabcolsep}{3pt}
\renewcommand{\arraystretch}{0.9}
\begin{longtable}{@{}llrrrr@{}}
\caption{All ProCraft--baseline follow-up comparisons after the task-wise ANOVAs. Each cell gives the mean rating difference (ProCraft minus baseline) and Holm-adjusted $p$; correction covers 12 comparisons per task.}\label{tab:listening-pairwise}\\
\toprule
Task & Attribute & Audio-Omni & SteerMusic & ZETA & DDIM inv. \\
\midrule
\endfirsthead
\toprule
Task & Attribute & Audio-Omni & SteerMusic & ZETA & DDIM inv. \\
\midrule
\endhead
Remove & Editing & +0.81 (0.341) & +0.50 (0.376) & +1.69 (0.047) & +1.69 (0.047) \\
 & Preservation & +2.69 ($<$0.001) & +1.19 (0.014) & +3.62 ($<$0.001) & +3.56 ($<$0.001) \\
 & Audio quality & +2.88 ($<$0.001) & +1.44 (0.005) & +2.31 (0.004) & +2.19 (0.007) \\
\midrule
Extract & Editing & -0.12 (0.791) & +0.75 (0.208) & +2.56 ($<$0.001) & +3.06 ($<$0.001) \\
 & Preservation & -0.25 (0.656) & +0.31 (0.656) & +3.12 ($<$0.001) & +3.31 ($<$0.001) \\
 & Audio quality & -0.62 (0.328) & +0.31 (0.656) & +1.75 (0.019) & +2.75 ($<$0.001) \\
\midrule
Accompany & Editing & -0.12 (1.000) & +0.31 (1.000) & -0.06 (1.000) & +0.06 (1.000) \\
 & Preservation & -1.94 (0.071) & -0.81 (1.000) & +1.44 (0.052) & +1.44 (0.052) \\
 & Audio quality & -0.81 (1.000) & +0.56 (1.000) & +0.12 (1.000) & -0.06 (1.000) \\
\midrule
Cover & Editing & -1.12 (0.090) & -0.75 (0.660) & +0.00 (1.000) & +0.62 (1.000) \\
 & Preservation & -1.75 (0.002) & -2.06 ($<$0.001) & +0.12 (1.000) & +0.56 (0.754) \\
 & Audio quality & -0.81 (0.338) & -0.69 (0.810) & -0.62 (0.819) & +0.31 (1.000) \\
\midrule
Instrument transfer & Editing & +0.06 (1.000) & +2.00 ($<$0.001) & +1.56 (0.001) & +2.81 ($<$0.001) \\
 & Preservation & -0.12 (1.000) & +1.38 (0.014) & +1.94 (0.002) & +3.06 ($<$0.001) \\
 & Audio quality & -0.38 (0.755) & +1.00 (0.090) & +0.81 (0.090) & +2.31 (0.004) \\
\midrule
Denoise & Editing & +1.56 (0.003) & +2.88 ($<$0.001) & +3.56 ($<$0.001) & +2.88 ($<$0.001) \\
 & Preservation & +1.69 (0.003) & +1.25 (0.010) & +3.81 ($<$0.001) & +3.69 ($<$0.001) \\
 & Audio quality & +1.69 (0.008) & +1.88 (0.008) & +3.12 ($<$0.001) & +3.19 ($<$0.001) \\
\bottomrule
\end{longtable}
\endgroup

\section{Baseline protocols}
\label{app:baselines}

\textbf{Instruct-MusicGen.}
The model fine-tunes a frozen MusicGen \citep{copet2023musicgen} with a LoRA \citep{hu2022lora} text-fusion module and a zero-initialized audio-fusion gate, and performs stem addition, removal, and extraction from templated instructions \citep{zhang2024instructmusicgen}.
We use the community-retrained checkpoint with MusicGen's default guidance of 3.0 and always provide its native training template.
The chunked rows use its native 5-second windows with 1-second overlap-add reconstruction, for remove and extract.
We do run it on accompany, where it returns a full mix rather than an isolated part, so we recover the added part with HTDemucs-6s before scoring and it is compared to the other systems on that separated stem.
Instrument match is left unscored for it there, since the separator rather than the model is what isolates the part.

\textbf{Instruct-MusicGen (retrained).}
We also retrain the same architecture on our corpus, following the released recipe except where the evaluation protocol requires a change.
We keep the released adapter configuration, rank-32 LoRA with $\alpha$ 64 on the cross-attention query and value projections, the cosine schedule to $10^{-6}$, and an effective batch of 32 sequences, and we train for the 20K steps ProCraft receives.
The retraining differs from the released setup in three ways.
Clips are 30 seconds instead of the released 5, since every test item in this paper is 30 seconds.
The model sees all seven tasks in one run instead of the three its recipe targets, and the corpus covers eight instrument families instead of four.
This comparison tests whether the architecture can handle the longer clips, wider instrument set and seven-task mixture used by ProCraft.

\textbf{AudioSep.}
We use its official weights and resample the text-queried 32\,kHz mask into our pipeline. For remove, we subtract the separated stem from the mix; for denoise, we query the noise and subtract it.

\textbf{MSDM and MusicGen-Stem}.
Both task-specific systems output an accompaniment stem.
MSDM runs its released Slakh source-imputation checkpoint over the four bass, drums, guitar and piano families at 22\,kHz, generating the withheld part conditioned on the others.
MusicGen-Stem runs the official \texttt{multistem} branch weights at 32\,kHz, prompted to add the requested part over the source; its stem vocabulary is bass, drums and other, so guitar and piano both fall in the last.
Both emit an isolated stem, as ProCraft does, so both can be scored on accompaniment metrics that a full-mix output would break.

\textbf{Pop2Piano and PiCoGen2.}
We use their official weights and deterministic decoding.
Both systems always emit solo piano, so an instrument-fidelity metric would score them perfectly on every item; those cells are masked.

\textbf{AP-Adapter and MuseControlLite.}
We run AP-Adapter with \texttt{ap\_scale} 0.5, 50 steps and guidance 7.5, and MuseControlLite with its defaults. We score both with per-conditioning alignment.

\textbf{DDPM-inversion}.
This zero-shot editor applies the inversion method of \citet{manor2024zeta} to a Stable Audio prior \citep{evans2024stableaudio}.
It inverts the source under the caption ``solo $\langle$source$\rangle$ playing'' and replays the inversion noise under ``solo $\langle$target$\rangle$'' to change the instrument without reference audio.

\textbf{SD-Codec} \citep{bie2024sdcodec}\textbf{.}
We use its official weights and denoise by decoding the music quantizer once at 16\,kHz, then resampling the result into our pipeline.
Its CLAP and FAD cells reflect that bandwidth limit.

\textbf{TokenSynth} \citep{kim2025tokensynth}\textbf{.}
We use TokenSynth's official weights and give it the ground-truth pianoroll plus a target-instrument exemplar taken from a different item.
It therefore reads the score and hears the timbre it is asked for, whereas ProCraft is given text alone, and it stands as a skyline rather than a like-for-like comparison.

\textbf{FxSearcher} \citep{ki2025fxsearcher} \textbf{and Text2FX} \citep{chu2024text2fx}\textbf{.}
Both baselines use their official repositories to fit effect parameters on the source against a CLAP objective. Their DSP processing preserves the untouched audio by construction.

\textbf{InstructFX2FX} \citep{yu2026instructfx2fx}\textbf{.}
Its language model reads the instruction and picks the effect chain, and its optimizer fits the parameters against a CLAP objective.
The planner is the same Qwen3 endpoint that wrote the corpus instructions.

\textbf{ZETA, DDIM inversion, and SteerMusic}.
We run these zero-shot editors over the AudioLDM2-music prior \citep{liu2024audioldm2} in their native caption-conditioned form. A source caption describes the mix, and a target caption states the arrangement and effects request.
ZETA uses edit-friendly DDPM inversion, the DDIM row uses standard DDIM inversion with regeneration, and SteerMusic uses its delta-denoising edit.

\textbf{HTDemucs-6s} \citep{rouard2023htdemucs}\textbf{.}
We run the official six-source weights once per clip, dropping the requested stem for remove and keeping it for extract.
It is also the judge that measures edit rate, edit strength and isolation strength, so those three cells are left blank for it: a system cannot be scored by itself.

\textbf{Banquet} \citep{watcharasupat2024banquet}\textbf{.}
We run the official weights with the instrument name as query, in one pass per clip at the model's native rate.
Its single-decoder separation copies samples from the input rather than generating new audio, as HTDemucs-6s does.

\textbf{Audio-Omni} \citep{tian2026audioomni}\textbf{.}
We run this free-form instruction editor with its official weights and default decoding on the same combined instructions as ProCraft. It provides a direct instruction-following comparison alongside the caption-conditioned inversion editors in Table~\ref{tab:fullcase-slakh}.

\end{document}